\documentclass{aa}
\usepackage{graphicx}
\usepackage[varg]{txfonts}
\usepackage{natbib,twoopt}
\usepackage[breaklinks=true]{hyperref} 
\makeatletter
\g@addto@macro{\UrlBreaks}{\UrlOrds}
\makeatother

\bibpunct{(}{)}{;}{a}{}{,} 
\makeatletter
\newcommandtwoopt{\citeads}[3][][]{\href{http://adsabs.harvard.edu/abs/#3}%
{\def\hyper@linkstart##1##2{}%
\let\hyper@linkend\@empty\citealp[#1][#2]{#3}}}
\newcommandtwoopt{\citepads}[3][][]{\href{http://adsabs.harvard.edu/abs/#3}%
{\def\hyper@linkstart##1##2{}%
\let\hyper@linkend\@empty\citep[#1][#2]{#3}}}
\newcommandtwoopt{\citetads}[3][][]{\href{http://adsabs.harvard.edu/abs/#3}%
{\def\hyper@linkstart##1##2{}%
\let\hyper@linkend\@empty\citet[#1][#2]{#3}}}
\newcommandtwoopt{\citeyearads}[3][][]%
{\href{http://adsabs.harvard.edu/abs/#3}
{\def\hyper@linkstart##1##2{}%
\let\hyper@linkend\@empty\citeyear[#1][#2]{#3}}}
\makeatother

\usepackage{lscape}
\usepackage{xcolor}

\begin{document} 

   \title{RedDots: Limits on habitable and undetected planets orbiting nearby stars GJ\,832, GJ\,674, and Ross\,128\thanks{The Appendices \ref{apdx:priors} to \ref{apdx:add_figures} are only available online at Zenodo via \url{https://zenodo.org/records/13626863}}}

   \author{
   F.~Liebing\inst{\ref{MPS}} \and S.\,V.~Jeffers\inst{\ref{MPS}} \and P.~Gorrini\inst{\ref{IAG}} \and C.\,A.~Haswell\inst{\ref{TOU}} \and S.~Dreizler\inst{\ref{IAG}} \and J.\,R.~Barnes\inst{\ref{TOU}} \and C.~Hartogh\inst{\ref{IAG}} \and V.~Koseleva\inst{\ref{IAG}} \and F.~Del~Sordo\inst{\ref{ICE}, \ref{IEEC},\ref{INAF}} \and P.\,J.~Amado\inst{\ref{IAA}} \and J.\,A.~Caballero\inst{\ref{CAB}} \and M.\,J.~L\'opez-Gonz\'alez\inst{\ref{IAA}} \and N.~Morales\inst{\ref{IAA}} \and A.~Reiners\inst{\ref{IAG}} \and I.~Ribas\inst{\ref{ICE}, \ref{IEEC}} \and A.~Quirrenbach\inst{\ref{LSW}} \and E.~Rodr\'iguez\inst{\ref{IAA}} \and L.~Tal-Or\inst{\ref{ArielUni}} \and Y.~Tsapras\inst{\ref{ZAH}}
   }
 
   \institute{Max-Planck-Institut-f\"ur Sonnensystemforschung,
              Justus-von-Liebig-Weg 3, 37077 G\"ottingen, Germany\\
              \email{liebing.astro@gmx.de}
              \label{MPS}
              \and
              Institut f\"ur Astrophysik und Geophysik (IAG), Universit\"at G\"ottingen,
              Friedrich-Hund-Platz 1, 37077 G\"ottingen, Germany
              \label{IAG}
              \and
              School of Physical Sciences, The Open University,
              Walton Hall, MK7 6AA, Milton Keynes, United Kingdom
              \label{TOU}
              \and
              Institut de Ci\`encies de l'Espai (ICE, CSIC),
              Campus UAB, Carrer de Can Magrans s/n, 08193 Bellaterra, Spain
              \label{ICE}
              \and
              Institut d'Estudis Espacials de Catalunya (IEEC),
              c/ Gran Capit\`a 2-4, 08034 Barcelona, Spain
              \label{IEEC}
              \and
              INAF - Osservatorio Astrofisico di Catania,
              Via Santa Sofia 78, 95123 Catania, Italy
              \label{INAF}
              \and
              Instituto de Astrof\'isica de Andaluc\'ia (IAA-CSIC),
              Glorieta de la Astronom\'ia s/n, 18008 Granada, Spain
              \label{IAA}
              \and
              Centro de Astrobiolog\'ia, CSIC-INTA, ESAC,
              Camino Bajo del Castillo s/n, 28692 Villanueva de la Ca\~nada, Madrid, Spain
              \label{CAB}
              \and
              Landessternwarte, Zentrum f\"ur Astronomie der Universt\"at Heidelberg,
              K\"onigstuhl 12, 69117 Heidelberg, Germany
              \label{LSW}
              \and
              Department of Physics, Ariel University,
              Ariel, 40700, Israel
              \label{ArielUni}
              \and
              Zentrum f\"ur Astronomie der Universit\"at Heidelberg, Astronomisches Rechen-Institut,
              M\"onchhofstr. 12-14, 69120 Heidelberg, Germany
              \label{ZAH}
             }

   \date{Received xxx / Accepted xxx}
 
  \abstract
   {The nearby ($d\,<$\,5\,pc) M dwarfs GJ\,832, GJ\,674, and Ross\,128 each host a single exoplanet, with Ross\,128\,b located within the optimistic habitable zone. Due to their low mass and close proximity, these three systems are prime candidates for further characterization studies.}
   {Using HARPS spectroscopic data obtained by the RedDots campaign, as well as archival data from HARPS and CARMENES, supplemented with ASH2 and T90 photometry, we aim to search for additional planets in the three systems. We also aim to determine limits on possible undetected, habitable planets. We investigate (i) the reliability of the recovered orbital eccentricities and (ii) the reliability of Bayesian evidence as a diagnostic for selecting the best model.}
   {We employed Markov-chain Monte Carlo, nested sampling, and Gaussian process (GP) analyses to fit a total of 20 different models comprising 0 -- 2 Keplerian signals and three different GP kernels for stellar activity. We used the residuals to create grids for injection-recovery simulations to obtain detection limits on potentially undiscovered planets.}
   {Our refined orbital elements for GJ\,832\,b, GJ\,674\,b, and Ross\,128\,b confirm (GJ\,832, GJ\,674) or increase (Ross\,128) prior eccentricity determinations. No additional planets were found in any of the systems. The detection limits obtained for all three systems are between 30 and 50\,cm\,s$^{-1}$ for orbital periods in the range of 1 to 10\,000 days. This corresponds to habitable planet masses of $< 1.5 M_\oplus$ for GJ\,832 and $< 1 M_\oplus$ for GJ\,674 and Ross\,128. Using N-body simulations, we find that undiscovered secondary planets are unlikely (Ross\,128) or incapable (GJ\,674) of having caused the observed eccentricities of the known planets. We find that the eccentricity of GJ\,832\,b is not significantly different from zero. }
  {GJ\,832\,b, GJ\,674\,b, and Ross\,128\,b retain their status as hosting lonely and (for the latter two) eccentric planets ($e=0.04, 0.24, 0.21$; respectively). This is unexpected in classical planet formation scenarios, which favor circular orbits and multi-planet configurations, demonstrating that planet formation in these cases is more complicated than traditionally thought. Additionally, the eccentricity of Ross 128 indicates that it spends some of its orbit outside of the optimistic habitable zone.   Finally, our results show that Bayesian evidence, when used in conjunction with GP, is not a robust diagnostic for selecting the best model in cases of low-activity stars.  In such cases, we advise an inspection of the shapes of the posterior distributions and to ensure that relevant simulations are performed to assess the validity of the perceived best model.}


   \maketitle
%
\section{Introduction}
\label{sec:intro}

With space missions such as JWST already operational, and others such as PLATO \citepads{2014ExA....38..249R} and Ariel (\citeads{2018ExA....46..135T}; \citeads{2022EPSC...16.1114T})  soon underway, it is a timely endeavor to perform a full census of our closest exoplanetary neighbors, so that these future missions can focus on the most promising planetary systems for further characterization. Currently, it is only in the Solar neighbourhood that radial velocity (RV) planet searches can plausibly approach the production of a (small) volume-limited sample. The aim of RedDots is to detect all of the closest terrestrial planets, where "close" is defined as within a distance of 5\,pc (\citeads{2016Natur.536..437A}; \citeads{2020Sci...368.1477J}). This 5\,pc sample comprises approximately 75\% M dwarf stars\footnote{\url{www.recons.org}}, which is in agreement with the 10\,pc sample \citetads{2021A&A...650A.201R} when brown dwarfs are excluded.  Due to their low mass and size, M dwarf stars are particularly amenable to the detection of small rocky planets, orbiting close-in to their host stars. The closest exoplanet hosting star is Proxima Centauri, which hosts an Earth-mass planet orbiting in its liquid-water Habitable Zone \citepads{2016Natur.536..437A} and shows evidence of a second (\citeads{2020SciA....6.7467D}; supported by \citeads{2020A&A...635L..14K} and \citeads{2020A&A...638A.120G}) and third \citepads{2022A&A...658A.115F} companion.\par

The lower luminosity and effective temperatures of M dwarfs also lead to a habitable zone that is much closer to the host than it is for G-type stars, such as our sun (\citeads{2013ApJ...765..131K}, \citeads{2014ApJ...787L..29K}). This further improves the chances of detecting an Earth-like, habitable planet and makes late-type stars the most promising targets for habitable planet searches as well. This is reflected in the Habitable Worlds Catalog\footnote{PHL @ UPR Arecibo (2024, 02, 20). The Habitable Worlds Catalog (HWC). \url{https://phl.upr.edu/hwc}}, which lists 29 conservatively habitable planets with 27 on orbits around M dwarfs and 2 around K dwarfs. Out of the optimistically habitable sample of 69 stars, only 4 super-Earths are listed as orbiting at the inner habitable edge around a G-type rather than a K- or M-type dwarf. It should also be noted that traditionally, the notion of habitability is focused around the presence of liquid water as a solvent for the development of life. Recently, the idea of alternative solvents has gained traction, which would lead to different habitability criteria, such as significantly lower or potentially much higher temperatures for liquid carbon-dioxide or pure sulfuric acid. The later would also require a nearly water free planet which could be achieved around M dwarfs where the habitable zone moves inward as the star ages. This would give delayed exposure of a planet that was initially baked dry to habitable conditions and allow a much wider range of habitable conditions. We refer to \citetads{2024arXiv240107296B} for a more detailed overview and discussion on the topic of alternative solvents for the development of life. Throughout this work, we keep our focus on traditional, liquid water habitability.\par

Stellar activity is currently the biggest obstacle in detecting small planets. It manifests in multiple ways spanning many timescales from minutes (granulation) to hours (supergranulation), days (rotation), and years (activity cycles). These timescales can interact in a complex way, as in convective blueshift (granulation timescale) suppressed (active region lifetime) and modulated (rotation) by active regions. Overall, with astronomical instruments having reached stability levels approaching 10\,cm\,s$^{-1}$, stellar activity, convective blueshift suppression in particular, has become the limiting factor for detecting small rocky planets on orbits close to their host stars (\citeads{2010A&A...512A..39M}; \citeads{2016MNRAS.457.3637H}). Convective blueshift suppression may be the biggest contributor to stellar activity related RV variations; however, despite significant effort, there has been no success thus far to make any corrections for its influence  (\citeads{2017A&A...607A...6M}; \citeads{2021A&A...654A.168L}; \citeads{2023A&A...673A..43L}).\par

To minimize sources of correlated noise, such as stellar activity, and to maximize the mitigating diagnostics, the RedDots search for terrestrial exoplanets uses a regular cadence observing strategy. We typically observe our target stars once per night over a timespan of a few months. This strategy has led to the detection of planets orbiting in the liquid-water habitable zone of our closest stellar neighbor: Proxima Centauri \citepads{2016Natur.536..437A} and the mid-M dwarf GJ\,1061 \citepads{2020MNRAS.493..536D}. Additionally, RedDots has detected a signal that could correspond to a habitable zone planet in the early M dwarf GJ\,887 \citepads{2020Sci...368.1477J}. The regular high cadence monitoring pursued by RedDots along with simultaneous photometry showed that the 37-day signal previously detected for GJ\,832 by \citetads{2014ApJ...791..114W} is a false positive.  Instead, \citetads{2022A&A...664A..64G} established that the 37-day signal is caused by stellar rotation.\par

We now know that the planetary occurrence rate is more than one Earth-mass or sub-Neptune mass planet per M dwarf. In the 1--10 Earth mass range for periods up to 100\,d, every M dwarf is likely to host 1.32$_{-0.31}^{+0.33}$ planets \citepads{2021A&A...653A.114S} or $1.37^{+0.24}_{-0.24}$ planets at up to 1\,000\,d \citepads{2023A&A...670A.139R}. This is up to seven times higher than for F type stars \citepads{2012ApJS..201...15H} and a factor of two to three more than for G types (\citeads{2011arXiv1109.2497M}; \citeads{2015ApJ...814..130M}), while Neptune-mass planets are under abundant by a factor of two \citepads{2015ApJ...814..130M}. Meanwhile hot Jupiters are even less common, at an occurrence rate of only 0.5\% for M dwarfs (\citeads{2004ApJ...617..580B}; \citeads{2006ApJ...649..436E}) compared to about 2.5\% \citepads{2005PThPS.158...24M} to 5\% \citepads{2003acfp.conf..359M} for FGK stars.  M dwarfs have been predicted to have either several orbiting planets in a multiplanetary system or no planets at all. This implies that when one planet forms and migrates into a warm orbit where it is detected, the planetary system usually also has a larger number of smaller bodies in resonant chains (\citeads{2019A&A...631A...7C}; \citeads{2019A&A...627A..83L}), such as the well known TRAPPIST-1 system (\citeads{2016Natur.533..221G}, \citeads{2017Natur.542..456G}, \citeads{2017NatAs...1E.129L}) or the recently discovered, resonant sextuplet around HD\,110067 \citepads{2023Natur.623..932L}. M dwarfs with single planets are not predicted by traditional planet occurrence models, although they are quite commonly observed. There are currently 1036\footnote{NASA exoplanet archive; 12 May 2023; \url{https://exoplanetarchive.ipac.caltech.edu/}} planets that have been discovered with the RV method. Of these, 41\% (424) are listed as the only planet around single stars, posing a challenge for traditional planet formation models. While the existence of some singular planets itself is not unexpected from those planet formation models, the number of planets without additional known companions is higher than expected. It was further thought that due to drag within the protoplanetary disk, small single planets should all have circular orbits (\citeads{1980ApJ...241..425G}; \citeads{2004A&A...418..325S}; \citeads{2007A&A...473..329C}). This stands counter to the majority of such observed planets, which we discuss further in Sect.~\ref{sec:discussion}.\par

In this work, we use new RedDots observations of the stars GJ\,832, GJ\,674, and Ross 128 to search for additional exoplanets. To date, one planetary companion has been reported for each of these stars: GJ\,832 has one sub Jupiter-mass planet on a very long orbit (>\,3\,000 days), while Ross\,128 and GJ\,647 have one Earth-mass and Neptunian-mass planet, respectively, on orbits below ten days. In each case, the planets are in non-circular orbits. From planet formation and planet occurrence models, we expect that these systems should have additional planetary companions. In this paper, we detail our search for additional planets orbiting these stars using the regular cadence RedDots data in addition to archival HARPS\footnote{High-Accuracy Radial velocity Planetary Searcher} data.  In Sect.~\ref{sec:targets}, we describe the three planetary systems. We present the observations and data processing in Sect.~\ref{sec:data}. In Sect.~\ref{sec:analysis}, we describe our methods and in Sect.~\ref{sec:results}, we present our results, including planet detection thresholds, refined fundamental parameters for the confirmed planetary companions, and a new signal at four days around Ross\,128. We discuss our findings in Sect.~\ref{sec:discussion} and our summary and conclusions in Sect.~\ref{sec:Conclusions}.

\section{Targets}
\label{sec:targets}

\begin{table*}
\centering
\caption{Fundamental stellar parameters for GJ\,832, GJ\,674 and Ross\,128 from the literature.}
\label{tab:Params_known}
\begin{tabular}{llccc}
\hline\hline
\noalign{\smallskip}
 Parameter & Unit & GJ\,832\tablefootmark{a} & GJ\,674\tablefootmark{b} & Ross\,128\tablefootmark{c}\\
\noalign{\smallskip}
\hline
\noalign{\smallskip}
Sp. Type & & M1.5\,V & M2.5\,V & M4\,V\\
$M_\bigstar$ & [M$_\odot$] & $0.45\pm0.05$ & 0.35 & $0.168\pm0.017$\\
T$_{\textrm{eff}}$ & [K] & 3657 & 3500 -- 3700 & $3192\pm60$\\
$R_\bigstar$ & [R$_\odot$] & 0.48 & $0.36^{+0.012}_{-0.011}$\tablefootmark{d} & $0.1967\pm0.0077$\\
$L_\bigstar$ & [L$_\odot$] & 0.026\tablefootmark{e} & 0.016 & $0.00362\pm0.00039$\\
$P_\mathrm{rot}$ & [d] & $37.5^{+1.4}_{-1.5}$\tablefootmark{f} & $34.85\pm0.03$ & 101 -- 123\\
$P_\mathrm{rot, alt}$ & [d] & & 33\tablefootmark{i} & $163\pm3$\tablefootmark{g}; $165.1\pm0.8$\tablefootmark{h}; 223\tablefootmark{i}\\
\noalign{\smallskip}
\hline
\end{tabular}
\tablefoot{Uncertainties are added where they were available from the original publications.}
\tablebib{\tablefoottext{a}{\citetads{2009ApJ...690..743B}}, \tablefoottext{b}{\citetads{2007A&A...474..293B}}, \tablefoottext{c}{\citetads{2018A&A...613A..25B}}, \tablefoottext{d}{\citetads{2021ApJ...918...40P}}, \tablefoottext{e}{\citetads{2013A&A...549A.109B}}, \tablefoottext{f}{\citetads{2022A&A...664A..64G}}, \tablefoottext{g}{\citetads{2019A&A...621A.126D}}, \tablefoottext{h}{\citetads{2016A&A...595A..12S}}, \tablefoottext{i}{Photometry; this work}}
\end{table*}

In this section, we summarize the properties of the three targets of this work: GJ\,832, GJ\,674, and Ross\,128.  Their known stellar parameters are listed in  Table~\ref{tab:Params_known}. 
The earliest M dwarf in our sample is GJ\,832 (HD\,204961), with a spectral type of M1.5\,V. It is located at a distance of 4.97\,pc (\textit{Gaia} DR3 parallax; \citeads{2016A&A...595A...1G}, \citeads{2023A&A...674A...1G}). It has a Jupiter-mass planet which was first detected by \citetads{2009ApJ...690..743B} with the orbit further refined by \citetads{2022A&A...664A..64G}. Semi-empirical models of the outer atmospheric layers of GJ\,832 show that it is moderately active with UV flux comparable to the Sun at activity maximum \citepads{2016ApJ...830..154F}. For an M dwarf, this is only a moderate level of stellar activity. Further details of GJ\,832 were refined by \citetads{2022A&A...664A..64G}.\par

The M2.5\,V dwarf GJ\,674 (CD-46\,11540) has one eccentric, Neptune-mass planet with an orbital period of 4.69\,days and a mass of 11.8 M$_\oplus$ \citepads{2007A&A...474..293B}.  GJ\,674 is located at a distance of 4.55\,pc (\textit{Gaia} DR3 parallax; \citeads{2016A&A...595A...1G}, \citeads{2023A&A...674A...1G}) with a rotation period of 34.8 days \citepads{2007A&A...474..293B} and shows persistently low stellar activity. From an analysis of the $S$-index, which is a measure for the chromospheric activity of a star based on the \ion{Ca}{II} H\&K lines, \citetads{2007A&A...474..293B} were able to present the first "closed-loop" pattern of stellar activity for GJ\,674. Closed-loop here refers to the roughly circular locus traced by the observations when an activity indicator, such as the $S$-index, is plotted against the measured RV. This pattern has since been observed for the more magnetically active M dwarf EV\,Lac, but only when a timespan of not longer than a few stellar rotation periods is used \citepads{2022A&A...663A..27J}.\par

\citetads{2019MNRAS.488..633V} extrapolated what is known for radiometric emission from the Solar System to all then known exoplanets.  They identify GJ\,674\,b as the most promising planet to detect star-planet interactions via radio emission within that sample, matched only by Proxima\,b and followed by YZ\,Ceti\,b, GJ\,1214\,b, and GJ\,436\,b. Since then, \citetads{2023NatAs.tmp...65P}  detected coherent radio bursts from YZ\,Ceti that \citetads{2023arXiv230500809T} identified as originating from star-planet-interaction. With its radio emission expected to be stronger than YZ\,Ceti's, GJ\,674 becomes a compelling target: detections of auroral emission allow for the determination of planetary magnetic field strengths.\par

Ross\,128 (or GJ\,447) is an M4.0\,V dwarf located at a distance of 3.37\,pc (\textit{Gaia} DR3 parallax; \citeads{2016A&A...595A...1G}, \citeads{2023A&A...674A...1G}). It has a 1.35 M$_\oplus$ planet orbiting with a period of 9.9 days \citepads{2018A&A...613A..25B}, just inside the inner edge of the optimistic liquid-water habitable zone. Ross\,128 is classified as a flaring star, with a suggested long rotation period of approximately 110 to 120 days \citepads{2018A&A...613A..25B} or 165 days \citepads{2016A&A...595A..12S}, both from ASAS data \citepads{1997AcA....47..467P}. Furthermore,  \citetads{2018A&A...613A..25B} supplemented their data with K2 data \citepads{2014PASP..126..398H} but were unable to further refine their result as the K2 data only covered 80 days, less than a full rotation period. \citetads{2019A&A...621A.126D} subsequently used MEarth data (\citeads{2008PASP..120..317N}; \citeads{2012AJ....144..145B}) in addition to ASAS and K2 and got results in agreement with \citetads{2016A&A...595A..12S}. The long-term magnetic activity of Ross\,128 was investigated by \citetads{2019A&A...628L...1I}. They reported that Ross\,128 has moderate-to-low levels of magnetic activity and quasi-periodic variability with cycle length of 5.4 or 5.8 years. The cycle lengths were derived from the $S_\mathrm{K}$ index, defined similarly to the classical $S$-index but focused only on the \ion{Ca}{II} K line, and \ion{Na}{I}. The original HARPS data that resulted in the detection of the 9.9-day planet were secured just after the maximum of the $S_\mathrm{K}$ cycle identified by \citetads{2019A&A...628L...1I}, starting in late in 2014 and spanning into early 2015. The RedDots data were obtained between 2020 and 2021, approximately 1.5 cycles later, during an S-index minimum. Ross\,128\,b was identified by \citetads{2023arXiv231117075P} as part of the golden sample of small, rocky exoplanets most suitable for future atmospheric characterization by the ANDES instrument at the Extremely Large Telescope (ELT).

\section{Observations and data processing}
\label{sec:data}

\subsection{Spectroscopy}
\label{subsec:spec_data}

\subsubsection{HARPS}
\label{subsubsec:HARPS}

\begin{table}
\centering
\caption{Observation runs used in this work.}
\label{tab:obs_runs}
\begin{tabular}{lllc}
\hline\hline
\noalign{\smallskip}
 Target & Run ID & PI & \#Spectra \\
\noalign{\smallskip}
\hline
\noalign{\smallskip}
GJ\,832 & 072.C-0488(E)\tablefootmark{a} & Mayor, M. & 54 \\
HARPS & 183.C-0437(A)\tablefootmark{a} & Bonfils, X. & 7 \\
 & 198.C-0838(A)\tablefootmark{a} & Bonfils, X. & 55 \\
 & 0104.C-0863(A)\tablefootmark{*, a} & Jeffers, S. & 67 \\
\noalign{\smallskip}
\hline
\noalign{\smallskip}
GJ\,674 & 072.C-0488(E)\tablefootmark{b} & Mayor, M. & 43 \\
HARPS & 077.C-0364(E)\tablefootmark{e} & Mayor, M. & 1 \\
 & 183.C-0437(A)\tablefootmark{e} & Bonfils, X. & 49 \\
 & 191.C-0873(B)\tablefootmark{e} & Bonfils, X. & 12 \\
 & 191.C-0873(D)\tablefootmark{e} & Bonfils, X. & 1 \\
 & 191.C-0873(A)\tablefootmark{e} & Bonfils, X. & 69 \\ 
 & 191.C-0873(E)\tablefootmark{e} & Bonfils, X. & 6 \\
 & 191.C-0873(F)\tablefootmark{e} & Bonfils, X. & 1 \\
 & 198.C-0838(A) & Bonfils, X. & 28 \\
 & 1102.C-0339(A) & Bonfils, X. & 26 \\
 & 0104.C-0863(A)\tablefootmark{*} & Jeffers, S. & 20 \\ 
\noalign{\smallskip}
\hline
\noalign{\smallskip}
Ross\,128 & 072.C-0488(E)\tablefootmark{c, d} & Mayor, M. & 6 \\
HARPS & 183.C-0437(A)\tablefootmark{c, d} & Bonfils, X. & 35 \\
 & 191.C-0873(D)\tablefootmark{c, d} & Bonfils, X. & 8 \\
 & 191.C-0873(A)\tablefootmark{c, d} & Bonfils, X. & 105 \\
 & 191.C-0873(E)\tablefootmark{c, d} & Bonfils, X. & 3 \\
 & 191.C-0873(F)\tablefootmark{c, d} & Bonfils, X. & 1 \\
 & 1102.C-0339(A) & Bonfils, X. & 11 \\ 
 & 106.21PJ.001\tablefootmark{*} & Jeffers, S. & 132 \\
 & 106.21PJ.002\tablefootmark{*} & Jeffers, S. & 6 \\
\noalign{\smallskip}
\hline
\noalign{\smallskip}
\begin{tabular}{@{}c@{}}Ross\,128 \\ CARMENES\end{tabular} & \multicolumn{2}{c}{GTO} & 58\\
\noalign{\smallskip}
\hline
\end{tabular}
\tablefoot{\tablefoottext{*}{Observations from the RedDots campaign.}}
\tablebib{Also used in: \tablefoottext{a}{\citetads{2022A&A...664A..64G}}, \tablefoottext{b}{\citetads{2007A&A...474..293B}}, \tablefoottext{c}{\citetads{2018A&A...613A..25B}}, \tablefoottext{d}{\citetads{2019A&A...628L...1I}}, \tablefoottext{e}{\citetads{2017MNRAS.468.4772S}}}
\end{table}

This work makes use of archival and new spectroscopic data (Table~\ref{tab:obs_runs}) secured with the HARPS instrument \citepads{2003Msngr.114...20M}. HARPS is a fiber-fed, cross-dispersed echelle spectrograph located in an evacuated and temperature stabilized chamber at the 3.6\,m telescope of La Silla Observatory, Chile. It covers the wavelengths from 380\,nm to 690\,nm at a resolving power of $\mathcal{R}$=115\,000 within 72 echelle orders and can reach a precision of at least 1\,m\,s$^{-1}$. Archival data and new data collected under the RedDots project have their radial velocity (RV) time series extracted from the DRS reduced spectra using the \texttt{serval\footnote{SpEctrum Radial Velocity AnaLyser}} code \citepads{2018A&A...609A..12Z} (See Sect. \ref{subsubsec:serval_reduction}).\par
The RV time series are shown in Fig.~\ref{img:new_RedDots_data}. They are separated by color into archival and new observations to highlight the quality and consistency of the RedDots observations. For GJ\,832, the RedDots observations contribute over one third of all available observations, with 67 observations secured over a timeframe of two and a half months from mid-October to end of December 2019, with a one-week break at the beginning of December. This averages to nearly one observation each night, while during one night GJ\,832 was observed twice. The archival data for GJ\,674 were supplemented with 20 RedDots observations in October 2019, also averaging close to one observation per night. For Ross\,128, RedDots has nearly doubled the number of datapoints, adding 138 in total between mid-December 2020 and mid-March 2021, while keeping a regular cadence over those four months with an average of just above one observation per night. The additional observations are distributed over 12 nights with two observations, 10 nights with three, 5 nights with 4, and 1 night with five observations, for a total of 87 observing nights. This cadence is intended to increase the chances of recovering very short period signals. The window functions corresponding to the three data sets are shown in Figs.~\ref{img:GJ832_wFunc}--\ref{img:Ross128_wFunc}.\par
There were two interruptions in the usually continuous operation of HARPS.  The first interruption occurred on 2 June 2015 for the exchange of the original optical fiber and the second on  23 March 2020 when the cooling of HARPS was temporarily suspended due to the COVID pandemic. The three blocks created by these discontinuities show different velocity zero points and jitter properties, so we treated each of them as a different instrument in our analysis. We refer to them as different "instrument seasons" for the rest of the paper and refer to them as the pre-fiber upgrade (pre), post-fiber upgrade (post), and post-COVID warmup (warmup) seasons.

\subsubsection{CARMENES}
\label{subsubsec:CARMENES}

\begin{figure*}
\resizebox{\hsize}{!}{\includegraphics{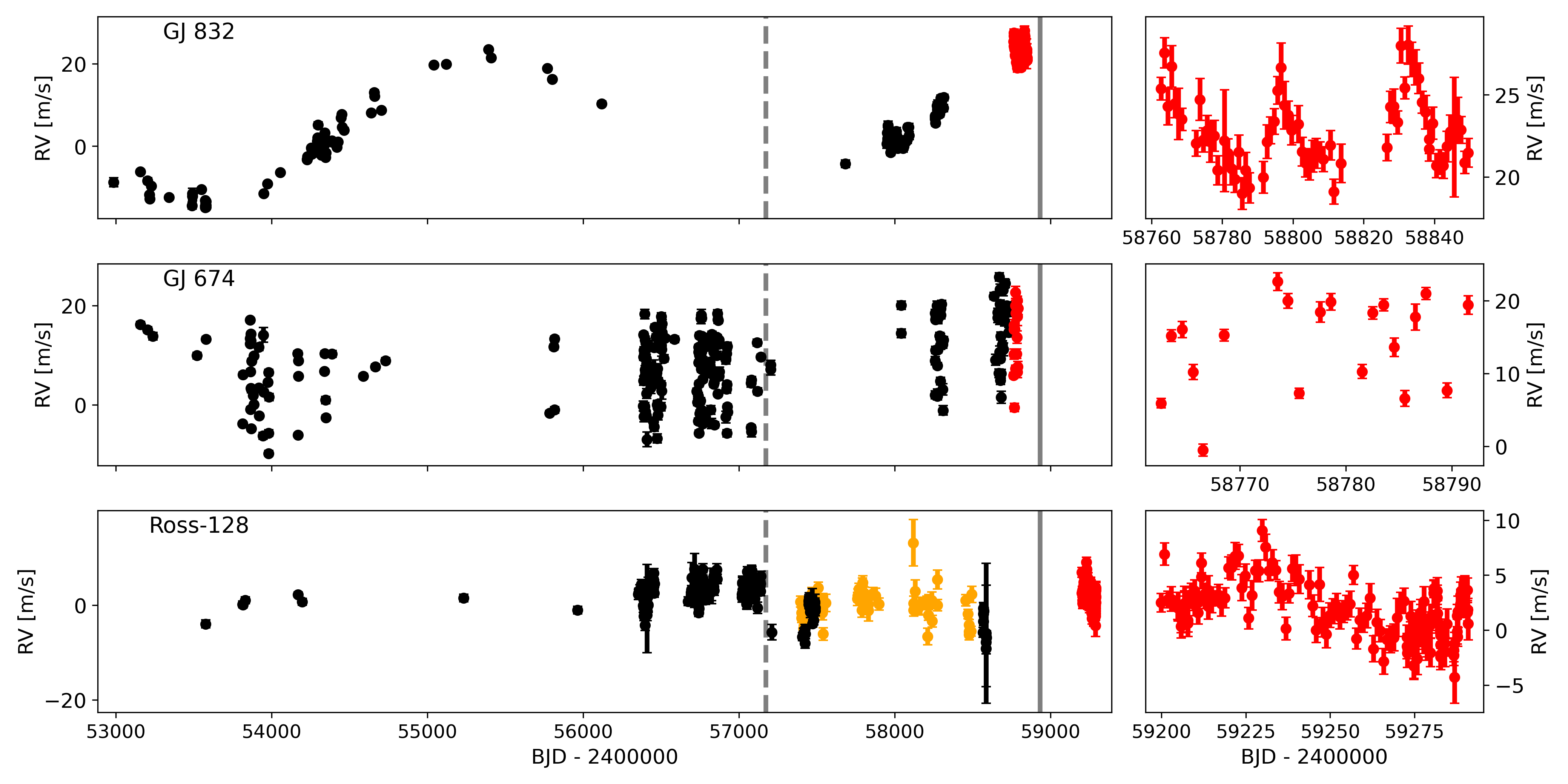}}
\caption{Archival (black) and new (red) RV extracted from HARPS observations and CARMENES DR1 RV (orange) for the stars named in the top left of each panel. The HARPS pre- and post-fiber upgrade (dashed gray line) and warmup transitions (solid gray line) are marked. The right panels are zoomed in on the HARPS RedDots observations for each target.}
\label{img:new_RedDots_data}
\end{figure*}

We supplemented our HARPS observations for Ross\,128 with 58 observations by the CARMENES\footnote{Calar Alto high-Resolution search for M dwarfs with Exoearths with Near-infrared and optical Echelle Spectrographs} instrument \citepads{2014SPIE.9147E..1FQ}. CARMENES is a fiber-fed echelle spectrograph installed at the Calar Alto Observatory's 3.5\,m telescope in Almer\'ia, Spain. The instrument consists of two channels covering the visual (520 -- 960\,nm) and near-infrared (960 -- 1710\,nm) ranges at $R=94\,600$ and $R=80\,400$, respectively, for near continuous spectral coverage. The first data release\footnote{\url{http://carmenes.cab.inta-csic.es/gto/jsp/dr1Public.jsp}} for the guaranteed time observations (GTO) was published by \citetads{2023A&A...670A.139R}, including reduced spectra, RV and activity indicator timeseries, periodograms, and detection maps. For our analysis, we used the nightly zero-point-corrected RVs shown in Fig.~\ref{img:new_RedDots_data}. They fill the gap in coverage of Ross 128 during the middle of the HARPS post-fiber change season.

\subsubsection{\texttt{serval} data reduction}
\label{subsubsec:serval_reduction}
The observations were uniformly processed with \texttt{serval}, which uses a template matching approach, to obtain consistent RVs. All spectra are first cross-correlated with the highest S/N spectrum in each data set to obtain an initial RV guess. These initial RVs are used to create a high-S/N coadded template spectrum from all observations. The RVs are then redetermined relative to the template spectrum, providing robust, relative (to the template) RV values independent of any choices of spectral line lists or synthetic library.. The \texttt{serval} algorithm further corrects for secular acceleration, provides a number of activity indices and the chromatic index, a measure of the wavelength dependence of the RV.

\subsection{Photometry}
\label{subsec:phot_data}
We used photometric observations of GJ\,674 by the 40\,cm robotic telescope ASH2 (Astrograph for South Hemisphere II) at the San Pedro de Atacama Celestial Explorations Observatory (SPACEOBS), Chile where the data analysis is described by \citetads{2020MNRAS.493..536D}. We further used photometric observations of Ross\,128 by the 90\,cm T90 telescope at the Observatorio de Sierra Nevada, Granada (Spain), with the data analysis performed following \citetads{2021A&A...650A.188A}.\par

ASH2 is equipped with a with a 2.7\,k\,$\times$\,4\,k-pixel STL1100 CCD camera with a field of view (FOV) of $54\,\times\,82$\,arcmin. For the present work, we used subframes with 40\% of the total FOV, that is, the de facto FOV covers $21.6\,\times\,32.8$\,arcmin$^2$. The observations of GJ\,674 were obtained on 45 nights during the period July to October 2019 using $B$ (817 observations) and $V$ (811 observations) filters. The time series for both filters are shown in Fig.~\ref{img:GJ674_photometry}.\par

T90 is a 90\,cm Ritchey-Chr\'etien telescope equipped with a 2\,k\,$\times$\,2\,k-pixel VersArray CCD camera with a resulting FOV of $13.2\,\times\,13.2$\,arcmin$^2$. The camera is based on a high quantum efficiency back-illuminated CCD chip, type Marconi-EEV CCD42-4, with optimized response in the ultraviolet \citepads{2021A&A...650A.188A}. Our set of observations, collected in Johnson $V$ and $R$ filters, consists of 52 epochs obtained between November 2020 and May 2021. Each epoch typically consists of 20 exposures in each filter per night, of 30\,s and 20\,s respectively, for a total of respectively 974 and 957 observations. The time series for both filters are shown in Fig.~\ref{img:Ross128_photometry}.\par

The CCD measurements for both telescopes were obtained by the method of synthetic aperture photometry using a $1\,\times\,1$ binning (meaning no binning). Each CCD frame was corrected in a standard way for dark and flat-fielding. Different aperture sizes were also tested in order to choose the best one for our observations. A number of nearby and relatively bright stars within the frames were selected as check stars in order to choose the best ones to be used as reference stars. We further averaged each night of observations, as shown in Figs.~\ref{img:GJ674_photometry} and \ref{img:Ross128_photometry}, before further analysis.

\section{Analysis}
\label{sec:analysis}
For a first look at any potential signals we used the Generalized Lomb-Scargle Periodogram (GLS, \citeads{2009A&A...496..577Z}, Fig~\ref{img:GLS_new}), which also computes the false-alarm-probability (FAP) from the periodogram power. The RVs obtained for each planetary system were analyzed using a total of 20 different models (see Table \ref{tab:modelEvidence}). These comprise zero to two Keplerian signals with circular or eccentric orbits combined with either no Gaussian process (GP) or one of three different GP kernels (simple harmonic oscillator (SHO), double SHO (dSHO), quasi-periodic (QP); see Sect. \ref{subsec:GP}) to account for the effects of stellar activity. Furthermore, one jitter term and offset per instrument season was applied within each model.

\subsection{Keplerian signal}
\label{subsec:kep_signal}
We parameterized the Keplerian signal as follows:
\begin{align}
RV &= K \cdot \left[\cos\left(\omega + f\left(e, P, t_0\right)\right) + e \cdot \cos\left(\omega\right)\right]. \label{eq:kep_rv}
\end{align}
The formulation uses, in order, the line-of-sight radial velocity \textit{RV}, RV semi-amplitude, $K$, argument of periastron, $\omega$ (the angle between the ascending node and the periastron), eccentricity, $e$, orbital period, $P$, and the time of periastron passage, $t_0$. Calculating the true anomaly $f$ requires solving the transcendental Kepler equation, which is done through a root finding algorithm. We employed the algorithm implemented in the python package \texttt{scipy.optimize.root}.

\subsection{Gaussian process}
\label{subsec:GP}
We used a Gaussian process (GP) to statistically model the rotational modulation of surface activity in a non-parametric way. Gaussian processes rely on a kernel (see Sects. \ref{subsubsec:SHO} -- \ref{subsubsec:QP} for our choices) which defines the correlation between individual data points based on a set of hyperparameters.
The specific GPs we used are implemented through \texttt{celerite2} (\citeads{2017AJ....154..220F}; \citeads{2018RNAAS...2a..31F}) and \texttt{george} \citepads{2015ITPAM..38..252A}. The algorithm of \texttt{celerite2} offers large performance advantages compared to reference codes because the algorithm is restricted to a specific class of covariance, or kernel, functions to allow for much faster matrix inversions (see \citeads{2017AJ....154..220F} for details). The \texttt{george} code instead offers a larger choice of kernel functions.

\subsubsection{SHO kernel}
\label{subsubsec:SHO}
One of the kernels provided by \texttt{celerite2} is the stochastically-driven, damped, simple harmonic oscillator (SHO) kernel (Eqn.~\ref{eq:SHO}):
\begin{align}
S\left(\omega\right) &= \sqrt{\frac{2}{\pi}}\frac{S_0\omega^4_0}{\left(\omega^2-\omega^2_0\right)^2+\omega^2_0\omega^2/Q^2}\label{eq:SHO},
\end{align}
with the following reparameterization:
\begin{align}
\rho &= 2\pi/\omega_0\label{eq:SHO_rho},\\
\tau &= 2Q/\omega_0\label{eq:SHO_tau},\\
\sigma &= \sqrt{S_0\omega_0 Q.}\label{eq:SHO_sigma}
\end{align}
Use of this kernel allows for very similar behavior to the QP kernel, particularly when using a mixture of two such kernels \citepads{2017AJ....154..220F}.
The SHO kernel, in its reparameterized form following Eqs.~\ref{eq:SHO_rho}--\ref{eq:SHO_sigma}, takes as parameters\footnote{\url{https://celerite2.readthedocs.io/en/latest/api/python/\#celerite2.terms.SHOTerm}} the period of the undamped harmonic oscillator, $\rho$, the damping timescale, $\tau,$ and the standard deviation of the process, $\sigma$. We could then try to make the connection that $\rho$ should relate to the stellar rotation period and $\tau$ to the surface evolution timescales of spots, while $\sigma$ captures stochastic variations inherent to stellar surface activity (see Sect.~\ref{subsec:rotation_discuss}).

\subsubsection{dSHO kernel}
\label{subsubsec:dSHO}
As an alternative we also tried the "RotationTerm" (\texttt{celerite2}) or "double SHO" (dSHO, \texttt{juliet}) kernel, which combines two SHO kernels at a 1:2 period ratio and is supposed to better capture rotation signals, analogous to the QP kernel. The amplitudes of the primary and secondary period are related by the fractional amplitude parameter, $f,$ which should be between zero and one in order for the primary period to remain stronger. In practice, this kernel and, particularly, the fractional amplitude did not behave as expected, as discussed in Sect.~\ref{subsec:juliet_vs_mcmc}.

\subsubsection{QP kernel}
\label{subsubsec:QP}
Lastly, we also used the QP kernel which is a popular choice in the literature (see for example: \citeads{2014MNRAS.443.2517H}, \citeads{2015MNRAS.452.2269R}). Its implementation is provided by \texttt{george} through \texttt{juliet}:
\begin{align}
k\left(\tau\right) = \sigma^2 \exp\left(-\alpha\tau^2 - \Gamma\sin^2\left(\frac{\pi\tau}{P_{\mathrm{rot}}}\right)\right).
\end{align}
The QP kernel is the multiplicative combination of a squared-exponential kernel with an exponential-sine-squared. It is characterized by the amplitude $\sigma$, the squared inverse $\alpha$ of a characteristic length scale $l$ with $\alpha=1/(2l^2)$, the harmonic complexity, $\Gamma$, and the periodicity, $P_\mathrm{rot}$. As the last parameter implies, this kernel is designed to capture stellar rotation signals that are not strictly periodic. The periodicity occurs on the timescale of $P_\mathrm{rot}$ and is allowed to decay on the characteristic scale, $l$. The harmonic complexity, $\Gamma$, is related to the amount of possible substructure within one period. High values allow for more complex intra-period structure, as if the stellar surface was highly spotted, while low values can be thought of as the equivalent to a single rotating spot or spot group. High values of $\Gamma$ are problematic as the additionally allowed frequency components may no longer be related to surface activity and even start to obfuscate important signals, such as planetary companions.

\subsection{Instrumental corrections}
\label{subsec:inst}
Season specific velocity offsets $\mu_X$, with $X$ denoting the season in question are added to the Keplerian models to account for different velocity zero points from each of the three HARPS observational seasons (pre-fiber change: H-pre; post-fiber change: H-post; post COVID warmup: H-warmup) as well as the CARMENES set of observations. Independent jitter terms $s_X$ are added in quadrature to the formal RV uncertainties to capture excess white noise in the data, with $X$ again denoting the season. The GP is also split into multiple parts, one per instrument and season, which share period $\rho$ and damping time $\tau$ as global stellar parameters but each with its own standard deviation $\sigma_\mathrm{GP, X}$ to account for changes in how HARPS and CARMENES might observe the variability plus any temporal changes the star might have undergone as part of an activity cycle.

\subsection{Parameter optimization}
The previous three sections have introduced the components of the models we use. To find the optimal (hyper-)parameters of each component, we used a nested-sampling approach with \texttt{juliet} (Sect. \ref{subsec:nested_sampling_setup}) for the initial model comparison and Markov-chain Monte Carlo (MCMC) inference (Sect. \ref{subsec:MCMC_setup}) for the detailed analysis.

\subsubsection{Nested sampling}
\label{subsec:nested_sampling_setup}
We used the nested sampling implementations from \texttt{juliet} \citepads{2019MNRAS.490.2262E}. \texttt{Juliet} internally uses the Gaussian Process implementations from \texttt{celerite2} and \texttt{george} \citepads{2015ITPAM..38..252A} as well as the nested sampler \texttt{dynesty} \citepads{2020MNRAS.493.3132S}. As a Bayesian process, nested sampling offers the advantage over classical fit algorithms that it is possible to obtain the full posterior distribution, including all correlations between all parameters. The best-fit can then be defined as the mode or more commonly the median of the marginalized posterior of each parameter. The uncertainty can be defined either as the width of the symmetric central 68th percentile or as the 16th and 84th percentiles to mirror the classical one-sigma Gaussian confidence intervals.\par

A nested sampler such as \texttt{juliet} further computes the logarithm of the Bayesian evidence, $\ln\mathcal{Z}$, for model comparisons. We interpret the values to mean that two models are evaluated as inconclusive at $\Delta\ln\mathcal{Z} \lesssim 5$ and one model preferred over the other at $\Delta\ln\mathcal{Z} \gtrsim 5$, orienting ourselves according to the evidence ladder by \citetads{2008ConPh..49...71T}.

\subsubsection{MCMC procedure}
\label{subsec:MCMC_setup}
We chose to use the MCMC implementation from \texttt{emcee} \citepads{2013PASP..125..306F}. Similar to nested sampling, MCMC is a Bayesian process and as such offers access to the full posterior distribution of the inferred parameters. Unlike nested sampling, MCMC does not compute the Bayesian evidence however and therefore does not allow for model comparisons.

As a Bayesian process, MCMC also requires the definition of priors and likelihood functions. \texttt{Emcee} uses the logarithmic probabilities for both and we defined our log-likelihood as the sum of the log-likelihoods returned by the GP (Sect.~\ref{subsec:GP}) with the Keplerian signal (Sect.~\ref{subsec:kep_signal}) and seasonal offsets as the mean function supplied to \texttt{celerite2}. For priors, we used exclusively uniform priors (see Tables~\ref{tab:priors1planet} and \ref{tab:priors2planet}), starting with wide windows to capture the initial posterior peaks and zoom in as necessary. This was done to avoid any potential biases through over-informative priors and is further discussed in Sect.~\ref{subsec:GLS_MCMC}. The procedure of narrowing the uniform priors is highlighted in Sect.~\ref{subsubsec:Ross128} and Fig.~\ref{img:Ross128_peak_zoom}. The enforced physical limits of $K > 0$, $P > 0$, $0 \leq e < 1$, and $-\pi \leq \omega < \pi $ are not expressly listed as priors. 

For the actual sampling, we used an ensemble of 500 walkers initialized uniformly within the hypercube described by the priors. The exceptions here are the eccentricity, $e$, and periastron argument, $\omega$. Caution is advised when running MCMC sampling close to a hard boundary in parameter space such as $e = 0$ as this can introduce biases through the fact that the walkers encounter limited movement options in their random walk and, in this particular case, an increasing degeneracy in $\omega$ as $e$ approaches zero. For this reason, we did not sample in $e$ and $\omega$ directly, but in the combined re-parameterization of $h$ and $k$ (Eqs.~\ref{eq:h}, \ref{eq:k}; see also Sect.~\ref{subsec:prior_choices}):
\begin{align}
h &= \sqrt{e}\sin\omega\label{eq:h},\\
k &= \sqrt{e}\cos\omega\label{eq:k}.
\end{align}
The inversion was performed following Eqs.~\ref{eq:e} and \ref{eq:w} to calculate the Keplerian model:
\begin{align}
e &= h^2 + k^2\label{eq:e},\\
\omega &= \mathrm{arctan2}\left(\frac{h}{\sqrt{e}}, \frac{k}{\sqrt{e}}\right).\label{eq:w}
\end{align}
The initial distribution of $h$ and $k$ was calculated from the uniform distribution of $e$ and $\omega$. The inversion was further applied to each step of the recorded chains to obtain the posterior distribution of $e$ and $\omega$ in addition to the sampled $h$ and $k$. Performing this for only the $h$ and $k$ estimate, instead of the full chains, would result in biased confidence intervals.\par

The sampling was run in 6 sections of 5\,000 steps with the first section rejected as burn-in to allow the walkers time for initial convergence towards the equilibrium distribution, resulting in a total of 25\,000 sampling steps used for inference. We verified this to be sufficient burn-in time by comparing the posterior distributions of the successive sections to each other. We found no significant differences between them, indicating that the chain has converged sufficiently within the first section. The full 25\,000 steps were used for the parameters inferred in Sect.~\ref{subsec:OrbElements}, the \texttt{corner} plots \citep{corner} in Figs.~\ref{img:GJ832_planet_corner}--\ref{img:Ross128_planet_corner_1P}, and \ref{img:Ross128_peak_zoom}.
For visualization purposes only, Figs.~\ref{img:GJ832_full_corner}--\ref{img:Ross128_full_corner_1P_carmenes} show only the last 5\,000 steps but are otherwise (visually) identical. The reduction was necessary due to computer memory limitations while creating the corner plots. The plots are also limited to the upper 90th percentiles, excluding the lowest 10th, to highlight the posterior maxima. The actual range is only limited by the used priors shown in Table~\ref{tab:priors1planet} and cover the full distribution.

\section{Results}
\label{sec:results}
We first searched for planetary signals orbiting GJ\,832, GJ\,674 and Ross\,128 using the Bayesian evidence $\ln\mathcal{Z}$ to compare the models and update the orbital parameters for the known planets (Sect.~\ref{subsec:OrbElements}). Finally, we determined the detection limits imposed by the available HARPS observations using injected planets and GLS derived FAP values as well as the Bayesian information criterion (BIC) in Sect.~\ref{subsec:glsGrids}.

\subsection{Planetary signals}
\label{subsec:OrbElements}
\begin{figure*}
\resizebox{\hsize}{!}{\includegraphics{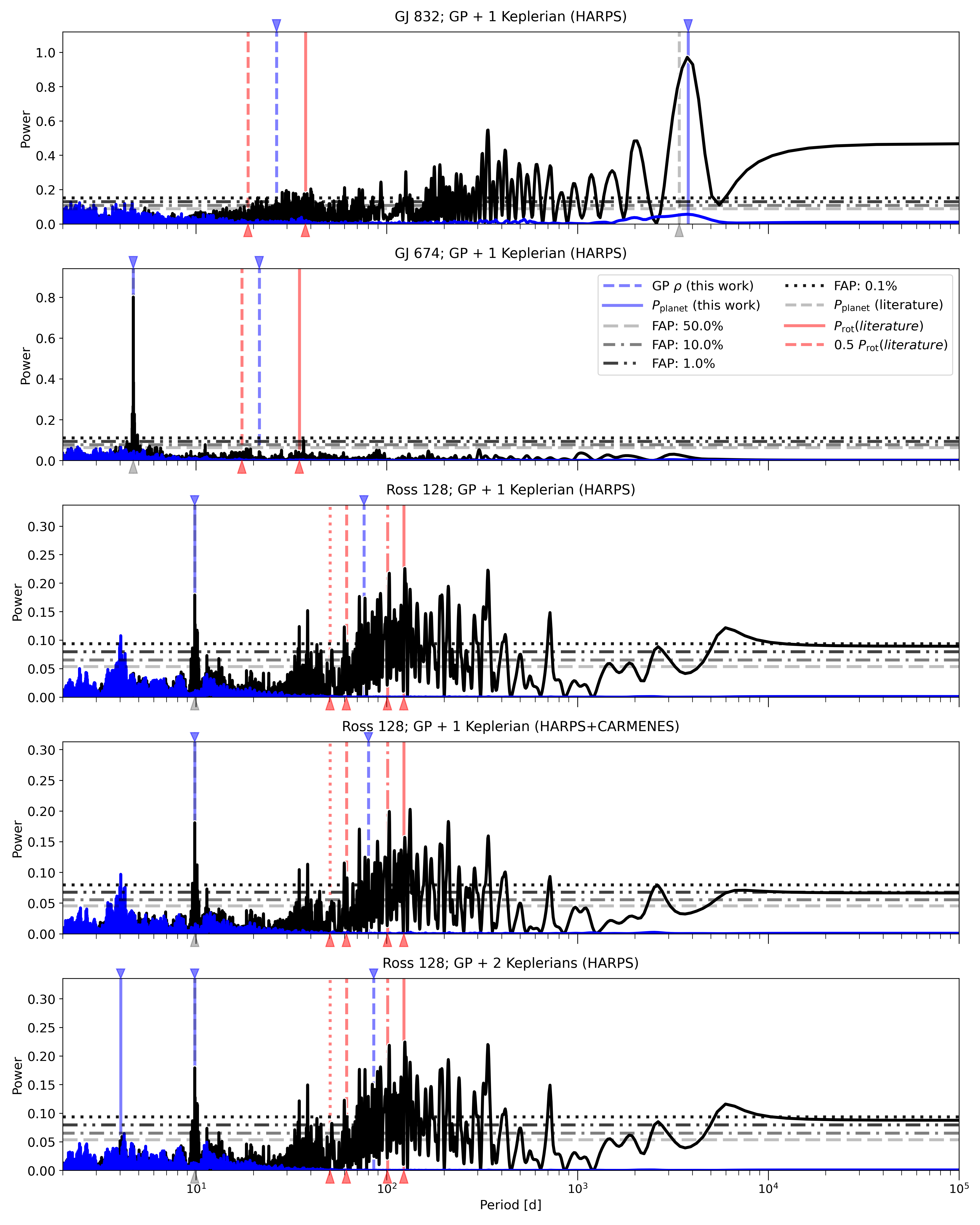}}
\caption{GLS periodograms of the three planetary systems' raw (black, fitted RV offsets are subtracted) and residual (blue, full model subtracted) RV for one planet (all) and two planet (only Ross\,128) models. Vertical lines mark the known (gray dashed) and inferred (solid blue) planetary periods $P_{\rm planet}$, literature rotation period $P_{\rm rot}$ (solid red), half period (red dashed), alternative literature rotation periods $P_{\rm rot,alt}$ (red dotted; see Table~\ref{tab:Params_known}), and GP periods, $\rho$ (blue dashed). Horizontal lines mark the false-alarm probability levels shown in the figure legend. The colored arrow heads correspond to the vertical lines of the same color for parameters of this work (located at the top) or the literature (at the bottom). The period axis is scaled to match the extent of the recovery grids presented in Sect.~\ref{subsec:glsGrids}.}
\label{img:GLS_new}
\end{figure*}

Figure~\ref{img:GLS_new} illustrates the GLS periodogram of the raw RV time series, only corrected for mutual offsets in RV, for GJ\,832, GJ\,674, and Ross\,128. The known planets can be identified at high significance for all three systems, as well as the approximate rotation period for Ross\,128. In the following Sects.~\ref{subsubsec:GJ832}--\ref{subsubsec:Ross128} we refined the initial periodogram recoveries using MCMC and nested sampling approaches for all three systems. As we state in Sect. \ref{subsec:MCMC_setup}, we started with wide, uniform priors in all planetary and instrumental parameters for all three systems and all 20 models.

\begin{table*}
\centering
\caption{Previously published orbital elements.} 
\label{tab:litElements}
\begin{tabular}{llcc@{\hspace{4\tabcolsep}}cc}
\hline\hline
\noalign{\smallskip}
 Parameter & Unit & GJ\,832\tablefootmark{a} & GJ\,832\tablefootmark{b} & GJ\,674\tablefootmark{c} & Ross\,128\tablefootmark{d}\\
\noalign{\smallskip}
\hline
\noalign{\smallskip}
$K$ & [m\,s$^{-1}$] & 14.9 $\pm$ 1.3 & $16.41^{+0.35}_{-0.34}$ & 8.7 $\pm$ 0.19 & 1.39 $\pm$ 0.18\\
$P$ & [d] & 3416 $\pm$ 131 & $3838.03^{+47.30}_{-49.23}$ & 4.6938 $\pm$ 0.007 & 9.8658 $\pm$ 0.0070\\
$e$ &  & 0.12 $\pm$ 0.11 & $0.04 \pm 0.02$ & 0.20 $\pm$ 0.02 & 0.116 $\pm$ 0.097\\
$M\,\sin\,i$ & & 0.64 $\pm$ 0.06\,$M_{\textrm{Jup}}$ & $0.74 \pm 0.06\,M_{\textrm{Jup}}$ & 11.09\,$M_{\textrm{Earth}}$ & 1.4 $\pm$ 0.21\,$M_{\textrm{Earth}}$\\
$a$ & [au] & 3.4 $\pm$ 0.4 & - & 0.039 & 0.0496 $\pm$ 0.0017\\
\noalign{\smallskip}
\hline
\end{tabular}
\tablebib{\tablefoottext{a}{\citetads{2009ApJ...690..743B}}, \tablefoottext{b}{\citetads{2022A&A...664A..64G}}, \tablefoottext{c}{\citetads{2007A&A...474..293B}}, \tablefoottext{d}{\citetads{2018A&A...613A..25B}}}
\end{table*}

\begin{table*}
\centering
\caption{Log evidence values for the different models.}
\label{tab:modelEvidence}
\begin{tabular}{lccccc}
\hline\hline
\noalign{\smallskip}
Model   & \multicolumn{5}{c}{$\ln\mathcal{Z}$}  \\
\noalign{\smallskip}
\hline
\noalign{\smallskip}
                &       GJ~832                          &       GJ\,674                         &       Ross\,128                       &    Ross\,128       &    Ross\,128                \\
                &       HARPS All                       &       HARPS All                       &       HARPS All                     &    HARPS RedDots   &  HARPS All + CARMENES    \\
\noalign{\smallskip}
0P & $-$702.819 & $-$872.623 & $-$732.539 & $-$329.860 & $-$879.788 \\
1P\textsubscript{ecc} & $-$417.245 & $-$721.134 & $-$726.180 & $-$329.607 & $-$861.687 \\
1P\textsubscript{circ} & $-$414.969 & $-$711.435 & $-$716.440 & $-$329.813 & $-$859.088 \\
2P\textsubscript{ecc} & $-$431.104 & $-$675.651 & $-$729.838 & $-$334.821 & $-$885.959 \\
2P\textsubscript{circ} & $-$408.309 & $-$714.733 & $-$725.025 & $-$308.788 & $-$868.615 \\
\noalign{\smallskip}
\hline
\noalign{\smallskip}
0P + GP-SHO & $-$427.620 & $-$856.580 & $-$638.419 & \textbf{$-$281.841} & $-$781.387 \\
0P + GP-dSHO & $-$433.437 & $-$720.755 & $-$659.400 & $-$290.842 & $-$805.854 \\
0P + GP-QP & $-$399.979 & $-$687.511 & $-$648.438 & \textbf{$-$284.849} & $-$791.900 \\
1P\textsubscript{ecc} + GP-SHO & $-$369.321 & \textit{\textbf{$-$593.106}} & \textit{\textbf{$-$624.030}} & \textit{\textbf{$-$284.425}} & \textit{\textbf{$-$761.586}} \\
1P\textsubscript{circ} + GP-SHO & \textit{\textbf{$-$365.269}} & $-$670.372 & \textbf{$-$623.183} & \textbf{$-$283.352} & \textbf{$-$760.829} \\
1P\textsubscript{ecc} + GP-dSHO & $-$373.048 & \textbf{$-$583.193} & $-$631.783 & $-$289.003 & $-$774.871 \\
1P\textsubscript{circ}+ GP-dSHO & $-$371.079 & $-$648.977 & $-$631.608 & $-$289.949 & $-$776.500 \\
1P\textsubscript{ecc} + GP-QP & $-$371.120 & $-$590.442 & $-$636.535 & $-$289.520 & $-$774.581 \\
1P\textsubscript{circ} + GP-QP & \textbf{$-$368.356} & $-$711.195 & $-$634.859 & $-$287.351 & $-$773.571 \\
2P\textsubscript{ecc} + GP-SHO & $-$372.301 & $-$797.946 & $-$650.356 & $-$286.935 & $-$786.794 \\
2P\textsubscript{circ} + GP-SHO & $-$370.827 & $-$667.461 & \textbf{$-$624.541} & $-$288.612 & \textbf{$-$763.897} \\
2P\textsubscript{ecc} + GP-dSHO & $-$382.833 & $-$824.353 & $-$685.895 & $-$290.482 & $-$814.590 \\
2P\textsubscript{circ} + GP-dSHO & $-$375.092 & $-$681.185 & $-$664.176 & $-$291.211 & $-$809.106 \\
2P\textsubscript{ecc} + GP-QP & $-$385.800 & $-$701.913 & $-$665.416 & $-$294.656 & $-$830.340 \\
2P\textsubscript{circ} + GP-QP & $-$372.643 & $-$682.551 & $-$678.288 & $-$292.517 & $-$781.612 \\
\noalign{\smallskip}
\hline
\end{tabular}
\tablefoot{We used models with zero, one, and two planets; with and without GP; using the SHO, dSHO or QP kernels. The highest evidence per column is marked in bold. Multiple entries are marked where the values are considered indistinguishable at $\Delta\ln\mathcal{Z} < 5$. The entries marked in bold italics refer to the models we prefer from Sects. \ref{subsubsec:GJ832} -- \ref{subsubsec:Ross128} even though for GJ\,674 this is a model with $\Delta\ln\mathcal{Z} > 5$. Our reasoning is given in Sect. \ref{subsubsec:GJ674}.}
\end{table*}

\begin{table*}
\centering
\caption{Inferred parameters from the MCMC runs.}
\begin{tabular}{llcc@{\hspace{4\tabcolsep}}ccc}
\hline\hline
\noalign{\smallskip}
 Parameter & Unit & GJ\,832 b & GJ\,674 b & Ross\,128 b & Ross\,128 b & Ross\,128 b\\
\noalign{\smallskip}
\hline
\noalign{\smallskip}
 &  & HARPS All & HARPS All & HARPS All & HARPS RedDots  & \begin{tabular}{@{}c@{}}HARPS All \\ + CARMENES\end{tabular}\\
\noalign{\smallskip}
\hline
\noalign{\smallskip}
$K$ & [m\,s$^{-1}$] & $17.2^{+0.6}_{-0.6}$ & $8.68^{+0.11}_{-0.12}$ & $1.41^{+0.14}_{-0.14}$ & $1.45^{+0.24}_{-0.24}$ & $1.41^{+0.13}_{-0.13}$\\
$P$ & [d] & $3808^{+111}_{-95}$ & $4.69502^{+0.00003}_{-0.00003}$ & $9.8556^{+0.0012}_{-0.0011}$ & $9.85^{+0.08}_{-0.08}$ & $9.8556^{+0.0011}_{-0.0011}$\\
$\sqrt{e}\sin\omega$ & & $-0.10^{+0.14}_{-0.10}$ & $0.327^{+0.020}_{-0.021}$ & $0.14^{+0.16}_{-0.19}$ & $0.27^{+0.23}_{-0.30}$ & $0.14^{+0.15}_{-0.19}$\\
$\sqrt{e}\cos\omega$ & & $0.07^{+0.12}_{-0.14}$ & $-0.367^{+0.019}_{-0.017}$ & $0.41^{+0.10}_{-0.15}$ & $0.35^{+0.18}_{-0.28}$ & $0.37^{+0.10}_{-0.16}$\\
$t_0$ & [BJD] & $2450989^{+773}_{-659}$ & $2453160.26^{+0.04}_{-0.04}$ & $2453584.5^{+0.8}_{-0.9}$ & $2459202.5^{+1.2}_{-1.2}$ & $2453584.5^{+0.8}_{-1.0}$\\
\noalign{\smallskip}
\hline
\noalign{\smallskip}
$e$ & & $0.036^{+0.030}_{-0.023}$ & $0.242^{+0.012}_{-0.013}$ & $0.21^{+0.09}_{-0.10}$ & $0.27^{+0.19}_{-0.18}$ & $0.18^{+0.09}_{-0.09}$\\
$\omega$ & [rad] & $-0.9^{+1.2}_{-1.0}$ & $2.41^{+0.05}_{-0.05}$ & $0.3^{+0.4}_{-0.5}$ & $0.6^{+0.6}_{-0.8}$ & $0.3^{+0.4}_{-0.5}$\\
$M\,\sin\,i$ & $M_\oplus$ & $247^{+ 8}_{- 8}$ & $10.95^{+0.14}_{-0.14}$ & $1.40^{+0.13}_{-0.13}$ & $1.39^{+0.21}_{-0.23}$ & $1.41^{+0.12}_{-0.13}$\\
$a$ & [au] & $3.66^{+0.07}_{-0.06}$ & $0.03867087^{+0.00000015}_{-0.00000015}$ & $0.049640^{+0.000004}_{-0.000004}$ & $0.0496^{+0.0003}_{-0.0003}$ & $0.049639^{+0.000004}_{-0.000004}$\\
\noalign{\smallskip}
\hline
\noalign{\smallskip}
$\rho$ & [d] & $26^{+10}_{- 5}$ & $21^{+ 3}_{- 3}$ & $76^{+20}_{-21}$ & $269^{+156}_{-152}$ & $80^{+22}_{-14}$\\
$\tau$ & [d] & $6^{+ 5}_{- 4}$ & $6.0^{+2.5}_{-1.9}$ & $20^{+32}_{-17}$ & $52^{+89}_{-49}$ & $27^{+35}_{-24}$\\
$\sigma_\mathrm{GP, H-pre}$ & [d] & $1.9^{+0.4}_{-0.3}$ & $2.75^{+0.28}_{-0.24}$ & $2.2^{+0.7}_{-0.4}$ &  & $2.2^{+0.9}_{-0.4}$\\
$\sigma_\mathrm{GP, H-post}$ & [d] & $2.0^{+0.8}_{-0.3}$ & $2.8^{+0.5}_{-0.4}$ & $3.8^{+2.0}_{-1.1}$ &  & $3.9^{+3.0}_{-1.2}$\\
$\sigma_\mathrm{GP, H-warmup}$ & [d] & &  & $2.3^{+1.6}_{-0.7}$ & $ 9^{+11}_{- 5}$ & $2.3^{+2.3}_{-0.8}$\\
$\sigma_\mathrm{GP, carmenes}$ & [d] &  &  &  &  & $1.8^{+0.8}_{-0.5}$\\
\noalign{\smallskip}
\hline
\noalign{\smallskip}
$s_\mathrm{H-pre}$ & [m\,s$^{-1}$] & $0.68^{+0.38}_{-0.21}$ & $0.48^{+0.16}_{-0.19}$ & $0.48^{+0.20}_{-0.25}$ &  & $0.51^{+0.20}_{-0.24}$\\
$s_\mathrm{H-post}$ & [m\,s$^{-1}$] & $0.36^{+0.20}_{-0.22}$ & $0.6^{+0.3}_{-0.3}$ & $0.5^{+0.4}_{-0.3}$ &  & $0.5^{+0.4}_{-0.3}$\\
$s_\mathrm{H-warmup}$ & [m\,s$^{-1}$] &  &  & $1.02^{+0.15}_{-0.17}$ & $1.05^{+0.16}_{-0.17}$ & $1.02^{+0.15}_{-0.17}$\\
$s_\mathrm{carmenes}$ & [m\,s$^{-1}$] &  &  &  &  & $1.4^{+0.4}_{-0.4}$\\
$\mu_\mathrm{H-pre}$ & [m\,s$^{-1}$] & $4.3^{+0.5}_{-0.5}$ & $5.7^{+0.4}_{-0.4}$ & $2.3^{+0.5}_{-0.6}$ &  & $2.4^{+0.5}_{-0.6}$\\
$\mu_\mathrm{H-post}$ & [m\,s$^{-1}$] & $9.2^{+3.0}_{-2.3}$ & $13.0^{+0.6}_{-0.6}$ & $-4.2^{+1.5}_{-1.8}$ &  & $-4.0^{+1.6}_{-1.8}$\\
$\mu_\mathrm{H-warmup}$ & [m\,s$^{-1}$] & &  & $2.2^{+1.2}_{-1.2}$ & $ 2^{+ 5}_{- 5}$ & $2.3^{+1.2}_{-1.2}$\\
$\mu_\mathrm{carmenes}$ & [m\,s$^{-1}$] &  &  &  &  & $0.1^{+0.6}_{-0.6}$\\
\noalign{\smallskip}
\hline
\end{tabular}
\tablefoot{In addition to the median of the posterior of the one eccentric planet + GP-SHO models, we give the 16th and 84th percentile confidence interval. The blocks are sorted into the directly inferred orbital elements, derived elements, GP hyperparameters, RV-jitter and -offset.}
\label{tab:inferredElements}
\end{table*}

\subsubsection{GJ 832}
\label{subsubsec:GJ832}

We started with the reanalysis of GJ\,832\,b, since this planet is well characterized by the data (see Fig.~\ref{img:new_RedDots_data}) and has already been the subject of a recent in-depth study utilizing the same RV measurements \citepads{2022A&A...664A..64G}. This made GJ\,832 a prime candidate to be used to verify our procedures before analyzing the other two planetary systems.\par

\paragraph{\textbf{Comparing Bayesian evidences $\ln\mathcal{Z}$:}}
We computed the Bayesian model evidences using \texttt{juliet} nested-sampling for the 20 models outlined in Sect. \ref{sec:analysis}. The two highest evidence values obtained are for the models comprising one circular planet and a GP with either SHO or QP kernel. While the evidence between the GP kernels is not decisive, the one planet determination shows a strong preference in $\ln\mathcal{Z}$. The preference for a circular over an eccentric orbit is negligible, as expected for the small value of $e=0.036^{+0.030}_{-0.023}$. The inferred planetary parameters are not affected by the choice of GP kernel and consistent to the MCMC results. A full listing of the $\ln\mathcal{Z}$ values is given in Table \ref{tab:modelEvidence} and an extended explanation about the evidences and the differences in the three GP kernels' periods and their variation from the photometric rotation period is given in Appendix \ref{apdx:lnZ_comparison}.

\begin{figure*}
\resizebox{\hsize}{!}{\includegraphics{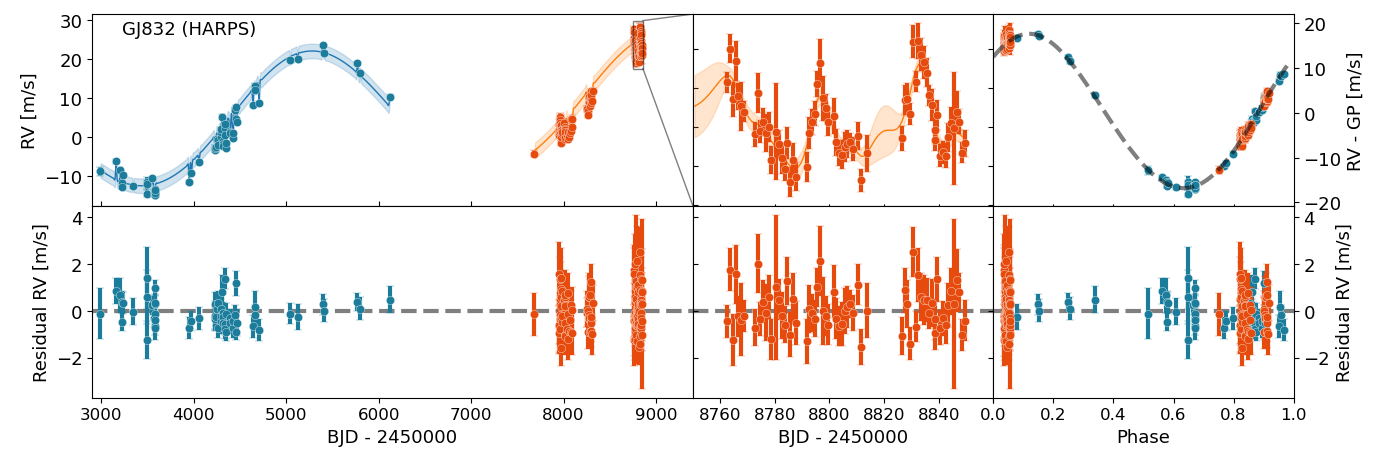}}
\caption{Best fit of the 1 eccentric planet + GP-SHO model (which includes instrumental offsets) in the left panel for GJ\,832 (top) and the residuals (bottom). The model and RV data are separated by a small difference in color shade and the uncertainty of the model fit indicated by the more transparent region. The colors indicate the pre-fiber change (blue) and post-fiber change (orange) HARPS instrument seasons. The model parameters can be found in Table~\ref{tab:inferredElements}. The gray box marks the extent of the zoomed sub-panel. Zoom into the region of densest observations marked by the gray box in the left panel (middle). Phase-folded, activity-subtracted, and instrumental offset-corrected RV (right).}
\label{img:GJ832_resid}
\end{figure*}

\begin{figure}
\resizebox{\hsize}{!}{\includegraphics{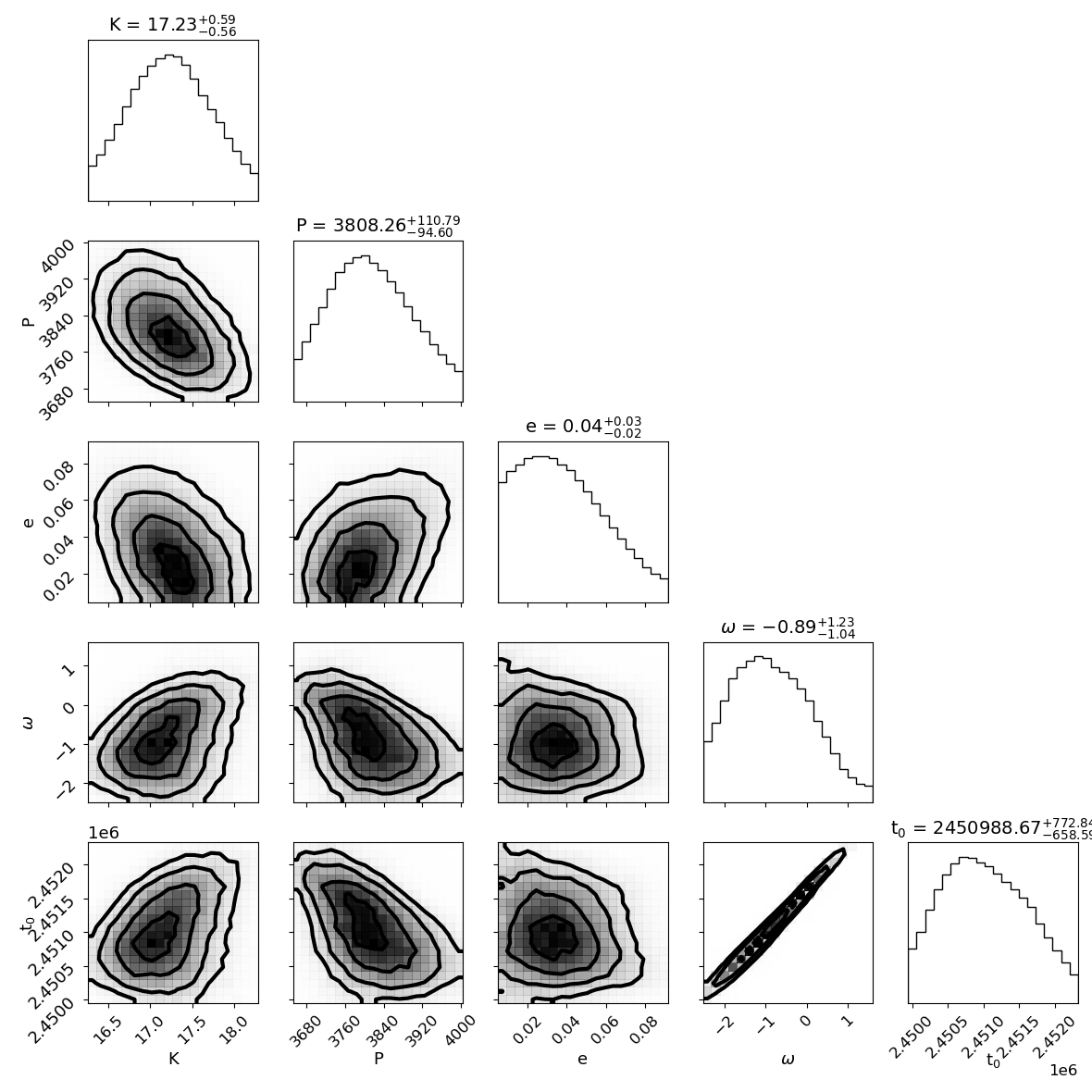}}
\caption{Corner contour and histogram plot for the MCMC derived planetary parameters for GJ\,832\,b after 25\,000 steps. The plotted ranges are limited to exclude the lowest 10 percentiles.}
\label{img:GJ832_planet_corner}
\end{figure}

\paragraph{\textbf{RV results:}}
The model parameters obtained by the MCMC analysis using a one planet + GP-SHO model are listed in Table~\ref{tab:inferredElements}, the posterior corner plot for the planetary orbital elements is shown in Fig.~\ref{img:GJ832_planet_corner} and the full corner plot in Fig.~\ref{img:GJ832_full_corner}. The best fit and its residuals are shown in Fig.~\ref{img:GJ832_resid}, together with the phase-folded and activity-subtracted RV. Our inferred parameters and those reported by \citetads{2022A&A...664A..64G} are in general agreement, though just slightly outside each other's $1\sigma$ uncertainties. In addition, our uncertainties are slightly larger than those previously reported by \citetads{2022A&A...664A..64G}. This might be due to a difference in data reduction since they used NAIRA \citepads{2017A&A...602A..88A} rather than \texttt{serval}. We also notice that the period, $\rho,$ of our GP does not match the rotation period or half the period, but lies in between. This is illustrated in Fig.~\ref{img:GLS_new} within the periodograms of the raw RV and the residuals but also shows that rotation is nonetheless accounted for, as shown by the lack of any remaining signal in the residuals. See the next paragraph and Sect.~\ref{subsec:rotation_discuss} for a deeper discussion.

\subsubsection{GJ 674}
\label{subsubsec:GJ674}
The second planetary system we investigated is GJ\,674, where we do not find any additional planetary signals.

\paragraph{\textbf{Comparing Bayesian evidences $\ln\mathcal{Z}$:}}
We performed Bayesian model comparison analogous to the GJ\,832 system using \texttt{juliet} and 20 individual models. The strongly preferred model comprises one eccentric planet and the dSHO GP kernel and shows decisively higher $\ln\mathcal{Z}$ to all other models. As we explain in Appendix \ref{apdx:lnZ_comparison} however, the fractional amplitude $f$ of the dSHO kernel trends to values above unity. This indicates a preference for the harmonic of the period and is inconsistent with the purpose of the kernel. For this reason we don't consider the dSHO to be a reliable choice. As for GJ\,832, the planetary parameter inference is not impacted by the choice of GP kernel and consistent to the MCMC results. A full listing of the $\ln\mathcal{Z}$ values is given in Table \ref{tab:modelEvidence} and an extended discussion of the Bayesian evidences $\ln\mathcal{Z}$ as well as GP periods obtained in Appendix \ref{apdx:lnZ_comparison}.

\begin{figure*}
\resizebox{\hsize}{!}{\includegraphics{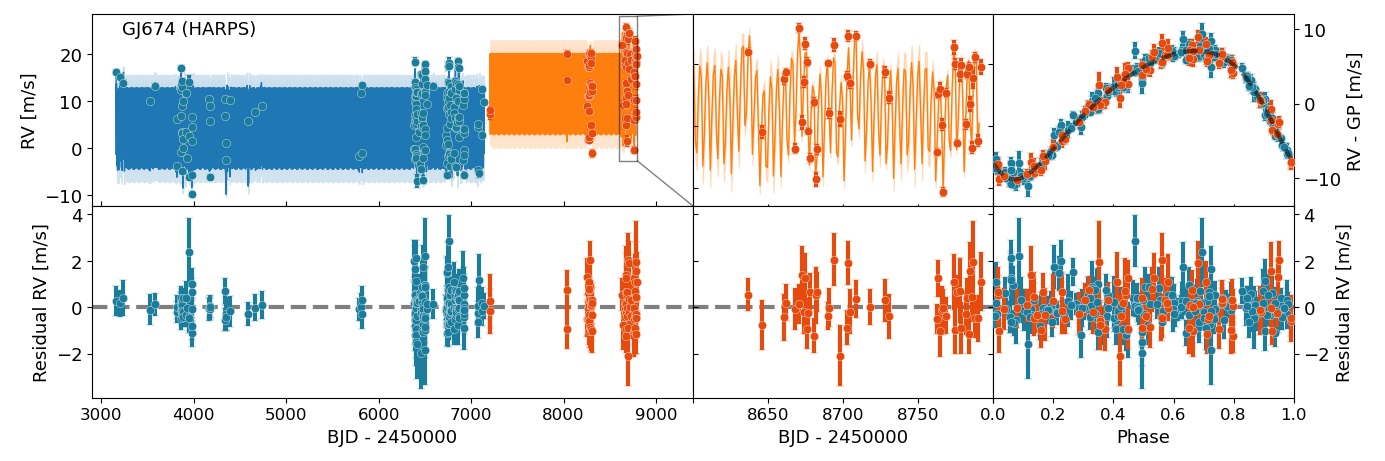}}
\caption{Best fit of the 1 eccentric planet + GP-SHO model (which includes instrumental offsets) in the left panel for GJ\,674 (top) and the residuals (bottom). The model and RV data are separated by a small difference in color shade and the uncertainty of the model fit indicated by the more transparent region. The colors indicate the pre-fiber change (blue) and post-fiber change (orange) HARPS instrument seasons. The model parameters can be found in Table~\ref{tab:inferredElements}. The gray box marks the extent of the zoomed sub-panel. Zoom into the region of densest observations marked by the gray box in the left panel (middle). \textit{} Phase-folded, activity-subtracted, and instrumental offset-corrected RV (right).}
\label{img:GJ674_resid}
\end{figure*}

\begin{figure}
\resizebox{\hsize}{!}{\includegraphics{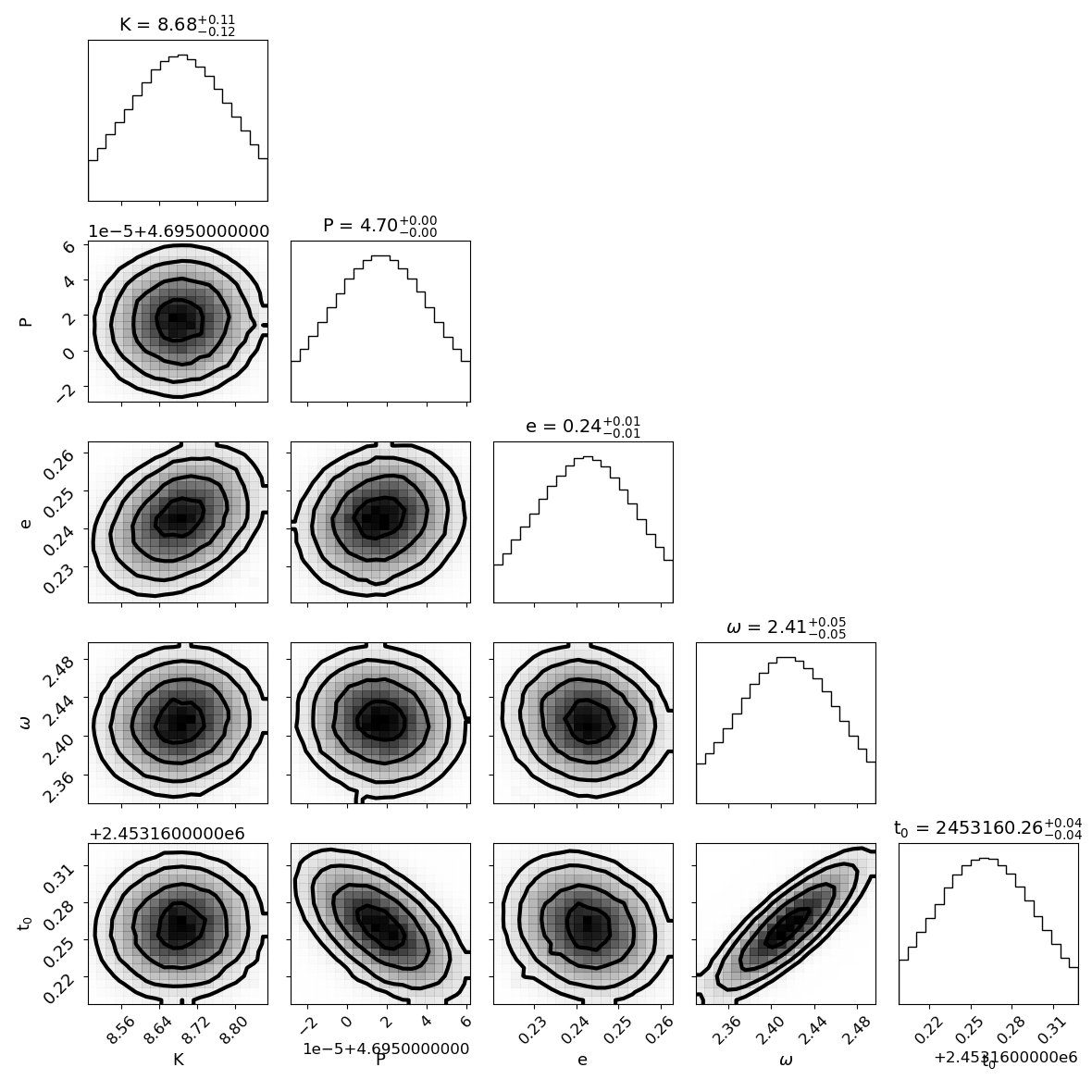}}
\caption{Corner contour and histogram plot for the MCMC derived planetary parameters for GJ\,674\,b after 25\,000 steps. The plotted ranges are limited to exclude the lowest 10 percentiles (see. Sect.~\ref{subsec:MCMC_setup}).}
\label{img:GJ674_planet_corner}
\end{figure}

\paragraph{\textbf{RV results:}}
The updated orbital parameters of the known planet, using MCMC and the one planet + GP-SHO model, are shown in Table~\ref{tab:inferredElements}. The planetary and full corner plots are shown in Figs.~\ref{img:GJ674_planet_corner} and \ref{img:GJ674_full_corner} respectively. The best fit and its residuals are shown in Fig.~\ref{img:GJ674_resid}, together with the phase-folded and activity-subtracted RV. The results for the RV semi-amplitude and orbital period are in good agreement with the previous findings of \citetads{2007A&A...474..293B} with significantly reduced uncertainties owing to the regular observational cadence of the additional RedDots data. However, our results show a planetary eccentricity value that is about two sigma higher than the eccentricity determined by \citetads{2007A&A...474..293B}.
\par
The difference in eccentricity could be because (1) we used a GP to fit the rotational modulation or (2) our analysis includes an additional 20 RV measurements. To investigate this further we analyzed the original RV data published by \citetads{2007A&A...474..293B} using the same methods described in Sect.~\ref{sec:analysis}. This resulted in an eccentricity of $0.22\pm0.04$ which is higher than the original value of $0.20\pm0.02$ but within the one sigma uncertainty. All other planetary parameters are in good agreement with the original orbital parameters of \citetads{2007A&A...474..293B}. Our results show that the differences in reconstructed eccentricity values is therefore most likely the result from the additional RedDots data points (see also a similar discussion in Sect.~\ref{subsubsec:Ross128} for Ross\,128).
\par
One difference between the results from the \citetads{2007A&A...474..293B} data and our results is in the recovered GP period $\rho$. From the \citetads{2007A&A...474..293B} RV data, the GP period is close to the $\sim35$ day stellar rotation period. From our extended RV data set, the GP converged to a period between the photometric full and half rotation periods from the literature (refer to Table~\ref{tab:Params_known}), as seen in Fig.~\ref{img:GLS_new}. This phenomenon of the GP period not matching the photometric rotation period of the star is further discussed in Sect.~\ref{subsec:rotation_discuss}.

\paragraph{\textbf{Photometric results:}}
To confirm the literature stellar rotation period from \citetads{2007A&A...474..293B} of 35\,d, we employed ground based $B$ and $V$ photometry (detailed in Sect.~\ref{subsec:phot_data}; shown in Fig.~\ref{img:GJ674_photometry_GLS}). The rotation period recovered from the periodogram confirms the literature value at 33\,d.
Contrary to GJ\,832 however, we can see peaks in the periodogram of the RV for GJ\,674 at both the 21\,d GP period and stronger peaks at the 33\,d photometric period. We additionally note that the photometric periodograms in Fig.~\ref{img:GJ674_photometry_GLS} show an unexpected peak at the location of the planet's orbital period. Though at low significance, the coincidence of the signal with the planetary orbital period together with the expectation by \citetads{2019MNRAS.488..633V} for a high chance of detectable star-planet interactions, caused us to take a closer look in Appendix~\ref{apdx:GJ674_add_signals}. There, we also discuss a spurious, 7\,m\,s$^{-1}$ RV semi-amplitude signal.

\subsubsection{Ross 128}
\label{subsubsec:Ross128}

The third planetary system that we analyzed is Ross\,128. Similar to GJ\,674, we did not find any additional planets. Since Ross\,128 has the highest number of RV data points, we analyzed the results in terms of HARPS+CARMENES, HARPS$_\mathrm{all}$, and HARPS$_\mathrm{RedDots}$ data sets.  Most notably the HARPS$_\mathrm{RedDots}$ data set was secured with regular cadence observations at an $S_\mathrm{K}$-index activity minimum \citepads{2019A&A...628L...1I} and covers almost nine densely sampled planetary periods.

\paragraph{\textbf{Comparing Bayesian evidences $\ln\mathcal{Z}$:}}
As for GJ\,832 and GJ\,674, we investigated 20 possible models for the Ross\,128 data using the Bayesian evidences obtained from \texttt{juliet}. The $\ln\mathcal{Z}$ values are shown in Table \ref{tab:modelEvidence} and we present a detailed explanation in Appendix \ref{apdx:lnZ_comparison}. The best fit parameters are shown in Tables \ref{tab:comparison_posteriors_inst_params} -- \ref{tab:comparison_posteriors_P2_params}.\par
For the HARPS+CARMENES set of observations, the three best models show comparable Bayesian evidence. These models are (i) two circular planets + SHO GP model, (ii) one circular planet + SHO GP, (iii) one eccentric planet + SHO GP. This is in full agreement with the HARPS$_\mathrm{all}$ set of observations.\par

We investigated each of the three cases. For case (i), we investigated the possibility of a second planet and find that there is a visible signal at 4 days in the one planet models' residuals. We attribute this signal to stellar activity because it is also present in the $H_\alpha$ periodogram. This is explained in more detail in Appendix \ref{apdx:Ross128_add_signals}. This signal is not visible in the RedDots subset of observations during the expected $S_\mathrm{K}$-index activity minimum (Sect. \ref{sec:targets}, \citeads{2019A&A...628L...1I}). For cases (ii) and (iii), the only difference between these two models is the planetary orbital eccentricity. We performed a detailed investigation in the following Sect. \ref{subsec:ecc_validation} and find it unlikely that the eccentricity recovered by the eccentric model is a result of data uncertainty.\par

Finally we note that for the HARPS$_\mathrm{RedDots}$ subset, the Bayesian evidence additionally indicates an equal probability of a solution with zero planets and SHO or QP GP. As we have noted, the HARPS$_\mathrm{RedDots}$ observations were secured during an $S_\mathrm{K}$-index minimum, according to \citetads{2019A&A...628L...1I}. In the absence of a significant contribution by stellar activity, the GP hyperparameters instead converged to mask the Keplerian component of the signal. The result of this is that the Bayesian evidence prefers the model without any planets due to its much lower complexity. We discuss the reliability of using $\ln\mathcal{Z}$ in more detail in the discussion (Sect. \ref{subsec:lnZ_discussion}). We conclude that the most accurate model for Ross\,128 is one eccentric planet + SHO GP.

\paragraph{\textbf{RV results:}}
After deciding on the one eccentric planet + SHO GP model, we restarted the MCMC analysis with only the HARPS$_\mathrm{all}$ data set to restrict the initial complexity. Previously, for GJ\,832 and GJ\,674, the planet parameters formed easily identifiable peaks in the posterior distributions. However, Ross\,128 initially showed three main peaks in the posterior for the planetary period (Fig.~\ref{img:Ross128_peak_zoom}, top-left). We explored this phenomenon in more detail in Appendix \ref{apdx:Ross128_add_signals} and find that the most likely true planetary orbital period is at 9.86\,d.\par

\begin{figure*}
\resizebox{\hsize}{!}{\includegraphics{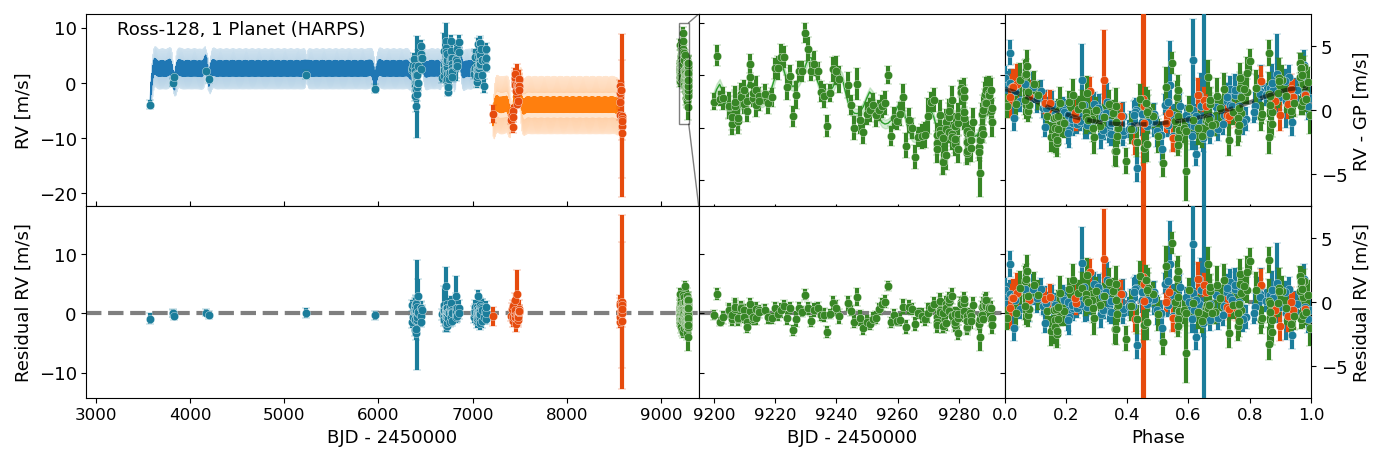}}
\caption{\textit{} Best fit of the 1 eccentric planet + GP-SHO model (which includes instrumental offsets) in the left panel for Ross\,128 (top) and the residuals (bottom). The model and RV data are separated by a small difference in color shade and the uncertainty of the model fit indicated by the more transparent region. The colors indicate the pre-fiber change (blue), post-fiber change (orange), and post-warmup (green) HARPS instrument seasons. The model parameters can be found in Table~\ref{tab:inferredElements}. The gray box marks the extent of the zoomed sub-panel. Zoom into the region of densest observations marked by the gray box in the left panel (middle). \textit{} Phase-folded, activity subtracted, and instrumental offset corrected RV (right).}
\label{img:Ross128_resid}
\end{figure*}

\begin{figure}
\resizebox{\hsize}{!}{\includegraphics{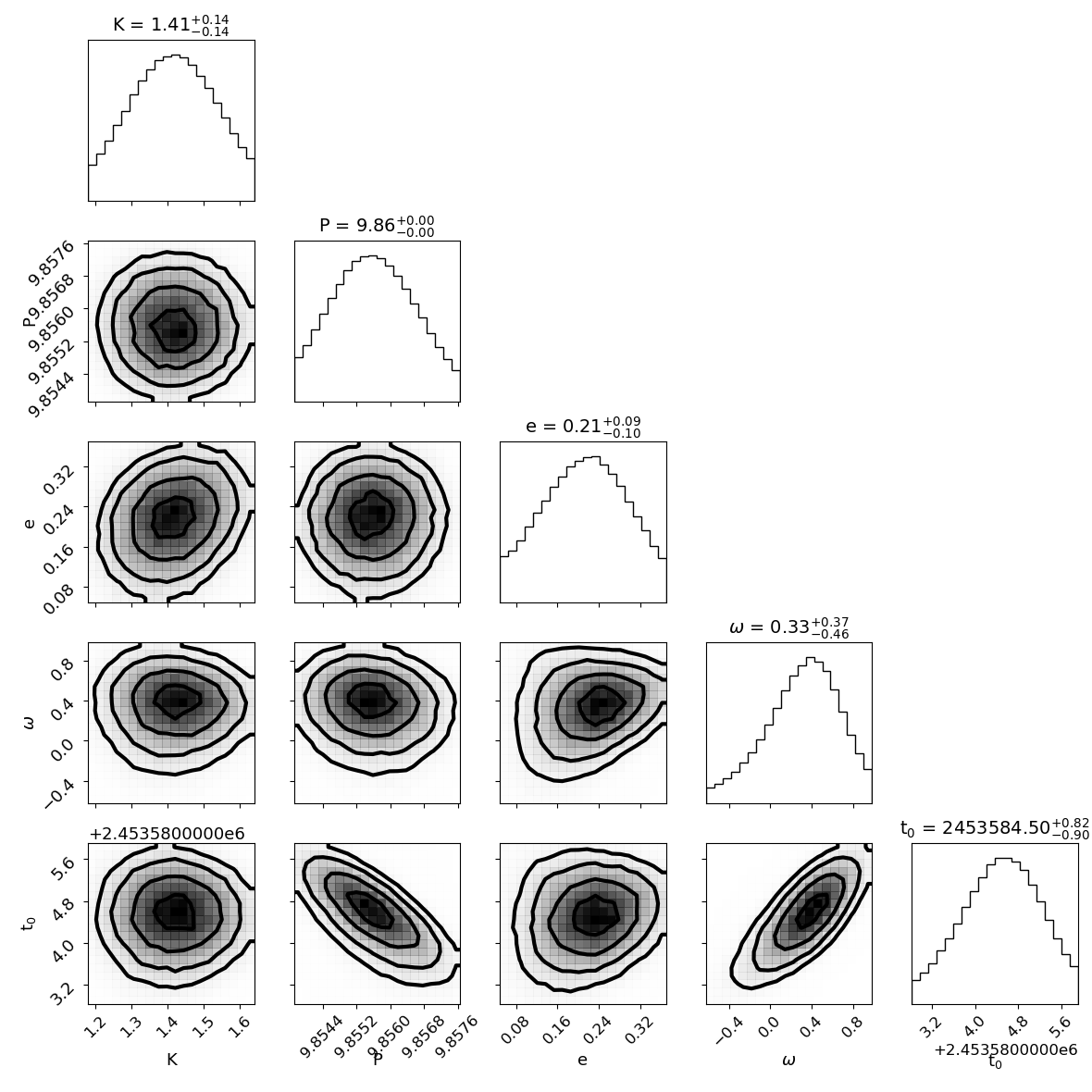}}
\caption{Corner contour and histogram plot for the MCMC derived planetary parameters for Ross\,128\,b using only HARPS data after 25\,000 steps. The plotted ranges are limited to exclude the lowest 10 percentiles (see. Sect.~\ref{subsec:MCMC_setup}).}
\label{img:Ross128_planet_corner_1P}
\end{figure}

We show the revised orbital parameters using MCMC, only HARPS$_\mathrm{all}$ data, and the model of one eccentric planet + GP-SHO in Table~\ref{tab:inferredElements} and the corner plots in Figs.~\ref{img:Ross128_planet_corner_1P} and \ref{img:Ross128_full_corner_1P}. The best fit and its residuals are shown in Fig.~\ref{img:Ross128_resid}, together with the phase-folded and activity-subtracted RV. The most notable difference between our results and the orbital parameters previously reported by \citetads{2018A&A...613A..25B} is the planetary eccentricity. Our results show an eccentricity of $0.21^{+0.09}_{-0.10}$ for the full HARPS dataset, whereas \citetads{2018A&A...613A..25B} derived a value of $0.116\pm0.097$, just outside the one sigma uncertainty. As with GJ\,674, we find additional spurious signals in the Ross\,128 data that we present in Appendix~\ref{apdx:Ross128_add_signals}.

\begin{figure}
\resizebox{\hsize}{!}{\includegraphics{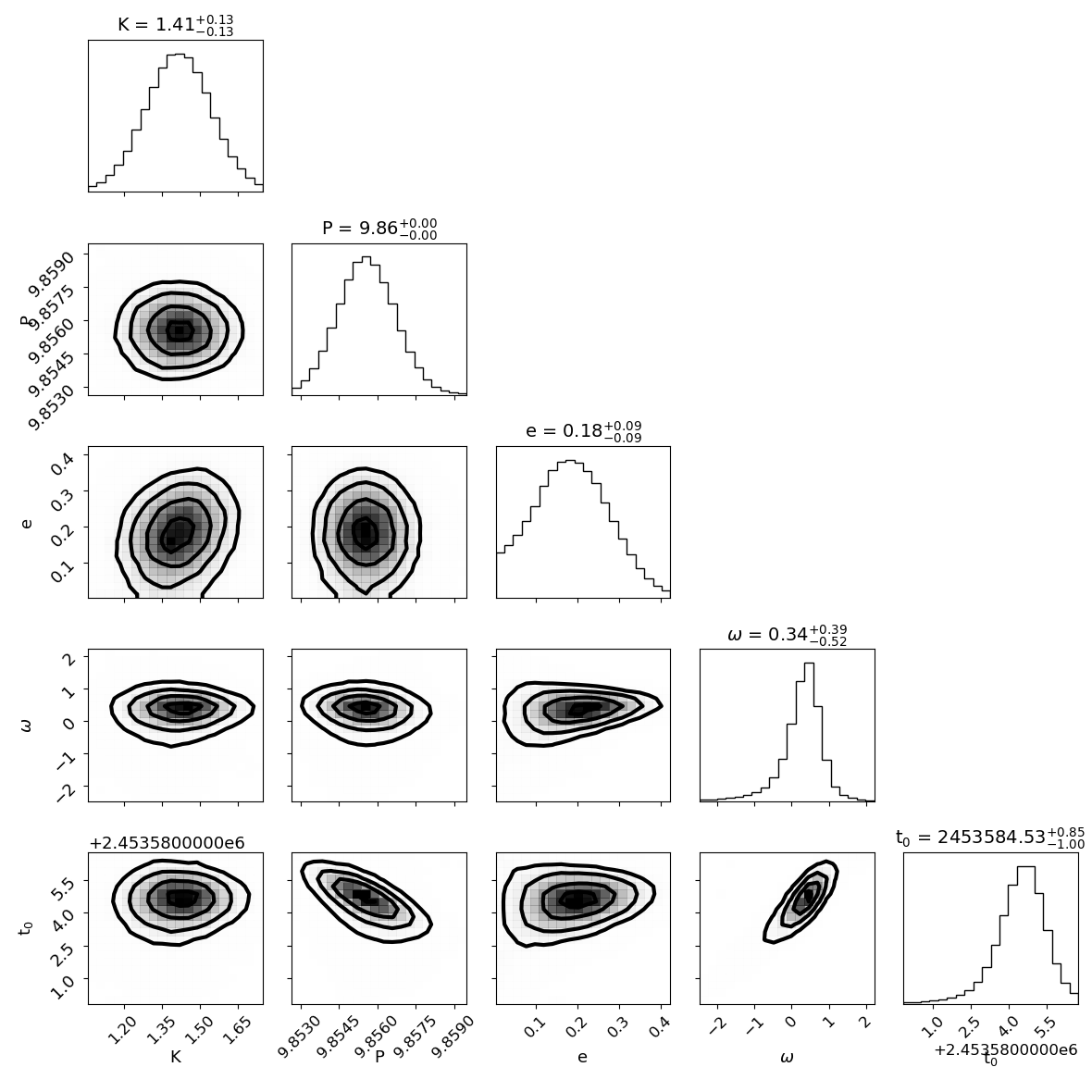}}
\caption{Corner contour and histogram plot for the MCMC derived planetary parameters for Ross\,128\,b  using HARPS and CARMENES data after 25\,000 steps. The plotted ranges are limited to exclude the lowest 10th percentiles (see. Sect.~\ref{subsec:MCMC_setup}).}
\label{img:Ross128_planet_corner_1P_carmenes}
\end{figure}

Adding the CARMENES observations (Sect.~\ref{subsubsec:CARMENES}) to the full set of HARPS RVs had no impact on the derived planet parameters (last column of Table~\ref{tab:inferredElements}) or any of the previously highlighted behavior beyond a minuscule reduction in uncertainty, a minor drop in eccentricity to $0.18\pm0.09$ that is well within the uncertainty, and a minor increase in the GP timescales, $\rho$ and $\tau$, again well within the error bars. The corner plots are shown in Figs.~\ref{img:Ross128_planet_corner_1P_carmenes} and \ref{img:Ross128_full_corner_1P_carmenes}. The fitted model and residuals are given in Fig.~\ref{img:Ross128_Carmenes_resid}, together with the phase-folded and activity-subtracted RV, while the model evidence $\ln\mathcal{Z}$ for different models is shown in Table~\ref{tab:modelEvidence}. Comparing the evidences, it seems that adding the CARMENES data increases $\Delta\mathcal{Z}$ in favor of the one planet plus GP models though lowers the overall evidence compared to the HARPS data alone. The combined data set also still shows a marginal preference towards the circular + SHO kernel model comparable to the full HARPS set.

\paragraph{\textbf{Photometric results:}}
Using the photometry detailed in Sect.~\ref{subsec:phot_data} for Ross\,128, obtained simultaneously to our RedDots RVs, we examined the stellar rotational modulations observable during this time frame. We obtained a wide (poorly sampled) but significant peak at 223 days (see Fig.~\ref{img:Ross128_photometry_GLS}) in the R and V filters. Not only is this inconsistent with either of the stellar rotation periods listed in the literature, it also seems to support the period recovered by the SHO GP kernel for the RedDots subset. While a lack of, or difference in, the photometric rotation signal would not be unexpected during an activity minimum, the coincidence with the GP period is noteworthy.\par

\subsection{Validating the Ross\,128 eccentricity}
\label{subsec:ecc_validation}
We aimed to validate the eccentricity value inferred in Sect. \ref{subsubsec:Ross128} for Ross\,128. To this end, we investigated the variation in orbital eccentricity between our results and those of \citetads{2018A&A...613A..25B} (compare Tables~\ref{tab:litElements} and \ref{tab:inferredElements}). This was done in several steps:\par

(i) We cross-checked our results using exactly the same RV values as published by \citetads{2018A&A...613A..25B} for their set of observations using our preferred MCMC model of one eccentric planet + SHO GP. This resulted in an eccentricity value of $0.15^{+0.12}_{-0.10}$.\par

(ii) Instead of using a GP, we included a linear instrumental drift of 36\,cm\,s$^{-1}$\,yr$^{-1}$, as performed by \citetads{2018A&A...613A..25B}. This resulted in an eccentricity value of $0.12^{+0.11}_{-0.08}$, compatible with the original determination by \citetads{2018A&A...613A..25B}.\par

(iii) We repeated the analysis with MCMC + SHO GP and using RVs reprocessed by \texttt{serval} from the observations used by \citetads{2018A&A...613A..25B}. This resulted in a planetary eccentricity of $0.14^{+0.12}_{-0.9}$ without use of the drift correction. As this is very similar to \citetads{2018A&A...613A..25B}, this excludes the data reduction procedure as the source for our increased planetary eccentricity and leaves only model differences. \par

(iv) We added the drift as a parameter instead of GP for the reprocessed \texttt{serval} RVs. This resulted in an eccentricity of $0.19^{+0.14}_{-0.12}$, compatible again with our initial determination.\par

(v) We analyzed the HARPS$_\mathrm{all}$ RV with both, an instrumental drift parameter and the SHO GP, simultaneously. Including the instrumental drift parameter did not make any difference to the planetary eccentricity.\par

(vi) We further utilized nested sampling to check our MCMC-based comparison of our obtained eccentricity values to the results by \citetads{2018A&A...613A..25B}. Using again the one eccentric planet + GP model, we essentially reproduced the results obtained by MCMC with an eccentricity value of $0.2\pm0.1$ for the full set with GP and $0.13^{+0.13}_{-0.09}$ without GP.\par

(vii) We also computed the planetary eccentricity using only the 138 observations over 87 nights of the HARPS RedDots dataset. This is nearly the same number of observations as the original sample by \citetads{2018A&A...613A..25B} (196 observations) but at a much denser, uninterrupted sampling of almost nine consecutive planetary orbits. These data were collected after \citetads{2018A&A...613A..25B} was published. Using the RedDots dataset we calculated a planetary eccentricity of $0.27^{+0.19}_{-0.18}$ using MCMC (the full set of fitted parameters is show in Table~\ref{tab:inferredElements}) and a compatible value of $0.31^{+0.41}_{-0.20}$ with nested sampling. This dataset, according to the findings by \citetads{2019A&A...628L...1I}, was secured at an $S_\mathrm{K}$-index activity minimum (see the discussion in Appendix \ref{apdx:lnZ_comparison}), and consequently will be less affected by spurious activity signals which may bias the eccentricity determination. Such a case was shown by \citetads{2021AJ....162..181L}. They find that during the activity minimum of $\epsilon$\,Eridani the recovered eccentricity of $\epsilon$\,Eridani\,b dropped from 0.6 to consistent with zero. While this is opposite to our results, which show an increased eccentricity during activity minimum, it illustrates the possibility for activity to influence the eccentricity measurement.\par

In summary, these tests support the conclusion that the eccentricity value is sensitive to the model. Overall, including the drift correction in the models along with the GP had no effect on any of the planetary orbital parameters. It appears that the GP is able to account for the drift, though systematically increases the eccentricity slightly but not enough to explain the difference of slightly more than one sigma compared to \citetads{2018A&A...613A..25B}. Finally, we consider these tests part of the validation that the true planetary eccentricity, according to our data, is likely close to or above 0.2 rather than the value of 0.1 reported by \citetads{2018A&A...613A..25B}. This leaves the possibility that the eccentricity is an effect of measurement uncertainty (statistical or due to measurement noise) from an otherwise circular orbit.\par

\citetads{1971AJ.....76..544L} show that proving a recovered eccentricity is non-zero is more difficult than one might expect. This is due to the mathematical boundary posed by $e=0$ and the corresponding asymmetry in the posterior probability space. \citetads{1971AJ.....76..544L} show that for the maximum-likelihood algorithm, a derived eccentricity with uncertainty $\sigma$ can only be considered distinct from zero at a separation of $2.45\sigma$ rather than the standard $2\sigma$ in order to reflect the expected 95\% confidence. Our obtained values for GJ\,832 and Ross\,128 (Table~\ref{tab:inferredElements}) fall below this Lucy-Sweeney limit. However, the Lucy-Sweeney limit was not derived for MCMC or nested-sampling or any other modern Bayesian inference procedure, especially as we did not sample the eccentricity itself (Sect.~\ref{subsec:MCMC_setup}). Comparing the Bayesian evidences (Table~\ref{tab:modelEvidence}), the result is inconclusive between the circular and eccentric models and for this reason we decided to check ourselves whether our recovered eccentricities could be a random result for an actually circular orbit.\par
Using 100 independent white-noise realizations (Sect.~\ref{subsec:glsGrids} and Appendix~\ref{apdx:striping}), we attempted to recover signals for circular orbits with period and semi-amplitude matching the known planet. We useed a full Keplerian inference, including eccentricity, as we did for the main analysis of Ross\,128. The distribution of recovered eccentricities is asymmetrical as expected, with an extended tail towards higher eccentricities. The mean recovered eccentricity is $0.053_{-0.014}^{+0.026}$, consistent with an injection-recovery from the residuals of our primary inference. This places the expected recovered value of a circular signal below the originally published value of $e=0.1\pm0.1$ (by $1.8\sigma$) \citepads{2018A&A...613A..25B} and significantly below our inference of $e=0.21^{+0.09}_{-0.10}$ (by $5.6\sigma$). It is important however that these significance values are not taken as metrics for the significance of the recovered eccentricity value itself. They quantify instead the significance of the hypothesis that the eccentricity is any non-zero value instead. This can be seen in that none of the simulations resulted in an eccentricity estimate comparable to our real result and only three simulations marginally exceeded the $e=0.1$ level, while the peak of the distribution is around $e\approx0.03$. From this we believe that our derived eccentricity for Ross\,128\,b, while not outside the Lucy-Sweeney limit and inconclusive in the Bayesian evidence (Table~\ref{tab:modelEvidence}; See our discussion in Sect. \ref{subsec:lnZ_discussion}), is still significant enough to cast strong doubts on the possibility of being the result of statistical effects hiding a circular orbit.\par
Since Ross\,128\,b, according to our analysis, is unlikely to have a low eccentricity and is located at the inner edge of the liquid-water habitable zone, this means that its overall habitability is strongly impacted by this finding. We discuss the implications of this in Section \ref{subsec:HZ_limits}. To accurately resolve the question of the true eccentricity of Ross\,128\,b, additional observations with cadences similar to the RedDots data set would be necessary, coupled with a more sophisticated modeling approach for the injection-recovery test beyond the white noise assumption employed here. These sets of observations would also ideally be spread over different parts of the seven-year S-index activity cycle of Ross\,128 \citepads{2019A&A...628L...1I} in order to answer the question of the possibility of low-activity observation times impacting the reliability of Bayesian evidence values with employed GP (see Sect. \ref{subsec:lnZ_discussion}).

\subsection{Detection limits from injected planets}
\label{subsec:glsGrids}

\begin{figure*}
\resizebox{0.32\hsize}{!}{\includegraphics{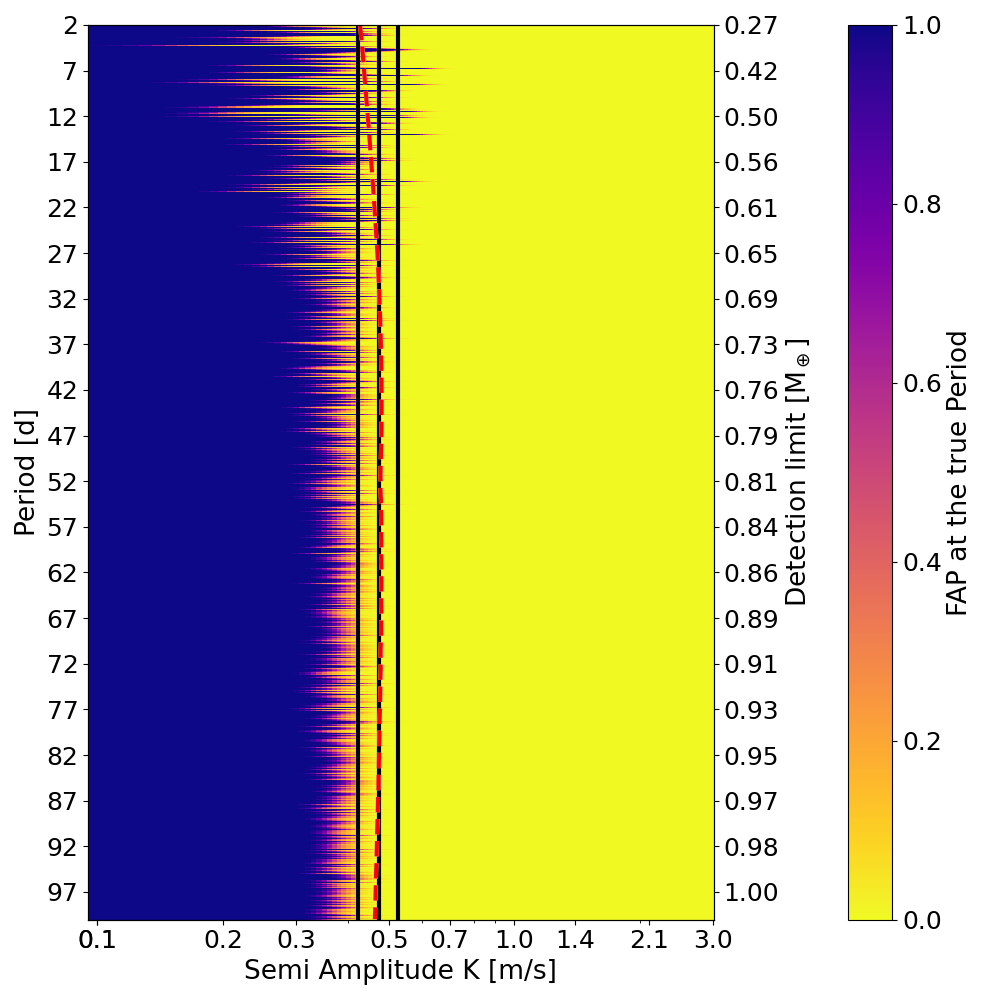}}
\resizebox{0.32\hsize}{!}{\includegraphics{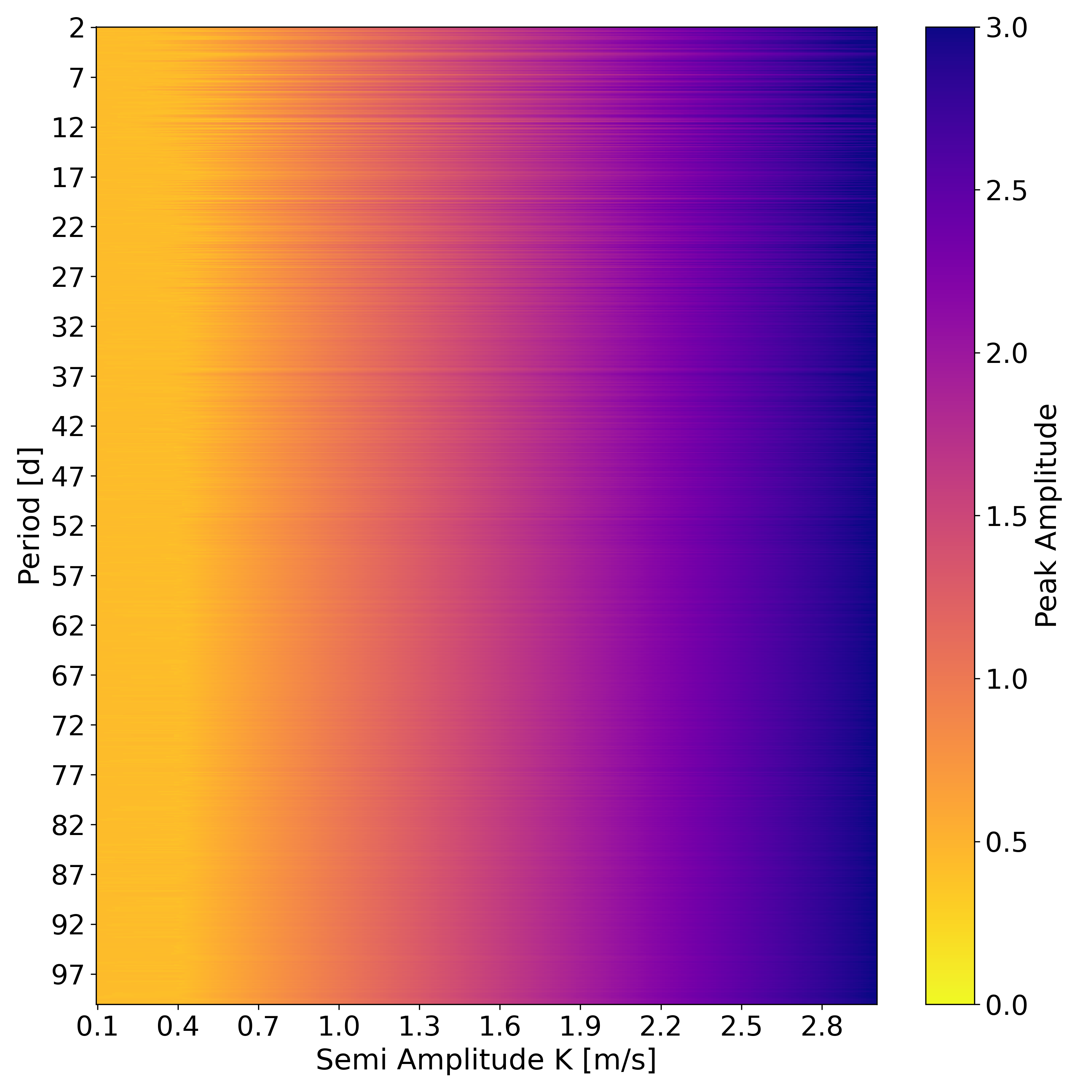}}
\resizebox{0.32\hsize}{!}{\includegraphics{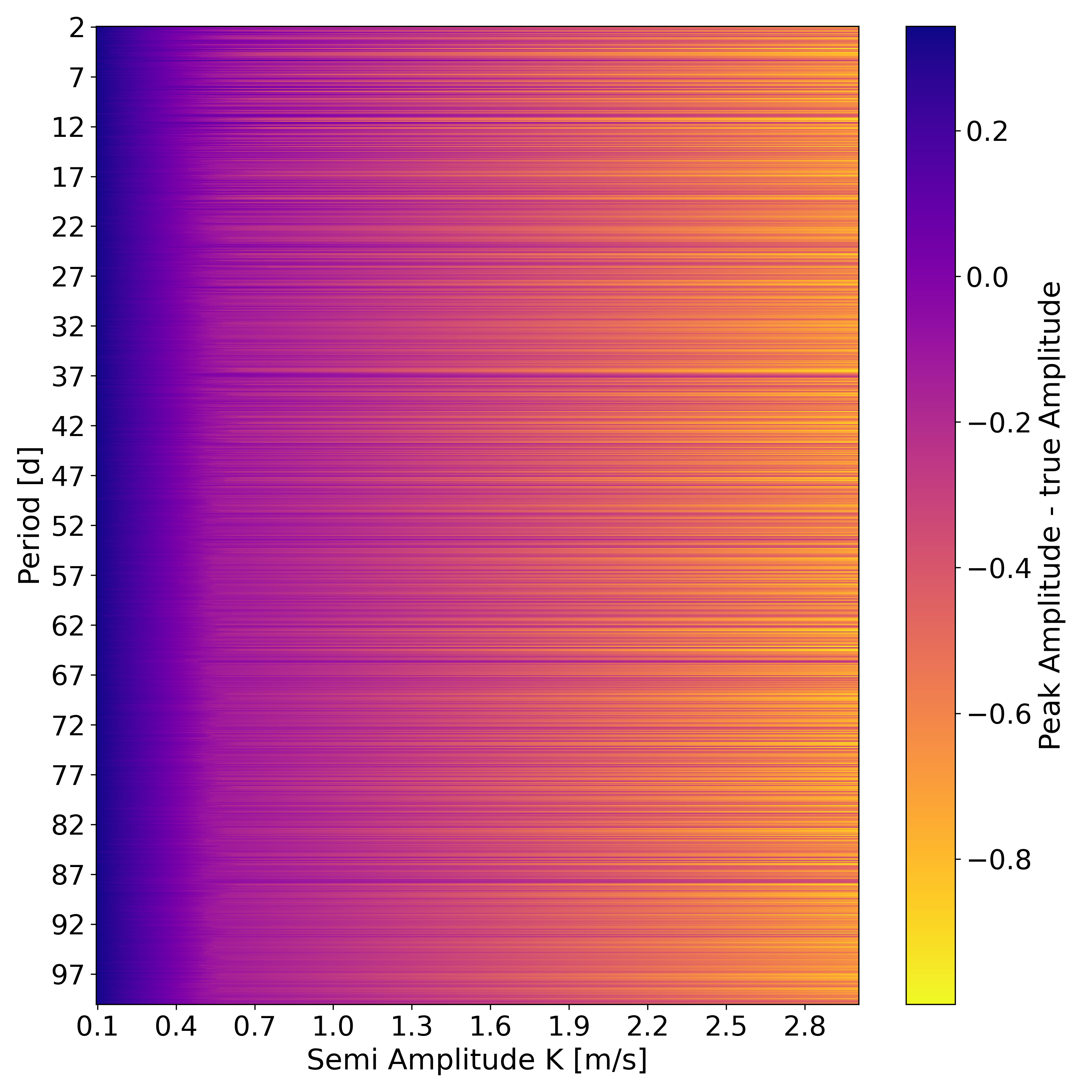}}
\caption{Grid of detections from GLS periodograms for planetary signals injected into the Ross\,128 residual RVs. \textit{} The left panel shows a mean detection boundary (black line plus-minus one standard deviation) and a SavGol smoothing (red line) are marked and the grid colored for the false-alarm-probability (FAP) at the true, injected period for zero eccentricity. The determined limit does not change significantly when the BIC is used in place of the FAP or on randomization of the orbital parameters within reasonable bounds (See Sect. \ref{subsec:glsGrids}). The zero eccentricity grid is colored for the RV Semi-amplitude of the strongest periodogram peak in each model (middle). A striped pattern of systematic offsets depending on the injected period is visible. The grid for $e=0.5$ is colored for difference of the RV Semi-amplitude of the strongest periodogram peak in each model and the injected semi-amplitude (right). A systematic offset of the recovered semi-amplitude is apparent.}
\label{img:GLS_detect_Ross128}
\end{figure*}

\begin{table}
\centering
\caption{Fixed orbital parameters for the injected synthetic signal.}
\label{tab:orbital_params}
\begin{tabular}{llc}
\hline\hline
\noalign{\smallskip}
Parameter & Unit & Value\\
\noalign{\smallskip}
\hline
\noalign{\smallskip}
Period $P$ & [d] & 2 -- 100\\
Semi-amplitude $K$ & [m\,s$^{-1}$] & 0.1 -- 3.0\\
Eccentricity $e$ & & 0.0 -- 0.9\\
Periastron argument $\omega$ & [rad] & $\frac{\pi}{2}$\\
Periastron passage $t_0$ & [d] & 0\\
\noalign{\smallskip}
\hline
\end{tabular}
\end{table}

To assess the detection limits for potential second planets for GJ\,832, GJ\,674 and Ross\,128, artificial Keplerian signals were added to the residuals from the models of Sect.~\ref{subsec:OrbElements} which gave the parameters shown in Table~\ref{tab:inferredElements}. This allowed us to retain the noise characteristics and windowing behavior due to the sampling and observation of the original data without needing any assumptions. We term the residual time series as noise in the following, under the assumption that there are no more signals left, as shown in Fig.~\ref{img:GLS_new}\par

The injected signal was parameterized in RV semi-amplitude, $K$, period, $P,$ and eccentricity, $e$, all other orbital parameters were kept constant as shown in Table~\ref{tab:orbital_params}. For each eccentricity value, a first grid was sampled in $K$ versus $P$ at $\delta_P=0.1$\,d and $\delta_K=0.01$\,m\,s$^{-1}$ for a total of $\approx$285\,000 points each. The corresponding Keplerian signal for each grid point was injected into the noise and analyzed with GLS periodograms to determine the highest-power period, its false-alarm probability (FAP) and RV semi-amplitude, and the FAP at the true, injected period. The FAP was calculated by the GLS script as detailed in \citetads{2009A&A...496..577Z}. We further calculated the Bayesian information criterion (BIC) for the circular Keplerian signal corresponding to the highest power periodogram peak and a fitted constant model.\par

We determined the detection limits based on the zero-eccentricity models under the assumption that for multi-planet systems the eccentricities should not be too large to remain dynamically stable and that for small eccentricities the recoveries from GLS periodograms will not deviate too much from the simplification of sinusoidal rather than Keplerian signals (see Sect.~\ref{subsec:GLS_MCMC} for validation).\par

The detection limit was then determined from the grids in $K$ versus $P$ using the FAP at the true period as the indicator (Figs.~\ref{img:GLS_detect_Ross128} left panel, \ref{img:GLS_detect_GJ832_boundary}, \ref{img:GLS_detect_GJ674_boundary}). A limit of $\mathrm{FAP}=0.01$ is set to obtain a preliminary detection boundary on a per-period basis, using the highest semi-amplitude each that still violates the FAP limit for a given period. This rough boundary was approximated using either the mean semi-amplitude, averaged over the sampled periods, and its corresponding standard deviation or a Savitzki-Golay (SavGol) filter of third order, which covers the full period range. The resulting smoothed boundaries are shown in Figs.~\ref{img:GLS_detect_Ross128} (left panel), \ref{img:GLS_detect_GJ832_boundary}, and \ref{img:GLS_detect_GJ674_boundary} as well. The difference between the SavGol filtering and median are small over nearly the full period range, showing that the limit is well characterized by RV semi-amplitude alone. The SavGol limit shifts towards lower periods, starting between 20 and 30 days. Below that, the SavGol curve corresponds to a detection limit up to one sigma lower than the average. This decrease can be traced to the dense, regular sampling of the RedDots data, which sample short period planets much better. From the mean limits we conclude that with the present data, the existence of any additional planets with RV semi-amplitudes above 0.35\,m\,s$^{-1}$ (GJ\,832), 0.29\,m\,s$^{-1}$ (GJ\,674), and 0.47\,m\,s$^{-1}$ (Ross\,128) can be rejected in the period range of 1 -- 100\,days. Assuming a canonical orbit of 10 days at zero eccentricity, this would correspond to minimum planet masses of 0.69, 0.48 and 0.46 Earth masses respectively for GJ\,832, GJ\,674, and Ross\,128. Mass limits per-period are given on the second axis of Figs.~\ref{img:GLS_detect_Ross128} (left panel), \ref{img:GLS_detect_GJ832_boundary}, and \ref{img:GLS_detect_GJ674_boundary}. Each mass value was calculated based on the mean detection limit in velocity semi-amplitude, the stellar mass from Table~\ref{tab:Params_known}, and zero eccentricity.\par

We validated our use of the FAP as the indicator for the detection of the injected signal's presence with the BIC. A comparison of the $\Delta\mathrm{BIC}=10$ limit, the threshold where the constant model is considered disfavored against the Keplerian one at a BIC difference of 10, results in nearly the same detection limits. For all three systems, the $\Delta\mathrm{BIC}$ boundary was 4-5\,cm\,s$^{-1}$ higher, consistent with a marginally stricter FAP limit definition. Increasing the eccentricity of the injected signal increased the detection limit, as we discuss in Sect.~\ref{subsec:GLS_MCMC}, by about 20\% at $e=0.5$ and 140\%  at $e=0.85$.
Application of a shift in phase or periastron argument of the injected Keplerian signals had no effect on the detection limits in either FAP or BIC at any eccentricity. Using white noise realizations of the residual noise, following the approach from Appendix~\ref{apdx:striping}, showed no deviation from the initial results.

In a second step, we extended the grids along the period axis to determine the limit in period up to which the constant detection limit will hold. To this end we set up another grid, logarithmically spaced in period between 100 and 100\,000\,days with 1\,500 points, grid points in semi-amplitude identical to the short period grid, and for fixed eccentricity, $e=0$. These periods exceed the time base of our available observations by a significant margin, precluding the exact determination of orbital parameters. However, since we only look at the FAP of the periodogram peaks and the $\Delta\mathrm{BIC}$ to a constant model, it is possible to draw conclusions on the presence of signals even at periods that are much longer than the observational base line.\par
For all three systems and both limit definitions these presence detection limits remained constant up to 10\,000\,days and only started to visibly increase around 20\,000\,days. The 10\,000 -- 100\,000\,day range is shown in Fig.~\ref{img:very_long_detect_grids}, where one can see each grid starting at limits identical to the short period grids and gradually increasing towards the longest periods, crossing the 1\,m\,s$^{-1}$ point around 60\,000\,days (GJ\,832 and GJ\,674) or 30\,000\,days (Ross\,128). This would correspond to planets with 36, 30, and 14 earth masses respectively. The correspondence between FAP limits and $\Delta\mathrm{BIC}$ remains unchanged at very long periods.\par

One peculiarity of the grids is shown in Fig.~\ref{img:GLS_detect_Ross128}, middle panel. While also weakly present in the FAP coloring, using the RV amplitude of the strongest periodogram peak for color distinctly shows a structure of horizontal stripes at constant periods. These stripes mirror the structures close to the detection boundary in FAP but further extend into the detected region as an offset in the recovered amplitude to the injected value of up to 20\,cm\,s$^{-1}$. These offsets are still comparatively small with a mean deviation of 3\,cm\,s$^{-1}$ and consistent with the expected background behavior.  The horizontal stripes are present in the grids of GJ\,832, GJ\,674, and Ross\,128, though at different periods each, warranting a closer look into their origins. We investigated this further in Appendix~\ref{apdx:striping}, using white noise realizations. The origin of the stripes can be traced as a purely statistical effect from the noise and is not due to windowing or any other systematic source.\par

\subsection{Limits for planets within the habitable zones}
\label{subsec:HZ_limits}

\begin{figure*}
\resizebox{0.27\hsize}{!}{\includegraphics{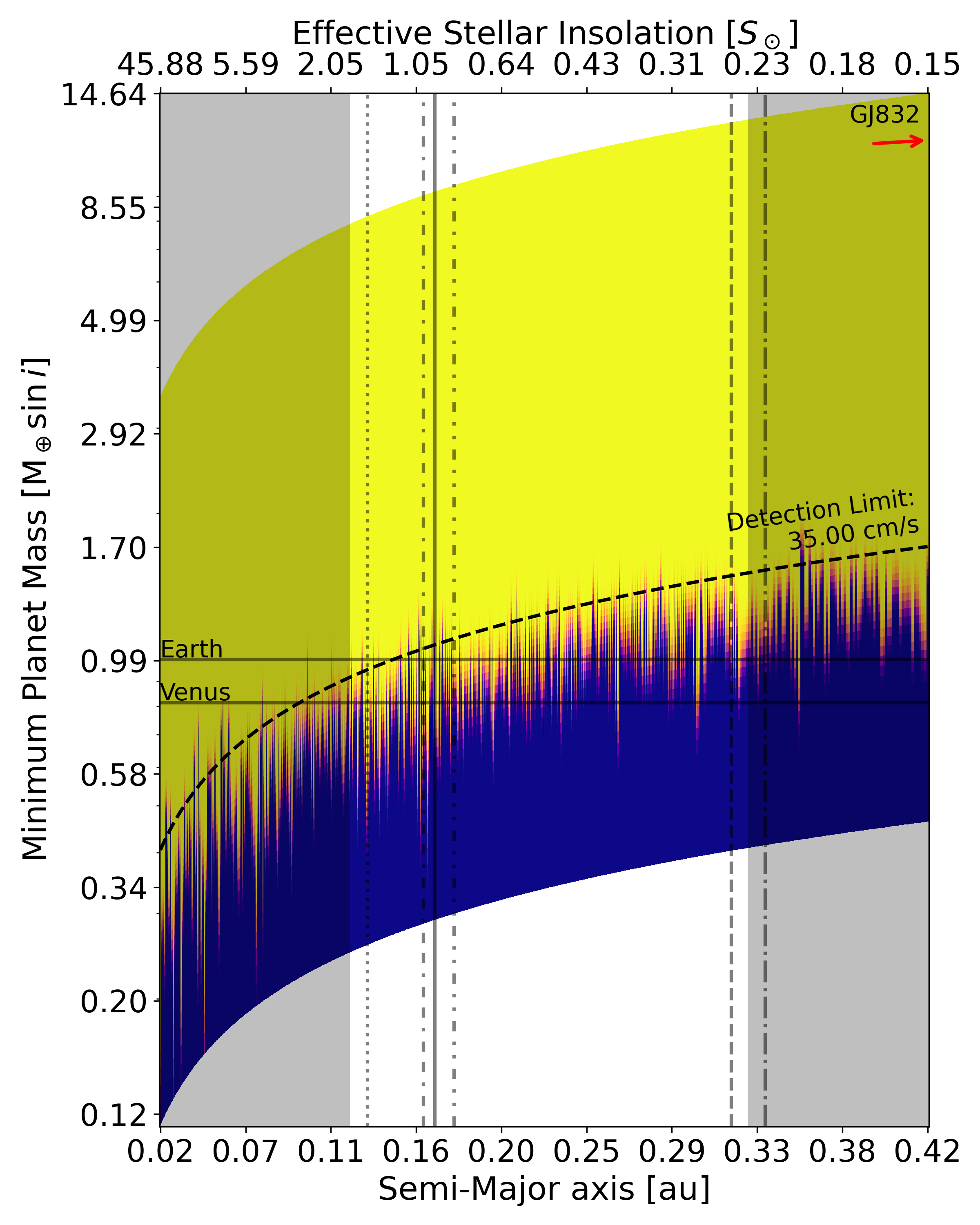}}
\resizebox{0.27\hsize}{!}{\includegraphics{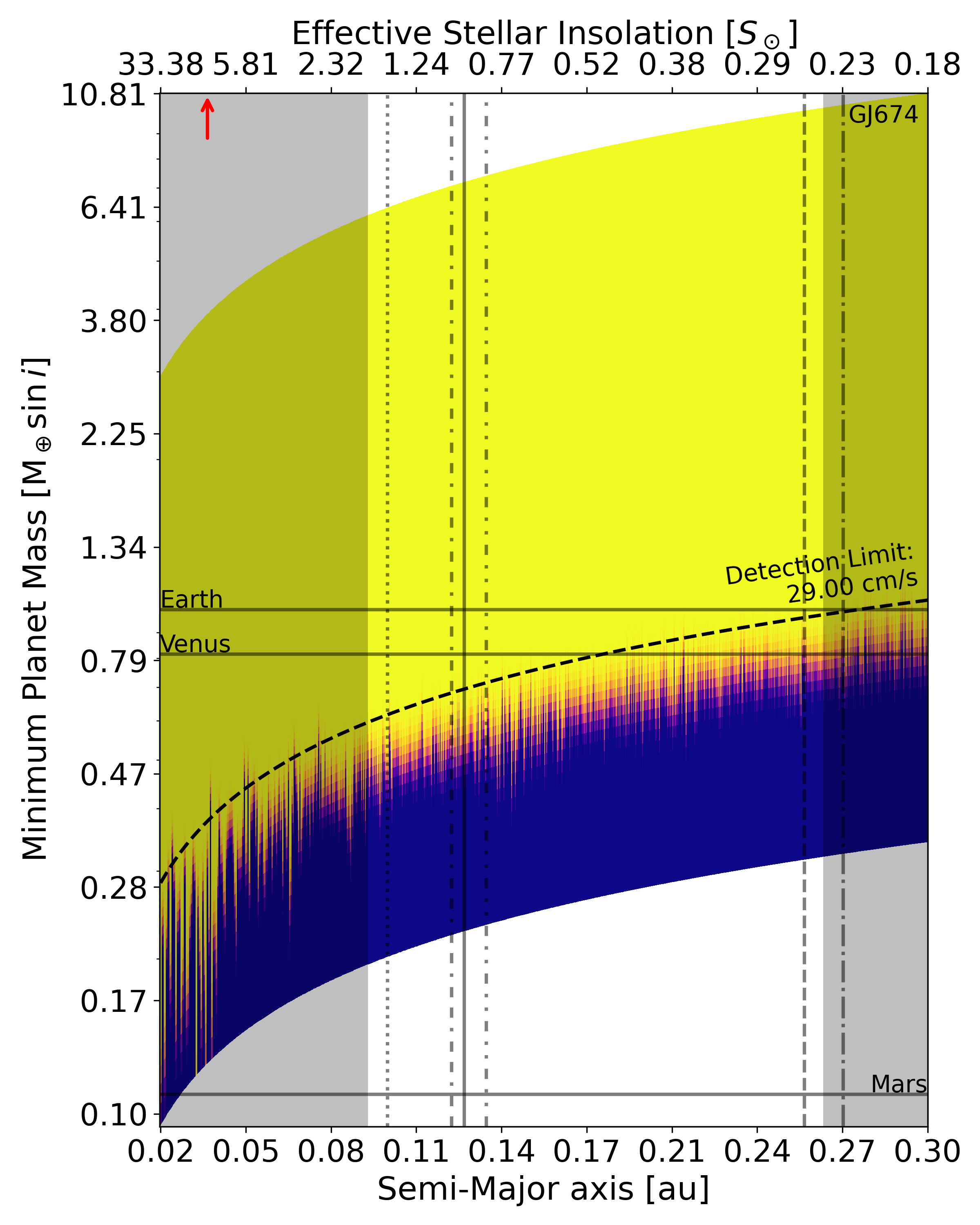}}
\resizebox{0.44\hsize}{!}{\includegraphics{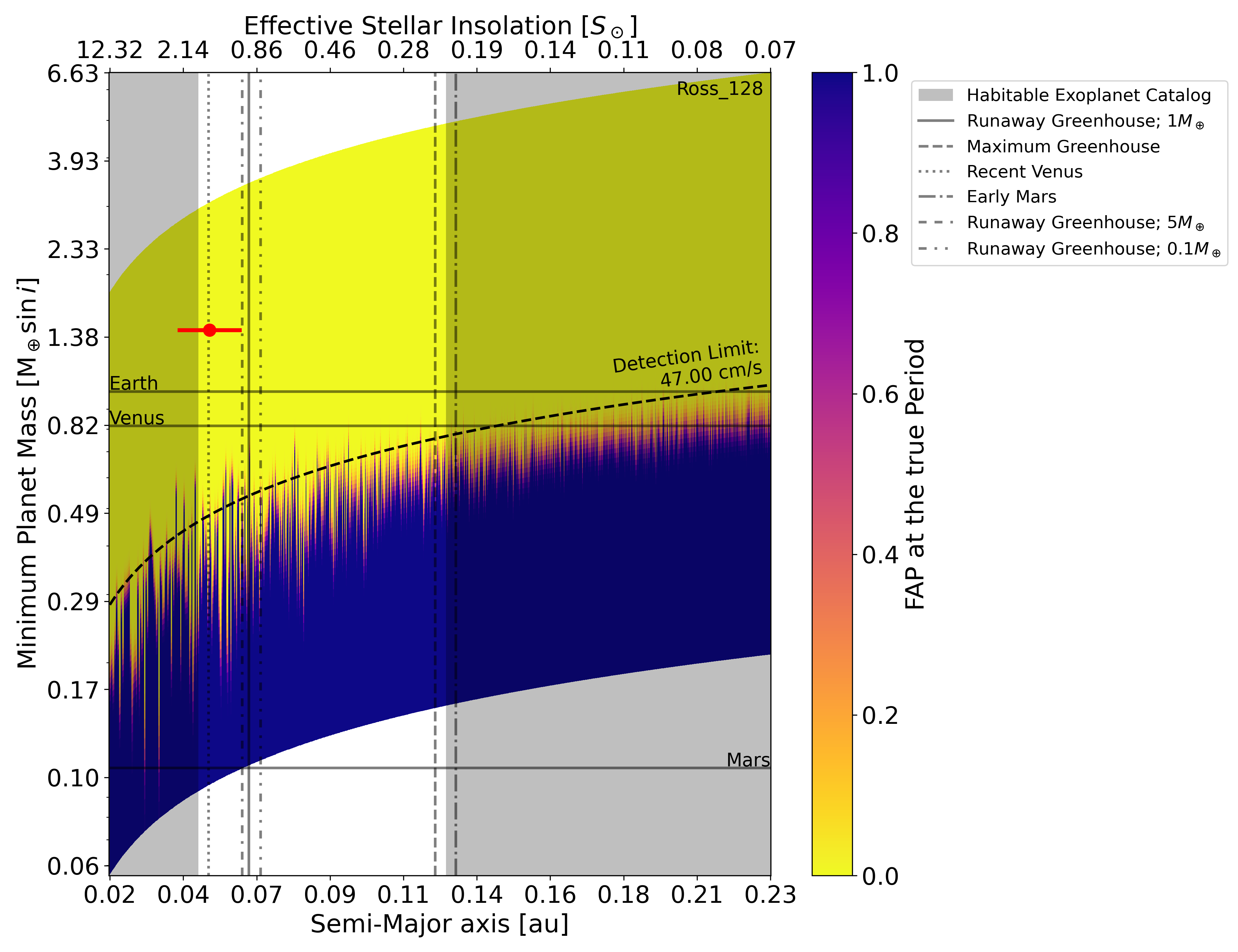}}
\caption{GLS detection grids (Figs. \ref{img:GLS_detect_Ross128}, \ref{img:GLS_detect_GJ832_boundary}, \ref{img:GLS_detect_GJ674_boundary}, and \ref{img:very_long_detect_grids}), transformed to the projected planet mass and semi-major axis for GJ\,832 (left), GJ\,674 (middle), and Ross\,128 (right). The habitable zones, following the models of \citetads{2013ApJ...765..131K} and \citetads{2014ApJ...787L..29K}, are indicated by vertical lines with Earth, Venus, and Mars masses marked by horizontal lines. The shaded regions mark the habitable zone as defined for the Habitable Exoplanet Catalog (HEC). Ross\,128\,b is indicated by the red marker, with the error bar indicating the orbital motion between periastron and apastron due to the orbital eccentricity of e=0.21 derived in this work. The location of GJ\,832\,b and GJ\,674\,b are indicated by the red arrows. The curved, black, dashed line is the RV detection limit obtained in Sect. \ref{subsec:glsGrids}. The effective stellar insolation shown on the upper axis was calculated from the Kopparapu et al. relations.}
\label{img:HZ_limits}
\end{figure*}

To assess the limits imposed by the detection limits determined in Sect. \ref{subsec:glsGrids} on potentially habitable planets around GJ\,832, GJ\,674, and Ross\,128, the grids from Figs. \ref{img:GLS_detect_Ross128}, \ref{img:GLS_detect_GJ832_boundary}, and \ref{img:GLS_detect_GJ674_boundary} were converted from the observational RV semi-amplitude and orbital period into the physical projected (minimum) planet mass and orbital semi-major axis. The resulting grids are given in Fig. \ref{img:HZ_limits} with the also transformed RV detection limits from the previous section and reference lines for the solar system planets of Earth, Venus, and Mars shown.\par
The boundaries of the habitable zone (HZ) were defined following two definitions. First, the definition included in the Habitable Exoplanet Catalog\footnote{\url{https://phl.upr.edu/library/labnotes/habitable-zone-distance-hzd-a-habitability-metric-for-exoplanets}} (HEC) based on \citetads{1993Icar..101..108K}, \citetads{2007A&A...476.1373S}, and \citetads{2003IJAsB...2..289U}. And second, the updated relations from \citetads{2013ApJ...765..131K} and \citetads{2014ApJ...787L..29K}. The HEC boundaries are an optimistic approximation of the HZ that take into account changes in luminosity and effective temperature of the host star and roughly correspond to 72\% to 177\% of the solar insolation on Earth. These values are based on the recent Venus and early Mars approximations by \citetads{1993Icar..101..108K}. The boundaries from \citetads{2013ApJ...765..131K} also use a re-derivation of these two cases for the optimistic HZ. The conservative boundaries are based on the onset of a runaway water greenhouse effect for the inner boundary and the maximum attainable water greenhouse effect for the outer boundary. \citetads{2014ApJ...787L..29K} expand the definition of the inner edge to super-earths of $5\,M_\oplus$ and Mars-like planets of $0.1\,M_\oplus$. They find the outer edge to be not significantly affected by planet mass.\par
For all three systems, we find that the presence of a planet of at least $1.5\,M_\oplus$ is excluded at $>$99\% within any of the HZ definitions. For GJ\,674 and Ross\,128 that exclusion limit is at $1\,M_\oplus$ and $0.8\,M_\oplus$, respectively, precluding the presence of a habitable Earth- or Venus-twin. We also find that the orbit of Ross\,128\,b, which is considered to be at the inner edge but still within the optimistic HZ according to its semi-major axis, passes within the inner optimistic boundary according to the eccentricity we derived in Sect. \ref{subsubsec:Ross128}. Such an orbit, as we point out, would be challenging at best with respect to the development of water-based life on Ross\,128\,b as it significantly impacts the planets ability to retain liquid water. A near-circular orbit would provide more favorable circumstances. It is therefore important to decide whether the eccentricity of Ross\,128\,b is truly as large as we are inferring, closer to a circular orbit as originally published by \citetads{2018A&A...613A..25B}, or even circular as preferred by the Bayesian evidence. We have discussed these possibilities in Sect. \ref{subsec:ecc_validation}.

\subsection{Reliability of GLS periodograms at higher eccentricities}
\label{subsec:GLS_MCMC}
The grids analyzed in the previous sections were all generated for zero eccentricity in the injected signal as a best case. The GLS algorithm only fits a sinusoidal function and as such its reliability at non-zero eccentricities needs to be investigated. It is expected that it will still show a peak in the periodogram power at the correct period, though the RV amplitudes are likely to be systematically offset and potential side signals due to the eccentricity become present. Where we are interested only in detecting the presence of the periodicity rather than deriving an accurate amplitude, GLS periodograms should nonetheless still provide reliable results.\par
Figure~\ref{img:GLS_detect_Ross128}, right panel, shows the grid of recovered semi-amplitudes for Ross\,128 at an injected eccentricity of 0.5 with the true, injected amplitude (horizontal axis) subtracted from the recovered semi-amplitude for each grid point. The striped structures visible in the right panel appear similar to the weak stripes that can be seen in the middle panel imposed on the expected gradient in velocity. The right panel further shows the systematic underestimation of the recovered velocity amplitudes due to the sinusoid approximation, as expected. An identical comparison for the recovered orbital period shows no such deviation for the case of $e = 0.5$. At $e = 0.5,$ the period determination is therefore stable, but with the amplitude underestimated.\par
\begin{figure}
\resizebox{\hsize}{!}{\includegraphics{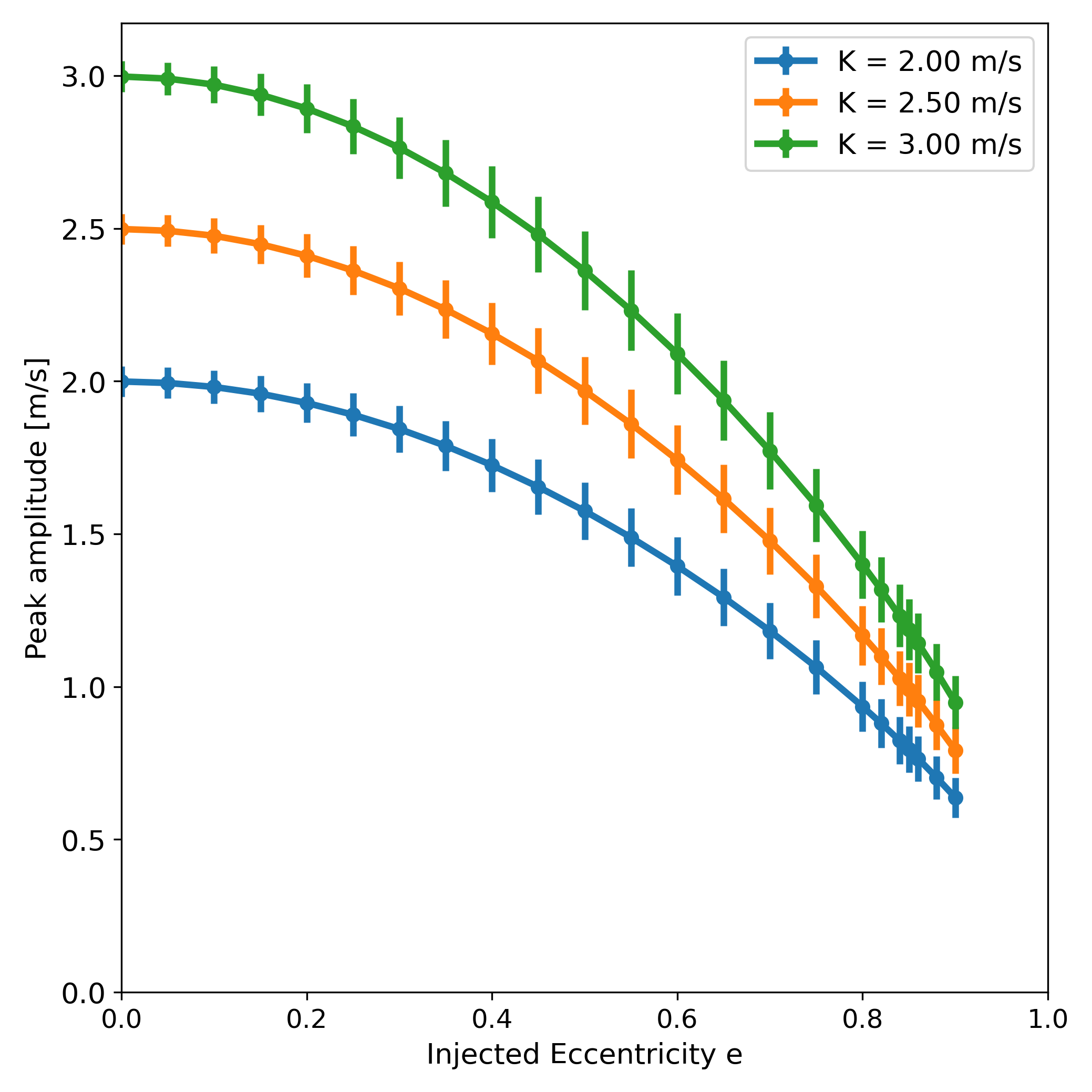}}
\caption{Deviations in recovered vs. injected amplitudes for different eccentricities, using Ross\,128 RV residuals. The recovered amplitude declines for increasing eccentricities due to the increasing deviation from a sinusoidal signal.}
\label{img:GLS_K_vs_e}
\end{figure}

Using the reduced resolution grids that were utilized in Appendix~\ref{apdx:striping} for a range of eccentricities, we plotted the recovered amplitudes for three different injected amplitudes, averaged over the periods sampled within the grid for the selected, injected amplitudes. This is shown in Fig.~\ref{img:GLS_K_vs_e} for the Ross\,128 residuals, though GJ\,832 and GJ\,674 are essentially the same, agreeing  within the error bars. The recovered amplitude is shown to decline monotonically for any eccentricity larger than zero, but remains approximately constant up to eccentricities of 0.1. 

\begin{figure}
\resizebox{\hsize}{!}{\includegraphics{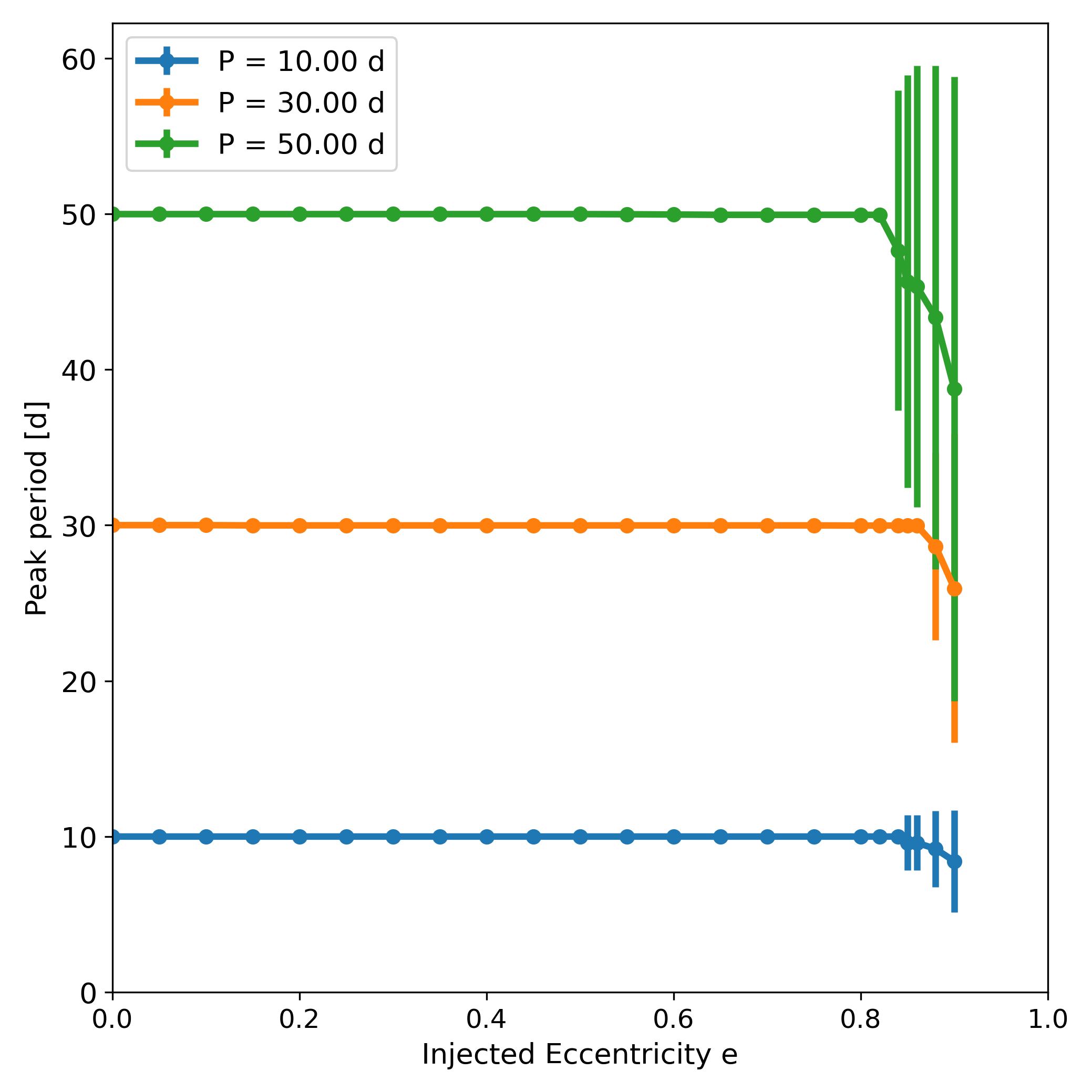}}
\caption{Deviations in recovered vs. injected periods for different eccentricities, using Ross\,128 RV residuals. The recovered period remains accurate until very high eccentricities despite the increasing deviation from a sinusoidal signal.}
\label{img:GLS_P_vs_e}
\end{figure}

We repeated the exercise for three different orbital periods, shown in Fig.~\ref{img:GLS_P_vs_e}. In contrast to the amplitude, the period is entirely unaffected up to very high eccentricities of 0.8, confirming that GLS periodograms can be used as a planet detection tool even at high eccentricities, though not for inference on amplitude. For this one would need to use a more versatile approach such as MCMC simulations as used in this work (Sect.~\ref{subsec:OrbElements}). Our results show however, that the choice of priors and parameterizations are non-trivial and can have large impacts on the results. We discuss this further in Sect.~\ref{subsec:prior_choices}.

\section{Discussion}
\label{sec:discussion}
In this work we have presented new HARPS data, obtained as part of the RedDots campaign, to confirm the status of GJ\,832\,b, GJ\,674\,b and Ross\,128\,b as single, lonely planets. We also confirmed (GJ\,832, GJ\,674) and strengthened (Ross\,128) the observational evidence that these lonely planets have orbital solutions with significant eccentricities. We further argued that the recovered eccentricities are valid in Sect.~\ref{subsec:ecc_validation}. We discussed the potential origins of lonely, eccentric planets and their importance in Sect.~\ref{subsec:lonely_eccentricities}. The possible consequences of undiscovered second planets, following our limits from Sect.~\ref{subsec:glsGrids}, and whether such planets could be responsible for the observed eccentricities of the known planets, were discussed in Sect.~\ref{subsec:second_planet_effects}. The discrepancy between the periods recovered with GP modeling and the literature rotation periods was discussed in Sect.~\ref{subsec:rotation_discuss}, our decision to analyze everything with MCMC as well as nested sampling in Sect.~\ref{subsec:juliet_vs_mcmc}; and our strict choice in MCMC prior distributions throughout this work in Sect.~\ref{subsec:prior_choices}.

\subsection{The limitations of blindly using Bayesian evidence values}
\label{subsec:lnZ_discussion}
The Bayesian evidence, $\mathcal{Z}$, is one of the standard metrics used throughout data analysis for model comparison. It offers a simple and, at first glance, robust way to decide whether a more complex model offers a sufficient improvement to the fit to counter the increase in complexity. Simply obtaining $\mathcal{Z}$ for two models and calculating their ratio, or as we state in Sect. \ref{sec:analysis}, taking the difference of the logarithmic values, $\Delta\ln\mathcal{Z}$, allows for a quick determination of which model is supposedly the better choice. The Bayesian evidence however is calculated as an integral over the entire parameter space and independent of specific parameter choices. This works well for classical fits of deterministic functions such as polynomials, which always have a fixed degree of flexibility. With the increased use of GP for data analysis and the possibilities offered by specific kernel functions, this point may no longer hold as stringently. The QP kernel, as an example, offers an intuitive parameterization (See Sect. \ref{subsec:GP}) that includes the harmonic complexity, $\Gamma$, and characteristic length scale, $l$. Depending on the values of $\Gamma$ and $l$ however, the GP can exhibit very different behaviors ranging from low complexity and close to strictly periodic to high complexity with very few remaining periodic characteristics. As this change in behavior is possible without changing the model and only depends on the value of the parameters themselves, it is not penalized in the Bayesian evidence. An unreasonably flexible GP by itself may therefore be preferable, according to the evidence, to a more reasonable choice that requires a more complex, analytical mean function in addition to the less flexible GP.\par

During the analysis in Sect. \ref{sec:results} we came across several instances where the Bayesian evidence did not behave as we would have expected. While these are individual occurrences and could potentially be specific to our sets of data, this is not a guarantee per the previous reasoning. As such, we took a closer look at the likely origin for the inconsistencies such that others can be made aware of the possibility for such non-obvious problems. A more general review of the reliability of Bayesian evidence would require a much larger set of stars or a mathematical approach, both of which are outside the scope of this work.\par

For the Ross\,128 RedDots subset, which covers almost nine consecutive orbital periods with very regular and near-nightly observations (see Sects. \ref{subsubsec:HARPS} and \ref{subsubsec:Ross128}), the Bayesian model evidence considers models without a planet equally probable to the one planet solutions. We attribute this to the low levels of stellar activity, expected from the findings by \citetads{2019A&A...628L...1I}, which the GP is supposed to model. In the absence of stellar activity, the GP instead is able to fit the Keplerian component of the signal well enough that the Bayesian evidence prefers the model without the complexity of an additional Keplerian function. This is clearly not the case based on the archival observations, the GLS periodogram (Fig. \ref{img:GLS_new}), and the shape of the posterior distributions (see Appendix \ref{apdx:lnZ_comparison}). We also found that the eccentricity appears to be statistically significant while the Bayesian evidence shows no difference to a circular orbit. For GJ\,674, we find the dSHO kernel to be strongly preferred by the evidence. However, as described in Sect. \ref{subsubsec:GJ674} and \ref{apdx:lnZ_comparison}, a critical investigation of the inferred parameters clearly indicates that the SHO kernel is the more realistic choice because a fractional amplitude parameter much greater than one is unreasonable for the dSHO kernel.\par

We were able to identify and address the problematic Bayesian evidence for Ross\,128 a posteriori, since we know that the planet exists. In general, one can attempt to mitigate such issues by restricting the GP to parameter spaces that disallow a fit to the planetary signal, such as the orbital period. This does bring the risk that the GP might be prevented from fitting the remaining, low levels of stellar activity and identifying the presence of additional planetary signals. Properly restricting the GP would also include enforcing a long damping timescale that does not allow high levels of inter-period variations. Effectively in this case, the GP would need to be close to, but not strictly, periodic. It would therefore be preferable to either be able to avoid this situation a priori or at least without the prior knowledge whether or not a planet exists. The only solution to this worst-case scenario that we are aware of is a visual inspection of the model fits to make a manual decision whether the phase-folded Keplerian model looks reliable. With one clear case of an unreliable Bayesian evidence, the question arises how reliable other model comparisons with more subtle implications are. In Sect. \ref{subsec:ecc_validation}, we discussed the eccentricity inferred for Ross\,128\,b in the context of the Lucy-Sweeney bias and conclude that our inference still holds as an argument for a non-zero eccentricity. This is in spite of the indifference of the Bayesian evidence between the circular or eccentric model. Another indicator that the Bayesian evidence may not work as intended here can be seen for the models without GP in Table \ref{tab:modelEvidence}. For the clearly eccentric GJ\,674\,b, the Bayesian evidence still prefers the 1P\textsubscript{circ} model to the 1P\textsubscript{ecc} one despite the phase-folded data in Fig. \ref{img:GJ674_resid} showing a definitive non-zero eccentricity. Similarly, the Ross\,128 RedDots only subset shows nearly identical evidence for the no planet models as well as the eccentric or circular one planet models. Taken together, we conclude that the Bayesian evidence alone cannot be blindly trusted as a diagnostic of the best model for the present data of Ross\,128 once a GP is involved for observations during low stellar activity. Further a posteriori analysis of the shapes of the posterior distributions and injection-recovery simulations, as we discuss in Sects. \ref{subsec:ecc_validation} and \ref{apdx:lnZ_comparison}, are required to make a decision between the models.

\subsection{Importance of the eccentricities of lonely planets}
\label{subsec:lonely_eccentricities}
\begin{figure*}
\resizebox{0.49\hsize}{!}{\includegraphics{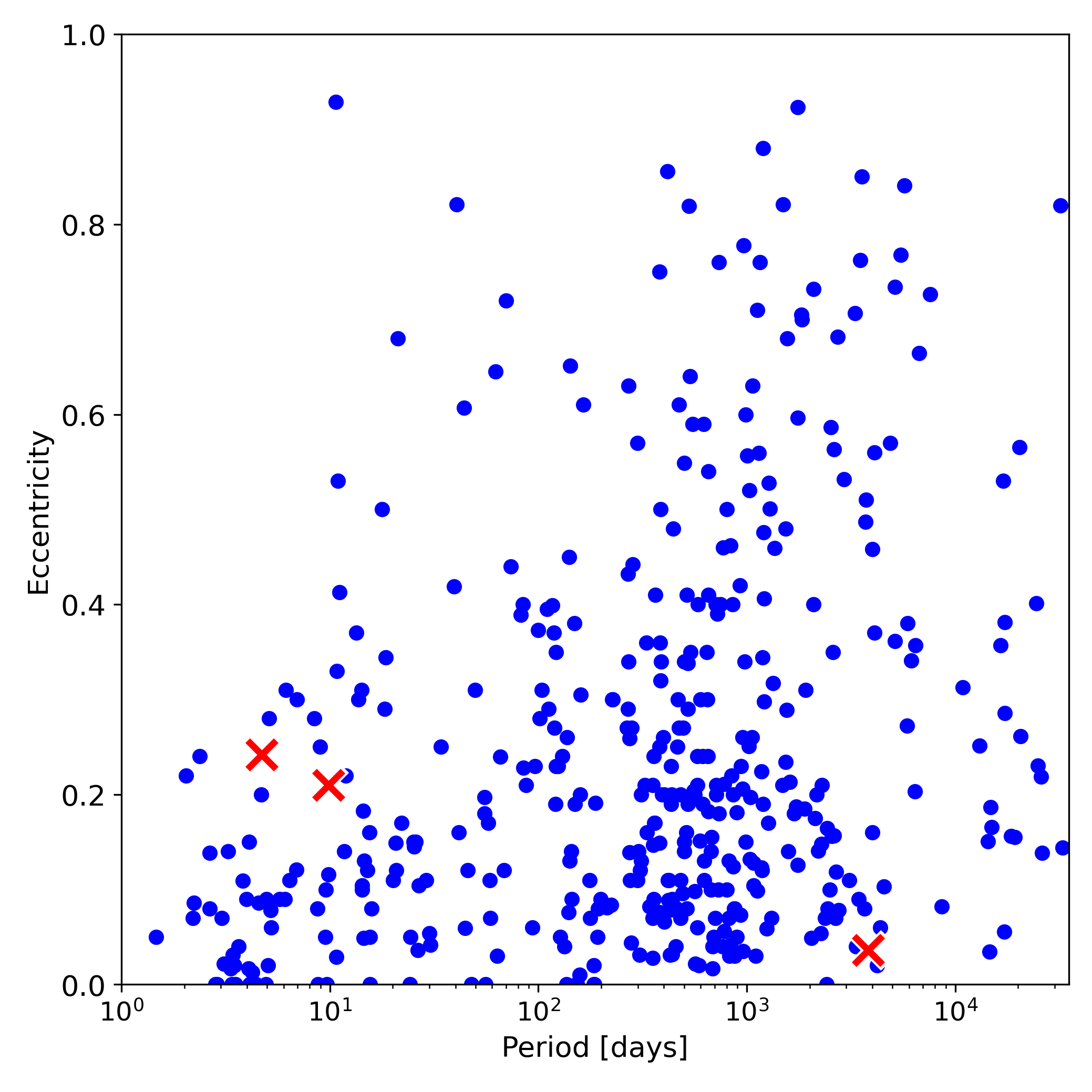}}
\resizebox{0.49\hsize}{!}{\includegraphics{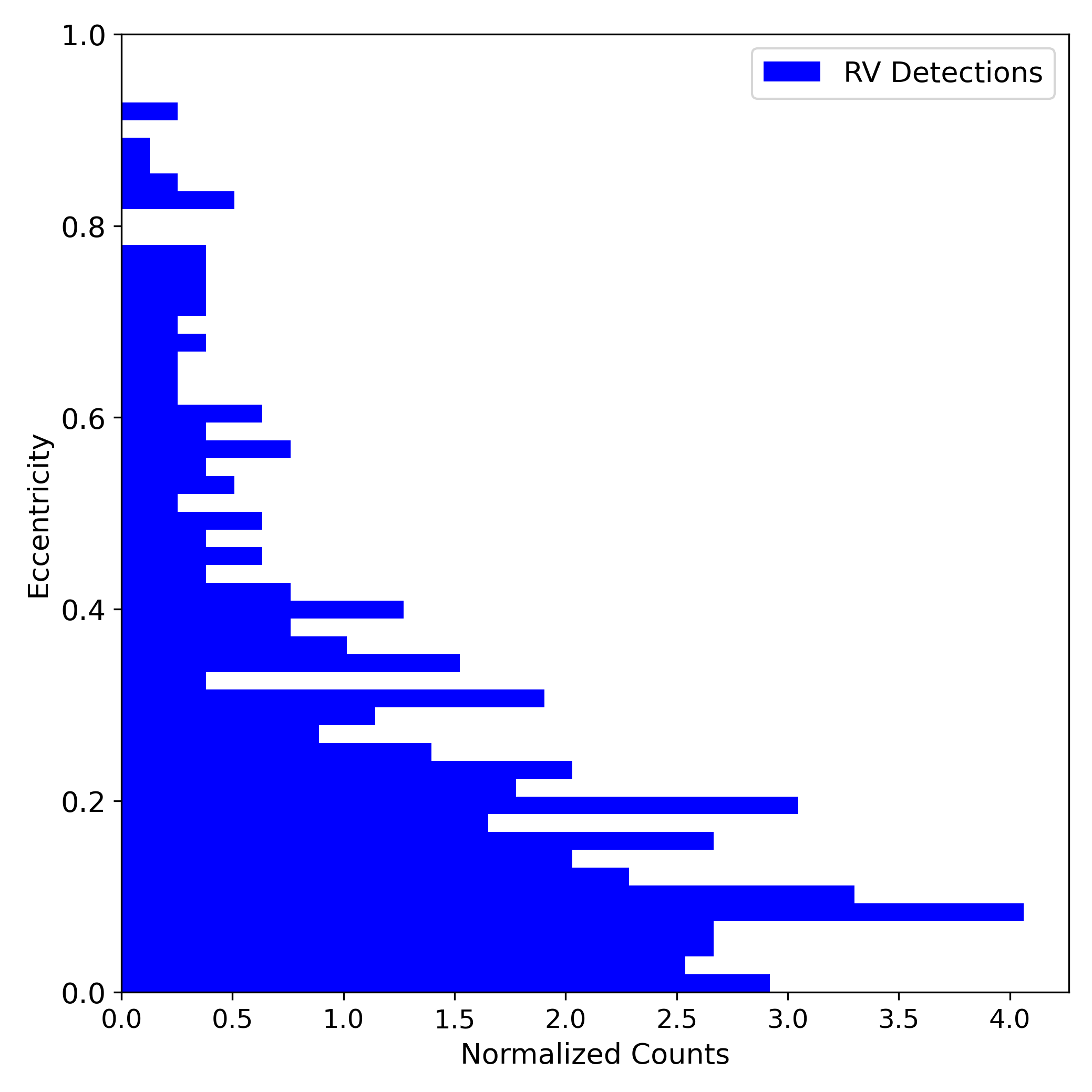}}
\caption{Distribution of all lonely planets detected with the RV method in eccentricity versus orbital period (blue circles) in the left panel. The three planets analyzed in this work are highlighted as red crosses. Normalized eccentricity distribution of all lonely planets detected with the RV method in the right panel. The histogram has its approximate maximum at a nonzero value for the planetary eccentricity of approximately 0.1.}
\label{img:ecc_hist}
\end{figure*}

We have discussed (in detail) the eccentricity of Ross\,128\,b in the context of the Bayesian evidences' reliability as a deciding metric (Sects. \ref{subsubsec:Ross128}, \ref{subsec:lnZ_discussion}) as well as its possible impact on the planets habitability (Sect. \ref{subsec:HZ_limits}). In addition to these important factors, Ross\,128\,b is also, according to our data, the only planet within the system, classifying it as a lonely planet. This leads to another important facet of the eccentricity that is worth a discussion: From a total of 1036\footnote{NASA exoplanet archive; 12 May 2023; \url{https://exoplanetarchive.ipac.caltech.edu/}} exoplanets detected with the RV method, 424 can be classed as lonely and 396 as eccentric. These classifications were made here based on the absence of additionally detected planets in the system (lonely) and whether the eccentricity estimate is larger than zero at one sigma (eccentric). The distribution of eccentricities for the known lonely planets detected with the RV technique and with published eccentricities is shown in Fig.~\ref{img:ecc_hist}. The left panel illustrates that the orbital solutions of the planets analyzed in this work are not unusual in their eccentricity and orbital period, compared to the overall known population. The right panel demonstrates the maximum for overall eccentricities around $e = 0.1$ and a mean eccentricity value of $e\sim0.25$. A tail of extremely eccentric planets stretches up to values as high as $e = 0.9$ (such as HD\,80869\,b at $e\sim0.86$, \citeads{2021A&A...653A..78D}). This demonstrates the significance of the lonely and eccentric type of planet despite the challenge these systems pose for traditional planet formation models. The understanding has been that especially low-mass planets which are unable to open a gap within the planet-forming disk would have any initial planetary eccentricity damped on short timescales. Only if  other forces are at work to excite the eccentricity to a higher value would $e$ be expected to deviate significantly from zero after dispersal of the disk (see for example \citeads{1980ApJ...241..425G}, \citeads{2004A&A...418..325S}, \citeads{2007A&A...473..329C}).\par

The new RV measurements and analysis presented herein provide additional confirmation for the lonely eccentric planets orbiting GJ\,674 and Ross\,128 through longer observational baselines, higher instrumental precision and larger datasets. We confirm that these planets are correctly classified as lonely and eccentric within the current observational constraints despite classical planet formation theory expecting strongly damped eccentricities for them. Thus, additional mechanisms must exist which keep the eccentricities of these planets at values higher than expected by traditional formation models.\par

A number of mechanisms have been proposed in the literature that may explain the observed high eccentricities for lonely planets.  An exhaustive overview on disk related effects has recently been presented by \citeads{2022arXiv221107305L}. One mechanism is that the planet originally formed in a multi-planet environment where planet-planet interactions increased the eccentricities until the companions were ejected either directly through planet-planet scattering \citepads{2001Icar..150..303F} or stellar encounters (\citeads{2020MNRAS.496.1453W}; \citeads{2023MNRAS.520..637R}). Far-out, undiscovered perturbers could also provide an explanation  \citepads{2014ApJ...781L...5D}, though they appear increasingly unlikely in the face of newer and higher precision observations. An example are the planetary detection limits computed in this work (Sect.~\ref{subsec:glsGrids}). 

The scenario of initial multi-planet formation followed by scattering and ejection relies only on known formation pathways but requires fine-tuning as to whether it can explain the number of observed lonely, eccentric planets. Alternatively, single-planet formation can led to significant eccentricities under the right circumstances. Since planet formation occurs while the system is still within its stellar nursery, encounters with neighboring stars are common and can boost the eccentricity without completely ejecting the planet. In the absence of strong interactions with other bodies, within or outside the system, planets with initially small eccentricities can also have their eccentricity increased. One way that this may happen is through interaction with the disk itself if the planet is massive enough to open a gap and saturate the corotation torque \citepads{2003ApJ...585.1024G}. Alternatively, the planet may remain locked on an eccentric orbit during quick, inwards migration should the disk evaporate before the periastron decreases sufficiently for tidal circularization to finish the migration. Finally, even initially circular orbits might become highly eccentric from resonant interaction with eccentric disk instabilities \citepads{2022arXiv221107305L}.\par

All these scenarios have different requirements on the stellar environment, planet mass, and disk evolution. To explain which case, if any, is dominant or involved to which degree, the empirical distribution of eccentricities must be sharpened in its uncertainties and cleaned, or at least restricted, of any potential multi-planet systems. This would then allow for the comparison of the predictions by the different mechanisms. This would, however, require a dedicated effort to re-observe hundreds of targets. But only through this would it be possible to obtain a more statistically significant estimate of the properties and formation pathways of the lonely planet population and, by extension, all planets. Examples of the work needed are \citetads{2023arXiv230502551T} or \citetads{2023arXiv230314570G}. Recently \citetads{2023arXiv230502551T} reanalyzed GJ\,3470 in the face of citizen-science reports on planet candidates c, d, and e with the latter two reportedly co-orbiting. They found no RV evidence in support of these claims and used injection modeling to determine detection limits on potential unobserved planets. However, their findings preclude the presence of any other higher-mass planets besides confirmed planet b, similar to this work. Additionally \citetads{2023arXiv230314570G} performed a detailed analysis of the highly eccentric ($e=0.75$), warm Jupiter TOI-4127 b. They attempted to ascertain whether the planet could be a hot Jupiter progenitor undergoing high-eccentricity migration. They show that the orbit as it is will remain stable at high eccentricity for the remaining main-sequence lifetime of the host star and not become a hot Jupiter. Further migration would necessitate the presence of a distant perturber, which \citetads{2023arXiv230314570G} cannot rule out below 10\,m\,s$^{-1}$ based on the current observations.

\subsection{Possible effects of an undiscovered second planet}
\label{subsec:second_planet_effects}
We showed in Sect.~\ref{subsec:ecc_validation} that it is unlikely that the significant eccentricity of $e=0.21^{+0.09}_{-0.10}$ of Ross\,128\,b we are seeing is the result of statistical effects. Similarly, the Bayesian evidence (Table~\ref{tab:modelEvidence}), inferred uncertainty (Table~\ref{tab:inferredElements}), and even a visual inspection (Fig.~\ref{img:GJ674_resid}) demonstrate that GJ\,674\,b definitively has a significant eccentricity of $e=0.242^{+0.012}_{-0.013}$. However, we have also discussed in Sect.~\ref{subsec:lonely_eccentricities} that the formation of singular, eccentric planets poses a challenge for classical formation scenarios. As such, in this section, we aim to clarify whether an undiscovered secondary planet would have the capability to excite the observed eccentricities through gravitational planet-planet interactions.\par
For this, we used the N-body simulator \texttt{rebound} \citepads{rebound} with the adaptive integrator \texttt{ias15} \citepads{reboundias15}. We constructed both systems around a central object with mass as given in Table~\ref{tab:litElements}. We then added the first, inner planet, representing a primordial version of the known planet, on a circular orbit with a period and mass matching our current results. A second, outer planet was added on an orbit with an inclination between 0 and 45 degrees, mass equivalent to twice our detection limit in RV semi-amplitude from Sect.~\ref{subsec:glsGrids}, corrected for the projection effect of the inclination, and orbital periods of 6 -- 30 days (GJ\,674) or 12 -- 60 days (Ross\,128) with one day spacing. Additional simulations were run at the 1:2 and 2:3 mean motion resonances. The simulations were run for 10\,000 years each.\par
GJ\,674\,b is entirely unaffected by the companion planet and shows no variations in its initial circular eccentricity. Ross\,128\,b is unaffected up to high inclination angles of 30 degrees. Above that angle, eccentricity oscillations of up to e=0.45 are possible at an inclination of 45 degrees. While 10\,000 years was not long enough for the oscillations to fully set in for the largest period differences in the Ross\,128 models, singular runs of 100\,000 years show that the expected eccentricity amplitudes are weaker than for the shorter models and extended runs for the full set are unnecessary.\par
Since the companions that are possible under the constraints of our detection limits are not able to excite the observed eccentricities (GJ\,674) or are unlikely to exist with the required inclinations (Ross\,128), we conclude that the planets' eccentricities are probably a result of the initial formation path.

\subsection{Rotation: GP versus reality}
\label{subsec:rotation_discuss}
The literature values for the rotation periods shown in Table~\ref{tab:Params_known} were primarily derived from photometric observations in all three cases, though \citetads{2022A&A...664A..64G} (GJ\,832) and \citetads{2007A&A...474..293B} (GJ\,674) validated theirs through chromospheric activity indicators. In this work we find that the activity periods inferred with MCMC for the SHO GP for GJ\,832, GJ\,674 and Ross\,128 purely from the RV timeseries fall between the full and half the rotation period reported in the literature. The exception appears to be Ross\,128, where alternate results from the literature (\citeads{2016A&A...595A..12S}; \citeads{2019A&A...621A.126D}) indicate a better match, with our GP period close to half the rotation values. This raises the question of how the parameters of a GP, which is a non-parametric, statistical algorithm instead of a physical model, relate to the underlying physical processes of the star and its planets. This has been previously discussed in the literature (\citeads{2021A&A...645A..58P}; \citeads{2023arXiv230413381S}), though a consensus has not been found thus far. An approach that is commonly found in the literature is to assume that the GP parameters are equivalent to  "effective model parameters"  with no immediate physical interpretation.  This was extensively discussed by \citetads{2023arXiv230413381S}. For general GP kernel formulations, such as squared-exponential, this is a sensible approach, keeping in mind that GPs are a statistical modeling tool. Any physical meaning is assigned by the user and dependent on the kernel choice and how well the physical processes correspond to the assumed behavior, as is true for any parametric model.\par
When implementing GP kernels such as the (d)SHO or QP however, the expectation is that parameters called period do have a physical meaning, and in RV analysis this is generally related to the stellar rotation period. It has been reported by \citetads{2023arXiv230413381S} that for the QP kernel, as long as the spot life time is at least on the order of the stellar rotation period, this expectation is correct. For shorter spot lifetimes, and therefore more rapidly evolving spots, the GP period instead often results in half-period determinations, particularly for modeled distributions with two active longitudes, though these have been shown to be physically unrealistic spot distributions \citepads{2009AIPC.1094..664J}. For spatially uncorrelated spot patterns the recovery of the stellar rotation period commonly failed and the GP returned unrelated periods. This is largely in agreement with findings by \citetads{2021A&A...645A..58P}. They also find good agreement between the recovered QP period and the stellar rotation periods from activity models in most cases, though were more optimistic about the relation between spot lifetime and QP decay time. They also analyzed the SHO kernel and found that for one and two modeled active longitudes the SHO undamped period matches the full rotation period and half the period, respectively. This is similar to the short-lifetime QP findings by \citetads{2023arXiv230413381S}. For random and polar spot distributions however, their results match our findings in that the SHO period lies in between the full and half-rotation period with the exact position dependent on the spot distribution. The findings by \citetads{2022MNRAS.515.5251N} mirror the agreement between QP period and decay time with stellar rotation period and spot lifetime respectively, but they state a warning that for a set of only RV data a QP GP runs the risk of overfitting the activity and becoming degenerate with the planetary RV variations. This matches our experience with the RedDots only data set for Ross\,128 described in Sect.~\ref{subsubsec:Ross128}.\par
A similar result was obtained by Barnes et al. (submitted to MNRAS), who investigated the use of the statistical moments of the cross-correlation function (CCF) of the spectra. They find that the third moment, the skew of the CCF, correlates better than the typically used bisector inverse span (BIS) with RV variations induced by activity modulated by rotation. The obtained periods can show the true rotation period while also frequently showing harmonics of the true rotation period, including in several cases 2 P$_\mathrm{rot}$ / 3 for models of high v\,sin\,i, high activity M dwarfs.  We find similar mixed harmonics of the photometric rotation period for GJ\,832, GJ\,674, and Ross\,128, even though all three are slowly rotating and low activity M dwarfs. This can be explained by GJ\,832, GJ\,674, and Ross\,128 having different spot patterns compared to the low activity models of Barnes et al.\par
We do not know if our SHO periods are subject to the findings by \citetads{2021A&A...645A..58P}, decoupling them from the physical rotation period or if the literature values for the rotation of GJ\,832 and GJ\,674 may not be accurate, as is possibly the case for Ross\,128. From the matches in derived planet parameters and the fit qualities (Figs.~\ref{img:GJ832_resid}, \ref{img:GJ674_resid}, \ref{img:Ross128_resid}, \ref{img:Ross128_Carmenes_resid}), we conclude that the GP still works well to capture the signature of activity on a statistical level, even if it does not result in easily interpretable, physical parameters.

\subsection{Comparing nested sampling, MCMC, and GP kernels}
\label{subsec:juliet_vs_mcmc}
We chose to analyze our data twice: Once with MCMC (\texttt{emcee}) and once with nested sampling (\texttt{juliet}). This was done primarily to exclude the potential of algorithm based error sources since with \texttt{emcee} and \texttt{juliet} we cover both of the most common Bayesian inference algorithms. Further, \texttt{juliet} operates on a fully independent code base, further validating our results. This allowed us to exclude the analysis itself as the source for the issues encountered with the eccentricity determination of Ross\,128\,b, since both algorithms were in agreement. Additionally, nested sampling was shown to be more robust against the side signals encountered in Sect.~\ref{subsubsec:Ross128}. This appears to be an advantage in this instance, as the resulting posterior peak was still in agreement with the literature and the MCMC result, but this may not always be the case. MCMC on the other hand always samples the entire log-likelihood space allowed by the priors, avoiding the risk of the sampling space contracting away from posterior signals that might be of interest.\par
Besides a different sampler, \texttt{juliet} was also set to use the dSHO and QP kernels for the GP, rather than a simple SHO one (see Sect.~\ref{subsec:GP}). Again, this served as verification for the inferred planetary parameters but also highlighted an important detail: While the recovered dSHO period, in both \texttt{juliet} and \texttt{emcee} analyses, was close to the photometric rotation periods from the literature, the fractional amplitude always showed posterior plateaus at values of $f >> 1$ while increasing linearly between zero and one. This means that the half period is strongly preferred over the full period in all cases, rendering the additional complexity of the dSHO kernel unnecessary compared to the simpler SHO one. Similarly, the QP kernel gave the correct rotation period in two out of three cases (GJ\,674 and Ross\,128), double the rotation period in the third (GJ\,832), but always at high harmonic complexities. Both the dSHO and QP kernels were further unable to provide an advantage to the inferred parameters, returning consistent results, and at only comparable Bayesian evidence levels.

\subsection{Choosing priors and parameterizations for MCMC}
\label{subsec:prior_choices}
While MCMC has become a commonly used technique in astrophysics in recent times, its nature as a Bayesian process is often underappreciated. Bayes rule states that the posterior distribution of a variable is the product of the likelihood and a prior distribution\footnote{The Bayesian evidence in the denominator is ignored for most MCMC implementations. This precludes model comparisons but has no effect on parameter inference so long as the data and model remain unchanged during the inference process.}. By design, the choice of prior is therefore up to the user as a way to include previous knowledge into the process but thereby also allows for biases to be introduced or propagated. There are common cases where this behavior is unintended, harmful, and often hard to detect a posteriori because the biased posterior appears to be completely in order. For example, the way Bayesianism is intended to be employed if one wanted to redetermine a parameter is to start with known values of that parameter. That and its uncertainty can then be used to define a Gaussian prior for the MCMC to determine the updated value in the face of new information. If the previous results contain an error however, then that error is propagated through the Bayesian process and results in a biased posterior. Unless the likelihood is overwhelmingly strong, this will result in values that are compatible with the prior within the uncertainty with only a small correction through the likelihood. An independent analysis which ignores the prior information might then find a significantly different result, unaffected by the error. For this reason we were careful to only use uniform, wide priors for this work and narrow those only after initial posterior peaks were found and confirmed to be compatible with previous findings in order to avoid any bias propagation. It is our belief that doing this, or at least verifying the unbiased nature of one's MCMC inferences, should become common practice in the literature. In practice this means re-running the sampler for any non-uniform priors with uniform priors as a check. We point to the summary by \citetads{2023arXiv230204703E} as a guide towards proper use of MCMC.\par
Another point raised in \citetads{2023arXiv230204703E} that is briefly touched at the end of Sect.~\ref{subsec:MCMC_setup} and discussed in depth by \citetads{2013PASP..125...83E} and \citetads{2006ApJ...642..505F}, is that uniform priors may not always remain that way. Under transformations or for hard edges in the parameter space (such as $e \geq 0$) they or their estimators may become biased if one does not carefully monitor them, similarly to the Lucy-Sweeney bias we investigated in Sect.~\ref{subsec:ecc_validation}.

\section{Conclusion}
\label{sec:Conclusions}
We analyzed archival spectroscopic data from HARPS and CARMENES as well as new HARPS observations recorded as part of the RedDots campaign for GJ\,832, GJ\,674 and Ross\,128, combined with photometry for GJ\,674 and Ross\,128. With GJ\,832 used as a verification case, we refined the parameters for the known single planets, determined detection limits for any other, previously unknown planets, and discussed the possibility of the undiscovered planets being habitable. We find:
\begin{itemize}
\item The refined orbital solutions are in agreement with previous results, confirming the eccentric nature of GJ\,674\,b at $>18\sigma$. The eccentricity of GJ\,832\,b remains consistent with zero at only $1.5\sigma$ significance.
\item The inferred eccentricity for Ross\,128\,b is significantly higher than previously published, at $\approx 0.2$ instead of $\approx 0.1$ at $2\sigma$. It is also highly unlikely ($>5\sigma$) to be a statistical effect of an underlying circular orbit.
\item An inspection of the model residuals shows a 4-day signal for Ross\,128, coincident with a signal in H$\alpha$. Therefore it is likely not a planet but stellar-activity related.
\item Injection-recovery simulations give limits of 0.47\,m\,s$^{-1}$ (Ross\,128), 0.35\,m\,s$^{-1}$ (GJ\,832), and 0.29\,m\,s$^{-1}$ (GJ\,674) for any additional, thus far undiscovered, planets. For periods up to 100 days this excludes any planet above 1.5 Earth masses around GJ\,832, and above one Earth mass around GJ\,674 and Ross\,128, coinciding with the limits for liquid-water habitable planets.
\item N-body simulations of possible undiscovered planets, utilizing the detection limits, are unable to explain the observed eccentricities of GJ\,674\,b and Ross\,128\,b as the result of planet-planet interactions within reasonable constraints.
\end{itemize}
We further discussed the peculiar nature of GJ\,674\,b and Ross\,128\,b as the only known planets within their systems while they still show statistically significant eccentricities, and the implications for planet formation scenarios. With an occurrence rate of 1.3 planets per M-dwarf, we can expect nearly two-thirds of all M-dwarf systems to fall into this category, further highlighting their importance. The relation of GP derived periods to physical rotation periods was debated. Finally we want to raise awareness of the Bayesian nature of MCMC, which is commonly overlooked in the literature, potentially leading to systematic biases, as well as a possible issue with the blind interpretation of Bayesian evidence values.

\begin{acknowledgements}
We thank Artie Hatzes and Eike G\"unther for fruitful discussions on planetary eccentricities and the anonymous referee for their constructive comments that has helped to improve the clarity and scientific integrity of the paper. FL, SVJ, and YT acknowledge the support of the DFG priority program SPP 1992 "Exploring the Diversity of Extrasolar Planets (FL, SVJ: JE 701/5-1; YT: TS 356/3-1). CAH and JRB were supported by STFC under grants ST/T000295/1 and ST/X001164/1. LT-O acknowledges the support of the Israel Science Foundation through grant 1404/22. We acknowledge financial support from the Agencia Estatal de Investigaci\'on (AEI/10.13039/501100011033) of the Ministerio de Ciencia e Innovaci\'on and the ERDF "A way of making Europe" through projects PID2021-125627OB-C31, PID2019-109522GB-C5[1:4], PID2019-107061GB-C64, PID2019-110689RB-100 and the Centre of Excellence "Severo Ochoa" Instituto de Astrof\'isica de Andaluc\'ia (grant CEX2021-001131-S funded by MCIN/AEI/10.13039/501100011033) and "Mar\'ia de Maeztu" awards to the Instituto de Astrof\'isica de Andaluc\'ia (SEV-2017-0709) and Institut de Ci\`encies de l'Espai (CEX2020-001058-M). Data were partly collected with the robotic 40-cm telescope ASH2 at the SPACEOBS observatory (San Pedro de Atacama, Chile) and the T90 telescope at the Sierra Nevada Observatory (SNO), both operated by the Instituto de Astrof\'isica de Andaluc\'ia (IAA, CSIC). The SIMBAD database\footnote{\url{http://simbad.u-strasbg.fr/simbad/}}, hosted at the CDS, Strasbourg, France, was used in this research. This research has made use of NASA's Astrophysics Data System Bibliographic Services\footnote{\url{http://adsabs.harvard.edu/}}. This work has made use of data from the European Space Agency (ESA) mission {\it Gaia} (\url{https://www.cosmos.esa.int/gaia}), processed by the {\it Gaia} Data Processing and Analysis Consortium (DPAC, \url{https://www.cosmos.esa.int/web/gaia/dpac/consortium}). Funding for the DPAC has been provided by national institutions, in particular the institutions participating in the {\it Gaia} Multilateral Agreement. This work has made use of observations collected at the European Organization for Astronomical Research in the Southern Hemisphere under ESO programmes: 072.C-0488(E), 183.C-0437(A), 198.C-0838(A), 0104.C-0863(A), 077.C-0364(E), 191.C-0873(B), 191.C-0873(D), 191.C-0873(A), 191.C-0873(E), 191.C-0873(F), 1102.C-0339(A), 106.21PJ.001, and 106.21PJ.002. The analysis was carried out using the programming language \texttt{Python3} (\url{https://www.python.org/}) Version 3.7.6 \citep{10.5555/1593511}, and the accompanying software packages: \texttt{Numpy} (\url{https://numpy.org/}) Version 1.18.1 \citep{harris2020array}, \texttt{Scipy} (\url{https://www.scipy.org/scipylib/}) Version 1.4.1 \citep{2020SciPy-NMeth}, and \texttt{Matplotlib} (\url{https://matplotlib.org/}) Version 3.1.3 \citep{Hunter:2007}.
\end{acknowledgements}

\bibliographystyle{bibtex/aa} 
\bibliography{bibtex/paper} 

\begin{appendix}

\section{Additional signals for GJ\,674}
\label{apdx:GJ674_add_signals}
We detected an additional signal within the photometric periodogram for GJ\,674 (Fig.~\ref{img:GJ674_photometry_GLS}) that coincides with the planets orbital period. This signals origin is unknown as it is distinct from the rotation period, which is itself distinctly visible. The peak also appears in both the nightly averaged and the full set of photometric observations. At the present sampling of slightly less than four nightly-averaged observations per period, this peak could be revealing a signal from a real phenomenon on the star with a periodicity the same of that of the orbit of the planet.
We consider it unlikely, that this surface effect could be the cause behind the strong signal we see in the RV periodogram however, which would invalidate the classification as a planet. If the unknown effect on the stellar surface was strong enough to cause an additional RV signal much stronger than the stellar rotation, we would expect its nightly averaged photometric trace to also be at least as strong as that of the stellar rotation.
Instead, the rotation signal still has FAP < 0.1\% in the nightly averaged photometry, whereas the 4.6-day peak has FAP > 50\%. One possible explanation could be that this signal might be an indication of star-planet-interaction. In that case, the RV curve would be composed of the continuous Keplerian signal and a trace of the activity enhanced SPI spot on the stellar surface appearing similar to a transit light curve. As the SPI signature would then be only transient, the GLS power could be significantly reduced compared to a quasi-periodic rotation signal.
\par
As we explain in Sect.~\ref{sec:analysis}, we started all our parameter determinations blind. This revealed the presence of another spurious peak in the posterior for GJ\,674 at an RV semi-amplitude of 7\,m\,s$^{-1}$. This peak has no correspondence within the periodogram and is also unstable when attempting to recover it using MCMC and narrow priors, shifting and disappearing at random. While it can be isolated as a second peak besides the primary, corresponding to a circular orbit at the same period, this also correlates to nearly double the jitter values for both seasons. In combination this appears to be a spurious signal within the log-likelihood space and not a true, planetary signal. As shown in Sect.~\ref{subsec:glsGrids}, a true signal of this amplitude would be reliably recovered. The 7\,m\,s$^{-1}$ is also not recovered by the \texttt{juliet} nested-sampling analysis.

\section{Additional Signals for Ross\,128}
\label{apdx:Ross128_add_signals}
From our first analysis of the Ross\,128 HARPS data set, we initially found three peaks in the planetary orbital period posterior at 9.86, 9.89, and 10.1 days. The 10.1-day period was unstable and is comparable in behavior to the 7\,m\,s$^{-1}$ signal for GJ\,674. The two remaining signals for Ross\,128 at 9.86\,d and 9.89\,d are stable and could be individually separated by prescribing sufficiently narrow priors (top-right and two bottom panels in Fig.~\ref{img:Ross128_peak_zoom} respectively). Both peaks were previously detected by \citetads{2018A&A...613A..25B}, who subsequently reported only on the combined maximum of the two peaks. They also report on the presence of "different posterior maxima" where they used an importance sampling estimator to reject any but the primary (bimodal) maximum, comprised of the 9.86\,d and 9.89\,d signals, and reinitialized the MCMC walkers. The 10.1\,d period is likely an example of these secondary peaks and further matches the yearly alias of the 9.86\,d signal. As can be seen in Fig.~\ref{img:new_RedDots_data}, the observations are taken in batches with an approximately yearly cadence, which is reflected in the window function (Fig.~\ref{img:Ross128_wFunc}) which shows a peak around 360\,d. Therefore, seeing a yearly alias is not unexpected and gives a first indication that the 9.86\,d signal is the true signal instead of the 9.89\,d one, which does not show a yearly alias.\par

While \citetads{2018A&A...613A..25B} only present results for the combined maximum peak, we attempted to identify and separate the two peaks. Our results show that, using MCMC, the relative strength between the 9.86\,d and 9.89\,d periods is highly variable depending on the choice of any of the priors, though the 9.86\,d signal always dominates. We take this as another indication that 9.86\,d is the true period. When separating the peaks by choosing very narrow, but still uniform, priors on the planet's orbital period (see Table~\ref{tab:priors1planet}) the complete orbital solutions retrieved are slightly different in other parameters as well as the period. This is most notable in the eccentricity, which is $0.21^{+0.09}_{-0.10}$ for the 9.86\,d case, but $0.28^{+0.09}_{-0.10}$ for the 9.89\,d case. Both values for the eccentricity are significantly higher than the previous values reported by \citetads{2018A&A...613A..25B} though just within the combined uncertainty intervals. Overall, the 9.86\,d peak we observe is closer to the period \citetads{2018A&A...613A..25B} computed from the combined maximum, while the second peak at a period of 9.89\,d is closer in amplitude to the results of \citepads{2018A&A...613A..25B}.\par

The 10.1\,d peak is a spurious signal coincident with the one year alias of the 9.86\,d signal and the 9.89\,d signal is consistently weaker than the 9.86\,d one. This lead us to consider the 9.86\,d signal as the true planetary orbital period. To verify this, we further analyzed the three peaks using the \texttt{juliet} nested-sampling code for verification. Using \texttt{juliet} and a one eccentric planet + GP model, neither of the two side peaks (with periods 9.89\,d and 10.1\,d) were recovered and the results show only a single peak at 9.86\,d. This is likely because the nested sampling algorithm excludes the other two side peaks during the contraction of the sampling space due to their lower evidence. We take this as verification that the correct solution is 9.86\,d as the \texttt{juliet} nested sampling results otherwise agree with our MCMC inference from the narrowed prior around 9.86\,d presented in Table~\ref{tab:inferredElements}.\par
A further examination of the residuals from Ross\,128 in Fig.~\ref{img:GLS_new} shows evidence for an additional signal with a period of approximately 4\,days. Previously, \citetads{2019A&A...623A..72K} found marginal evidence from \textit{Gaia} DR2 proper motion anomalies for a second planet around Ross\,128, though at a separation of 1 -- 10\,au and with roughly one Saturn mass, which does not match our tentative signal. The GLS periodogram of the H$\alpha$ indicator, output by \texttt{serval}, in Fig.~\ref{img:Ross128_Halpha_GLS} also shows a peak at approximately 4\,days. This suggests that this signal could be related to stellar activity rather than a second planet. The photometric periodogram does not show this signal. Nonetheless, treating the signal as if it were a planet would correspond to a 50\,cm\,s$^{-1}$ amplitude, 4.03\,d planet with an eccentricity near zero while the known planet's orbital parameters remain the same.

\section{Detailed comparisons of model evidence values}
\label{apdx:lnZ_comparison}
In this section we discuss in detail the Bayesian model evidence values $\ln\mathcal{Z}$ obtained from \texttt{juliet} during the model fits in Sects. \ref{subsubsec:GJ832} -- \ref{subsubsec:Ross128}. All individual $\ln\mathcal{Z}$ values are given in Table \ref{tab:modelEvidence} for the models including 0 -- 2 planets on circular or eccentric orbits, no GP, and GP with SHO, dSHO, or QP kernels.

\paragraph{\textbf{GJ\,832}}
The $\ln\mathcal{Z}$ values show a strong preference for models comprising one planet and a GP ($\Delta\ln\mathcal{Z} > 5$) as well as a preference of the SHO kernel over dSHO ($\Delta\ln\mathcal{Z} > 5$). There is no significant preference between SHO and QP kernels ($\Delta\ln\mathcal{Z} < 5$) or between circular and eccentric planet models ($\Delta\ln\mathcal{Z} \lesssim 4$).\par
The planetary parameters obtained by nested sampling for the eccentric, one planet models are in full agreement with our results obtained with MCMC. The results from the circular models are consistent with each other. The only significant difference between the different GP kernels that became apparent is that, as stated in Sect. \ref{subsubsec:GJ832}, the value of the SHO GP period falls between the full photometric rotation period (Table~\ref{tab:Params_known}) and half the period. The dSHO kernel, in this case, resulted in a period compatible with the photometric value. However, the posterior of the fractional amplitude parameter $f$ linearly increases towards values above unity. This indicates a strong preference for the half period component included in the dSHO kernel, putting the kernel choice into question. The period of the QP kernel results in double the rotation period and high harmonic complexity $\Gamma \sim 5$. We performed an extended discussion on the use of GP periods in the determination of stellar rotation periods in Sect.~\ref{subsec:juliet_vs_mcmc}.

\paragraph{\textbf{GJ\,674}}
The model evidences show a strong preference for the one eccentric planet + GP-dSHO model with $\Delta\ln\mathcal{Z} = 7$ compared to the next best choice, the also eccentric, one planet + GP-QP model. Circular orbits and models with zero or two planets are strongly disfavored ($\Delta\ln\mathcal{Z} > 65$).\par
The planetary orbital solutions for all models are consistent within the uncertainty with each corresponding other and, for the eccentric models, with the MCMC results from Table~\ref{tab:inferredElements}. As we see for GJ\,832, the only significant difference between the parameters inferred by the different eccentric models is the GP period. The SHO kernel results in a well constrained period that is neither the photometric stellar rotation period or half period, similar to the case of GJ\,832, though this time better compatible with the half period within one sigma. The dSHO kernel results in a GP period consistent with the rotation period at 36.6\,d but, also similar to the GJ\,832 results, with a fractional amplitude $f$ tending towards values much greater than one. This again strongly favors the half period component of 18\,d, which is consistent with the SHO kernels result and puts the kernel choice into question, although it is preferred by the evidence ($\Delta\ln\mathcal{Z} = 10$). The QP kernel performs better in this regard, giving a GP period consistent with the photometric period while at a significant harmonic complexity ($\Gamma \sim 3$). However, the QP kernel is disfavored by the Bayesian evidence at $\Delta\ln\mathcal{Z} = 7$ compared to the dSHO one. We performed an extended discussion on the use of GP periods in the determination of stellar rotation periods in Sect.~\ref{subsec:juliet_vs_mcmc}.

\paragraph{\textbf{Ross\,128}}
In Table~\ref{tab:modelEvidence}, we show that for Ross\,128, using the full HARPS data set, the one planet models are preferred at $\Delta\ln\mathcal{Z} > 17$, except for the two planet + GP-SHO model which has comparable evidence (see Appendix~\ref{apdx:Ross128_add_signals}). The one planet SHO models are preferred over the dSHO ones at $\Delta\ln\mathcal{Z} \sim 7$ and the dSHO kernel is also preferred over QP at $\Delta\ln\mathcal{Z} \sim 5$ while the eccentric and circular one planet + GP-SHO models are near identical ($\Delta\ln\mathcal{Z} < 1$), showing no preference. The orbital solutions, as we saw for GJ\,832 and GJ\,674, are consistent between the models that differ only in the GP kernel.\par

When comparing the models that differ only in the choice of GP kernel, we find a similar picture than for GJ\,832 and GJ\,674 before. The SHO GP period falls between the literature values for the stellar rotation period of 101-123\,d in Table~\ref{tab:Params_known} and half of the rotation periods. Unlike the other two planetary systems however, for Ross\,128, a GP period of 76\,d interpreted as a half rotation period closely matches the periods listed in the Carmenes input catalog \citepads{2019A&A...621A.126D} of 163\,d and \citepads{2016A&A...595A..12S} of 165\,d.\par

We compared this to the RedDots subset which has a regular, almost nightly cadence and should therefore allow a much better constraint of the correlated noise. Using only this data the GP period both attempts to converge towards the planetary period and simultaneously becomes less constrained. This presents itself as a weakening peak at shorter periods atop a wide plateau extending to long periods (see Sect.~\ref{subsec:rotation_discuss} and \citeads{2023arXiv230413381S}). When a more restrictive uniform prior is used to preclude convergence towards the planetary orbital period, a long period of 269\,d is recovered instead (see Table.~\ref{tab:inferredElements}). A possible explanation is that Ross\,128, according to \citetads{2019A&A...628L...1I}, would have entered an activity minimum during the time of the RedDots observations, while the previous majority of datapoints were taken around a maximum. This would also match the overall significantly increased evidence ($\Delta\ln\mathcal{Z} > 350$) we see for this subset. The temporal aspect is supported by the long-term photometric periodicity also reported by \citetads{2016A&A...595A..12S} at 4.1\,yrs, which is shorter than the 5.4\,yrs from \citetads{2019A&A...628L...1I} but still in general agreement. This interpretation is supported by the nearly tripled damping timescale estimate (see Table~\ref{tab:inferredElements}) and the change from well constrained to nearly unconstrained with a posterior distribution extending into hundreds of days. This indicates a temporally stable signal within the RedDots data that is atypical for activity.\par

When using the dSHO kernel for either the full dataset or restricted to the archival data, the GP period also converges to a 150\,d period, however with a fractional amplitude that again heavily favors the half period, similar to our results for GJ\,832 and GJ\,674. The GP period resulting from the QP kernel differs between the eccentric and circular models. For the circular planet model, the GP period behaves similarly to the eccentric SHO case when restricted to the RedDots subset, combined with the dSHO case. We see a strong posterior peak at the planetary orbital period, as with the restricted SHO model, that also correlates with a very low harmonic complexity ($\Gamma < 1$). A second, weaker but much wider, maximum in the GP period posterior around 150\,d mirrors the dSHO model, at a high harmonic complexity of $\Gamma \sim 5$. For the one eccentric planet + GP-QP case, the GP period also forms a wide maximum around 150\,d, but without the short period peak and also at a high harmonic complexity of $\Gamma \sim 5$. We performed an extended discussion on the use of GP periods in the determination of stellar rotation periods in Sect.~\ref{subsec:juliet_vs_mcmc}.\par

The behavior of the QP GP kernel for circular orbits further points toward an explanation for the slight preference of the planet-free SHO model by the Bayesian evidence shown in Table~\ref{tab:modelEvidence} when restricted to the RedDots data set over the circular planet + SHO model that is the preference within the full data set. The low harmonic complexity for the short period QP solution mirrors the long damping timescale mentioned above for the SHO kernel and indicates a stable, regular signal. It appears that for the RedDots subset in particular the correlated noise is weak enough, and the GP flexible enough, to mask the Keplerian signal on top of the remaining activity (see also \citetads{2022MNRAS.515.5251N}). Since the Bayesian evidence is penalized based on model complexity, the ability of the GP to adequately model both signals without the added Keplerian results in the apparent negation of the planet hypothesis. Since the combined data set as well as the archival one strongly supports the one planet hypothesis, we do not doubt its validity and instead take this as a warning to not overly rely only on the Bayesian evidence as an indicator when a GP is part of the model. Visual inspection of the posteriors and careful consideration of the GP hyperparameters appears essential to avoid discarding a bona-fide exoplanet when faced with low levels of activity. This suspicion of the Bayesian evidence in this context further motivated our discussion of $\ln\mathcal{Z}$ as a blind metric in Sect. \ref{subsec:lnZ_discussion} and the detailed investigation of the significance of the eccentricity in Sect.~\ref{subsec:ecc_validation}.

\section{Investigating the stripes}
\label{apdx:striping}
We investigated the horizontally striped structures uncovered in Sect.~\ref{subsec:glsGrids}, present in all three systems and in all parameters at varying strengths. We examined whether they are random structures from the noise, due to the window function, or have a systematic origin. First, we calculated the difference between the recovered amplitude and the injected signal. Then, we set a cutoff value to convert the grid into a binary map of high and low recovery deviations to highlight the stripes. This is shown in Fig.~\ref{img:GLS_offsets_vs_noise_Ross128} for the Ross\,128 residuals with a cutoff at 0.03\,m\,s${-1}$. The stripe locations within the injection grid were compared to the periodogram of the pure noise, before a signal is injected. The regions of denser stripes coincide with the regions of residual power in the periodogram while the lowest power periods show a lack of stripes. This indicates that the stripes are due to structure within the noise. \par
To distinguish windowing effects from statistical ones, the noise was re-created by drawing white noise realizations according to the RV uncertainties of the original data. Any signals that are due to the specific noise realization would thereby appear at different periods, while windowing related effects would remain unchanged. To run the required number of grids within a reasonable time frame, the grid resolution was lowered to $\delta_K=0.1$\,m\,s$^{-1}$ for $K \in \left[1.0, 3.0\right]$\,m\,s$^{-1}$. Since the striping appears at specific periods with little dependence on the injected amplitude this is not expected to affect the results. To verify, the $e = 0.0$ and $e = 0.5$ grids were run at full and reduced resolution and their structure and statistics were compared. We found that the striping is unaffected in location, strength and distribution by the lower resolution. The individual realizations again contained a striped structure but at different periods and with different deviation strengths. Drawing a total of 100 realizations shows that these patterns, while always present, have statistical origin because their strength and location in period is different between the realizations. Averaging over all 100 realizations reduced the difference between true injected semi-amplitude and recovered semi-amplitude to less than 2\,cm\,s$^{-1}$, one-tenth the original (Fig.~\ref{img:GLS_white_Noise_avg_Ross128}). This confirms the statistical nature and excludes windowing as a source. \par
The statistical properties of the realizations are as expected. The spread is always within three standard deviations of the original (Fig.~\ref{img:GLS_detect_Ross128}, middle panel). As expected, 68\% of the realizations are within one standard deviation.\par
\vspace{1cm}
The Appendices \ref{apdx:priors} to \ref{apdx:add_figures} with tables \ref{tab:priors1planet} and \ref{tab:priors2planet}, \ref{tab:comparison_parameters} and \ref{tab:comparison_priors}, \ref{tab:comparison_posteriors_inst_params} to \ref{tab:comparison_posteriors_P2_params}, and Figures \ref{img:GJ832_wFunc} to \ref{img:very_long_detect_grids} are only available online at \url{https://zenodo.org/records/13626863}.


\refstepcounter{section}\label{apdx:priors}

\refstepcounter{table}\label{tab:priors1planet}
\refstepcounter{table}\label{tab:priors2planet}

\refstepcounter{section}\label{apdx:comparison_priors}

\refstepcounter{table}\label{tab:comparison_parameters}
\refstepcounter{table}\label{tab:comparison_priors}

\refstepcounter{section}\label{apdx:comparison_posteriors}

\refstepcounter{table}\label{tab:comparison_posteriors_inst_params}
\refstepcounter{table}\label{tab:comparison_posteriors_baseGP_params}
\refstepcounter{table}\label{tab:comparison_posteriors_seasonGP_params}
\refstepcounter{table}\label{tab:comparison_posteriors_P1_params}
\refstepcounter{table}\label{tab:comparison_posteriors_P2_params}

\refstepcounter{section}\label{apdx:add_figures}

\refstepcounter{figure}\label{img:GJ832_wFunc}
\refstepcounter{figure}\label{img:GJ674_wFunc}
\refstepcounter{figure}\label{img:Ross128_wFunc}
\refstepcounter{figure}\label{img:Ross128_Carmenes_resid}
\refstepcounter{figure}\label{img:GJ832_full_corner}
\refstepcounter{figure}\label{img:GJ674_full_corner}
\refstepcounter{figure}\label{img:Ross128_full_corner_1P}
\refstepcounter{figure}\label{img:Ross128_full_corner_1P_carmenes}
\refstepcounter{figure}\label{img:Ross128_Halpha_GLS}
\refstepcounter{figure}\label{img:Ross128_peak_zoom}
\refstepcounter{figure}\label{img:GJ674_photometry}
\refstepcounter{figure}\label{img:Ross128_photometry}
\refstepcounter{figure}\label{img:GJ674_photometry_GLS}
\refstepcounter{figure}\label{img:Ross128_photometry_GLS}
\refstepcounter{figure}\label{img:GLS_detect_GJ832_boundary}
\refstepcounter{figure}\label{img:GLS_detect_GJ674_boundary}
\refstepcounter{figure}\label{img:GLS_offsets_vs_noise_Ross128}
\refstepcounter{figure}\label{img:GLS_white_Noise_avg_Ross128}
\refstepcounter{figure}\label{img:very_long_detect_grids}

\end{appendix}

\end{document}